\documentclass[superscriptaddress,onecolumn,aps]{revtex4-1}
\usepackage{amssymb,amsmath,amsthm,wrapfig,esint,afterpage,marginnote,afterpage,lipsum,color,tabularx}
\usepackage{times}
\usepackage{bm}
\usepackage{lineno}
\usepackage{siunitx}
\usepackage{graphicx}
\usepackage{lipsum}
\usepackage{color,sidecap,fancyhdr}
\usepackage[dvipsnames]{xcolor}
\usepackage{mathpazo}
\usepackage{textcomp}
\usepackage{nicefrac}
\usepackage{xfrac}
\usepackage[sort&compress]{natbib}
\usepackage{dcolumn,bm}
\usepackage{hyperref}
\usepackage{cleveref}
\usepackage{mathtools}
\usepackage{epsfig}
\usepackage{epstopdf}
\usepackage[authormarkuptext=name]{changes}
\usepackage{changes}
\definechangesauthor[color=blue, name={AY}]{AY}

\newcommand{\Bra}[1]{\left[#1\right]}

\begin{document}
	
	\title{Why A Large Scale Mode Can Be Essential For Understanding Intracellular Actin Waves}
	
	\author{Carsten Beta}
	\affiliation{Institute of Physics and Astronomy, University of Potsdam, 14476 Potsdam, Germany}
	
	\author{Nir S. Gov}
	\affiliation{Department of Chemical and Biological Physics, Weizmann Institute of Science, Rehovot 76100, Israel}
	
	\author{Arik Yochelis}\email{yochelis@bgu.ac.il}
	\affiliation{Department of Solar Energy and Environmental Physics, Blaustein Institutes for Desert Research (BIDR), Ben-Gurion University of the Negev, Sede Boqer Campus, Midreshet Ben-Gurion 8499000, Israel}
	\affiliation{Department of Physics, Ben-Gurion University of the Negev, Be'er Sheva 8410501, Israel}
	
	\received{\today}
	
	\begin{abstract}
		During the last decade, intracellular actin waves have attracted much attention due to their essential role in various cellular functions, ranging from motility to cytokinesis. Experimental methods have advanced significantly and can capture the dynamics of actin waves over a large range of spatio-temporal scales. However, the corresponding coarse-grained theory mostly avoids the full complexity of this multi-scale phenomenon. In this perspective, we focus on a minimal continuum model of activator-inhibitor type and highlight the qualitative role of mass-conservation, which is typically overlooked. Specifically, our interest is to connect between the mathematical mechanisms of pattern formation in the presence of a large-scale mode, due to mass-conservation, and distinct behaviors of actin waves.
	\end{abstract}
	
	
	\maketitle

	\section{Introduction}

Biological pattern formation refers to the emergence of complex spatiotemporal variations in living systems that are typically far from thermodynamic equilibrium~\cite{Cross1993,pismen06,CG09}. Even though these systems can differ in composition and scales, they share many similarities and generic phenomena that are observed in a wide variety of natural settings, such as stationary periodic patterns of pigments on animal skins, spiral waves in biological cells and cardiac arrhythmia, or swarming phenomena in bacterial colonies and in flocks of birds or fish. The theoretical study of biological pattern formation can be roughly divided into two time periods, namely: (\textit{i}) The second half of the twentieth century, following the seminal works by Turing on morphogenesis~\cite{tu52} and by Hodgkin and Huxley (HH) on action potentials in the giant squid axon~\cite{hodgkin1952quantitative}, and (\textit{ii}) the beginning of the twenty-first century, where an ever increasing amount of quantitative biological data provided the basis for more detailed mechanistic models of biological systems.

During the first period, theoretical studies were largely limited to a few prototypical reaction--diffusion (RD) or activator--inhibitor (AI) models~\cite{murray2001mathematical}, such as the FitzHugh--Nagumo (FHN)~\cite{fitzhugh1961impulses,nagumo1962active}, Gierer--Meinhardt~\cite{gm72}, and Keller--Segel equations~\cite{keller1971model}. Based on the relative simplicity of these models, e.g. the FHN model as compared to the HH equations, and their relation with models of inanimate matter, e.g. the Swift-Hohenberg model of thermal fluid convection~\cite{swift1977hydrodynamic} and the Gray--Scott model of chemical reactions~\cite{gray1983autocatalytic,gray1984autocatalytic,pearson1993complex}, several pattern formation methodologies have been advanced~\cite{Cross1993}, such as weakly nonlinear and singular-perturbation methods. These models provided deep insights into universal aspects of pattern formation phenomena and generic relations to applications were substantiated, such as frequency locking and spiral waves in the cardiac system. The second time period has manifested a gradual shift of research interests towards specific detailed biological and medical systems~\cite{ksp98,ksp08}, including, for example, micron-scale intracellular waves, the development of tissues and organs, sound discrimination in the auditory system, and pathologies such as cancer metastasis. In particular, systems in these contexts are generally described by elaborate, system-specific models that are less amenable to mathematical analysis than earlier toy models of pattern-formation. 

Consequently, despite the common pattern-formation thread that connects these different biological and medical applications, it became difficult to navigate through the vast number of distinct models and approaches, particularly in cases where technical jargon makes it difficult to adopt cross-disciplinary integration between different communities including biophysics, computational biology, mathematical biology, biological chemistry, dynamical systems, and numerical analysis. Even though the formulation of complete models for biological systems is currently unrealistic, uncovering partial mechanisms that drive pattern-formation phenomena remains of utmost importance for understanding functional aspects of living systems and for developing technological and medical applications, such as drugs or implants. Moreover, mechanistic studies of pattern forming systems are also fertile sources of new mathematical questions that advance the development of analytical and numerical methods~\cite{tyson1988singular,GoS2002,Hoyle,pismen06,CH82,Collet90etalbook,kuz04,mei2000,ps08,SU17}, which, in turn, contribute new insights into the original applications.

In this perspective, we will focus on intracellular actin waves (IAW), a topic that recently gained much interest not only in the biological context but also as an inspiring showcase of active matter. More specifically, we are interested in IAW that are affected by a large scale mode~---~a situation that arises due to conservation of actin monomers (over the time-scale of the IAW phenomenon).
{We note that phenomena such as $\text{Ca}^{2+}$ waves are, in general, beyond the scope of this perspective as they involve the transport of ions between the cell interior and the extracellular space (which acts as an infinite reservoir)~\cite{dupont2014modeling,falcke_deterministic_2003,kulawiak2019active}, unless conservation can be accounted for~\cite{radszuweit2013intracellular}}. Moreover, we emphasize that we aim to provide a perspective and not a comprehensive review, as such reviews are already available, e.g.,~\cite{onsum2009calling,Allard2013,Blanchoin2014,inagaki2017actin,beta2017intracellular,halatek2018self,deneke2018chemical}.

The perspective is organized as following. In Section~\ref{sec:cell_waves}, we introduce the rich phenomenology of IAW and the modeling aspects that are associated with mass conservation. Then, we present in Section~\ref{sec:physchem} the theoretical aspects of a large scale mode in the context of physicochemical settings and also indicate its significance to IAW {as a blueprint for conservation of actin monomers on the times scales at which many actin dynamics processes operate}. Finally, we discuss in Section~\ref{sec:disc} why incorporation of mass-conservation is {a plausible} qualitative step in unfolding the {robustness of} IAW mechanisms, and in Section~\ref{sec:concl} we conclude by emphasizing the theoretical strategies for modeling and control of wave persistence as a potential roadmap toward applications in synthetic biology.

\section{Inracellular actin waves}\label{sec:cell_waves}

The functions of many cells are tied to their ability to dynamically change their shape, mostly via the spatiotemporal organization of their actin cytoskeleton. Examples of this include diverse cell types, such as human neutrophils, fish keratocytes, or the social amoeba {\it Dictyostelium discoideum}. Among the most prominent dynamical patterns in the actin cytoskeleton are IAW that have attracted much of attention over the past decade~\cite{Allard2013}. These waves are assumed to play a role in several essential cellular functions, among them cell locomotion, cytokinesis, and phagocytic uptake of extracellular matter. Many competing models at different levels of complexity have been developed to describe cortical actin waves, mostly relying on coupled nonlinear AI equations.
Even though intracellular actin waves involve a large number of interacting molecular species as well as multiple local and global interactions,  prototypical AI models have been shown to capture many features of the overall dynamics.
However, important effects due to mass conservation constraints have been hitherto largely neglected.

\subsection{Phenomenology from experiments}

\label{sec:exp}

Actin waves are characterized by propagating of cytoskeletal regions that are enriched in filamentous actin and actin-related proteins. Depending on the cell type, IAW may differ in their biochemical composition and dynamics, including different wave morphologies and propagation speeds. One of the earliest examples of IAW was reported from cultured neurons that show propagation of fin-like actin-filled membrane protrusions along their axon~\cite{ruthel_actin-dependent_1998}. They were found to depend on actin polymerization and have been associated with neural polarization~\cite{toriyama_shootin1:_2006,tomba2017geometrical}.
Similar fin-like actin wave also emerge in non-neural cell types when cultured on thin fibers~\cite{guetta-terrier_protrusive_2015}. Also adherent cells that are attached to flat substrates may display traveling wave-like protrusions of their cell shape. They are particularly prominent when moving laterally along the cell border, such as in mouse embryonic fibroblasts~\cite{dobereiner_lateral_2006} or at the leading edge of fish keratocytes~\cite{barnhart2017adhesion}.
\begin{figure}[tp]
	\begin{center}
		\includegraphics[width=0.9\textwidth]{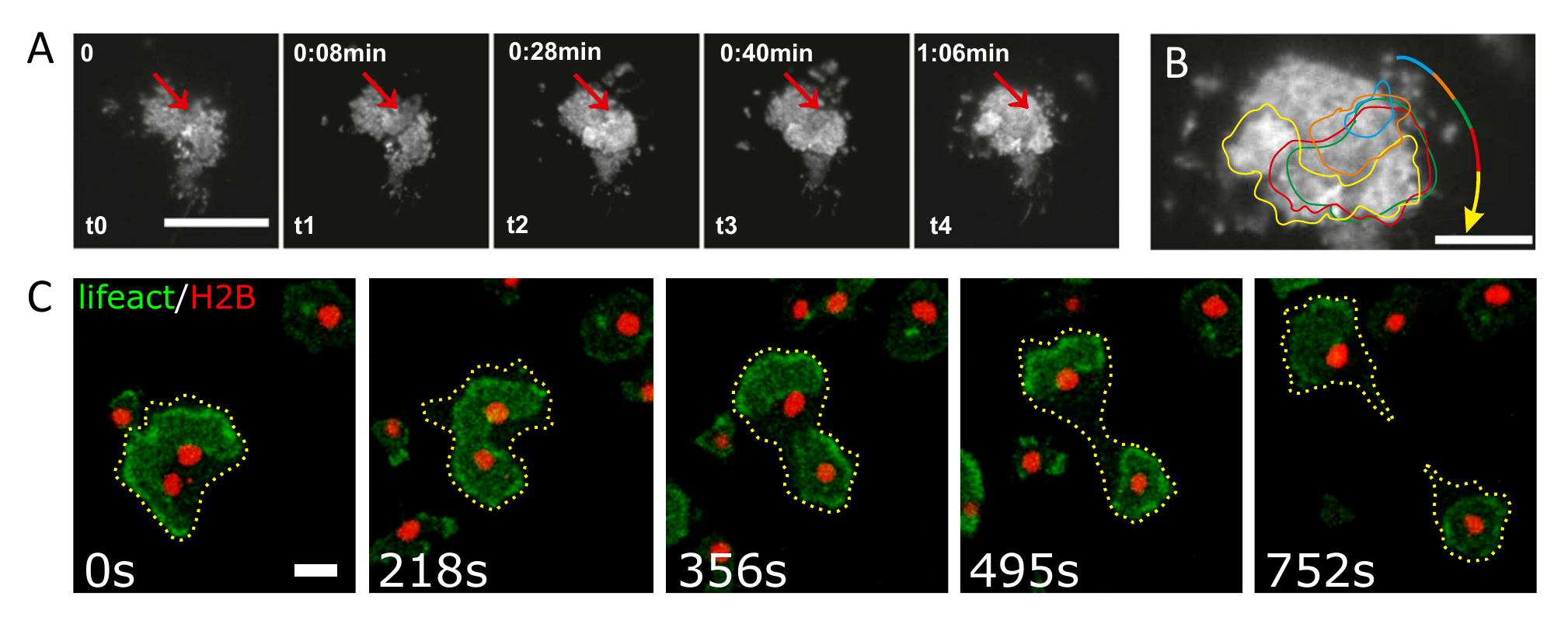}
	\end{center}
	\caption{Examples of intracellular actin waves.
		(A)~Actin wave nucleation and propagation in a migrating immature dendritic cell.
		Red arrows indicate origin of the wave, scale bar 20~$\mu$m. (B)~Overlay of contours of representative actin waves shown in (A) during propagation.
		(C)~Wave-mediated binary cytofission in a {\it Dictyostelium discoideum} cell, scale bar 10~$\mu$m.
		An actin wave in a cell with two nuclei becomes unstable and splits into two independent segments that move in opposite directions and induce a cytofission event.
		(A)~and (B)~are reproduced from~\cite{stankevicins_deterministic_2020}, (C)~is reproduced from~\cite{flemming_how_2020}. Copyright 2020 National Academy of Sciences.}\label{fig:cytofission}
\end{figure}

Traveling actin waves have also been observed at the dorsal and ventral sides of adherent cells. In neutrophils, small dynamic wave fragments emerge that organize cell polarity and leading edge formation~\cite{weiner_actin-based_2007}. Larger ring-shaped waves were found to travel across the substrate-attached bottom membrane of {\it D. discoideum} cells~\cite{gerisch_mobile_2004}. They enclose a region that is structurally distinct from the cortical area outside the actin ring~\cite{schroth-diez_propagating_2009,jasnin2019architecture} and their dynamics often shows rotating spiral cores and mutual annihilation upon collision~\cite{gerhardt_actin_2014,miao_altering_2017} but they could not be initiated by external receptor stimuli~\cite{gerhardt_signaling_2014}.
While understanding the rich dynamics of IAW is challenging on its own right, there are prominent applications and functional properties that stimulate further studies of IAW in different contexts: 
\begin{description}
	\item[Motility.] Recently, clear evidence was reported that actin waves directly impact the motility of immune cells, see Fig.~\ref{fig:cytofission}A,B.
	In particular, dendritic cells that move in an amoeboid fashion and search the human body for pathogens, display a random walk pattern that can switch between diffusive and persistent states of motion, a direct consequence of the intracellular actin wave dynamics~\cite{stankevicins_deterministic_2020};\\
	
	\item[Cell division.] In oocytes and embryonic cells of frog and echinoderms, excitable waves of Rho activity in conjunction with actin polymerization waves were observed shortly after anaphase onset, providing an explanation for the sensitivity of the cell cortex to signals generated by the mitotic spindle~\cite{bement_activator-inhibitor_2015}. Similarly, in metaphase mast cells, concentric target and spiral waves of Cdc42 and of the F-BAR protein FBP17 were found to set the site of cell division in a size-dependent manner~\cite{xiao_mitotic_2017}. IAW can also act as the force-generating element that directly drives the division process in a contractile ring-independent form of cytofission. This was observed in {\it D. discoideum} cells beyond a critical size, where waves that collide with the cell border not only induce strong deformations of the cell shape but also trigger the division into smaller daughter cells~---~a cell cycle-independent form of wave-mediated cytofission, see Fig.~\ref{fig:cytofission}C~\cite{flemming_how_2020};\\
	\item[Macropinocytosis.] While functional roles in phagocytosis and motility have been proposed~\cite{gerisch_self-organizing_2009,miao2019wave}, recent genetic studies suggest a relation to macropinocytosis~\cite{veltman_plasma_2016}. This is supported by similarities between the basal actin waves and circular dorsal ruffles (CDR)~\cite{bernitt_dynamics_2015,bernitt2017fronts}. The latter also adopt a ring-shaped structure but meander across the apical membrane, where they induce membrane ruffles that were related to the formation of macropinocytic cups~\cite{buccione_foot_2004};\\
	\item[Cancer.] Macropinocytosis has been also identified as an important mechanism of nutrient uptake in tumor cells~\cite{Commisso2013}. Specifically, inability of cells to undergo efficient macropinocytosis, e.g., thorugh disordered IAW behavior or suppressed activity via pinning of IAW to cell boundaries~\cite{bernitt2017fronts}, has been associated with cancerous phenotypes~\cite{Itoh2012,Hoon2012}.
\end{description}
Despite intense studies over the past years, the molecular details of IAW mechanisms remain largely unclear and most likely vary between different cell types.


\subsection{Modeling approaches of actin waves}\label{sec:modeling}

Following the numerous experimental observations of IAW in different cell types and during different cellular functions, many model equations have been proposed to describe this phenomenon. Here, we will briefly describe the main types and features of theoretical models that have been employed while referring the reader to~\cite{ryan2012review,Allard2013,sept2014modeling,beta2017intracellular,halatek2018self} for more details.

The growth of the cortical actin network within IAW is a complex dynamical process that involves many components that perform a coordinated set of functions, giving rise to the formation of a three-dimensional network of actin filaments, that propagates along the cell membrane. This process involves the activation of actin associated proteins some of them membrane bound, that initiate the nucleation of actin polymerization, branching of actin filaments, cross-linking and bundling, as well as severing and depolymerization. There are very few theoretical models that attempt to give a molecular-scale description of the IAW phenomenon where all of these processes are described. One example for such a model that describes the waves at the scale of the individual actin filaments is given in \cite{carlsson2010dendritic}. While providing detailed pictures of the actin network, it is difficult and time-consuming to use such modeling to extract understanding regarding the large-scale dynamics of the IAW. Such modeling efforts could in the future include more molecular components \cite{huber2008growing,khamviwath2013continuum}, on larger length and time scales, and provide a platform for theoretical advances in this field, that works in conjunction with filament-scale experimental data \cite{jasnin2019architecture}. 

Since the IAW have widths in the range of hundreds of nanometers, propagate over tens of microns and persist over hours, it is natural to describe them using coarse-grained models that avoid prescribing the molecular-scale details of the actin network. As will be shown, many of these models agree with some qualitative or even quantitative features of the observed IAW in cells. It is therefore difficult at present to reach a clear consensus regarding the validity of these models. Comparisons in between such models is complicated since they often include different components and it is not clear if and which of those components play a fundamental role in the emergence of IAW or can be neglected otherwise.

Among the coarse-grained models we can find a small class of models that contain biophysical elements, such as forces and/or the membrane shape, which play a key role in the mechanism that drives the propagation of the IAW. One example is well demonstrated by Gholami \textit{et al.}~\cite{gholami2012membrane}, who show that the dynamics of the actin polymerization/depolymerization drive the oscillatory propagation of waves. When actin filaments polymerize against the cell membrane, they exert a protrusive pressure on the membrane, which pushes the membrane forward and the actin network backwards. The interplay between the rate of actin polymerization and the rate at which the actin filaments are cross-linked into a stable gel-like network, determine if the cortical actin is stable or whether it exhibits an unstable oscillatory regime.

Another group of biophysics-based models contain curved membrane proteins that nucleate the cortical actin polymerization \cite{gov2006dynamics,shlomovitz2007membrane,veksler2009calcium,peleg2011propagating,naoz2020cell}. In these models, the curved proteins flow/adsorb to the membrane regions that have a curvature similar to their intrinsic shape, and their concentration is therefore affected by the membrane deformations that are induced by the forces exerted by the actin cytoskeleton. These forces include the protrusive force of actin polymerization, as well as contractile forces due to myosin-II mediated contractility. 
Recently, also models combining an RD kinetics coupled to mechanical properties through the impact of curved actin nucleators and/or membrane shape and tension were introduced \cite{barnhart2017adhesion,wu2018membrane}. Other models combine the RD dynamics with a physical effect, such that the directed or random lateral actin polymerization can physically drive the treadmilling of the IAW components along the membrane \cite{katsuno2015actin}. The advantage of the biophysical class of models is that they can naturally account for the observed effects of physical parameters on the IAW, such as membrane tension \cite{gholami2012membrane,wu2018membrane} or the contractile forces of myosin-II motors \cite{chen2009three}.

In many cases however, RD equations that include both positive and negative feedback loops, are sufficient to demonstrate the formation of propagating waves, fronts, or localized pulses. These models exhibit different levels of complexity and different numbers of components.
{In the simplest cases, generic activator-inhibitor models of FHN-type were proposed.
	In particular, they were used together with a local-excitation, global-inhibition (LEGI) mechanism to account for the response of the receptor-mediated signaling pathway and the downstream actin cytoskeleton to external cues~\cite{beta_bistable_2008,xiong_cells_2010,devreotes_excitable_2017}.
	Other basic RD-models} describe the actin dynamics, including the monomeric and filamentous species, and one form of an actin activator, using the filamentous actin itself as a source of negative \cite{carlsson2010dendritic} or positive \cite{sambeth2001autocatalytic} feedback. More complex models include different numbers of activators of actin polymerization, inhibitors, and their complex network of interactions \cite{Dreher2014,miao2019wave,cao2019plasticity}. Yet, in general, RD equations are not subjected to conservation of mass although often some of the components are conserved, for example when they represent two different forms of the same protein \cite{Mata2013}. In other cases, the actin is conserved as it is converted from monomeric to filamentous forms and back, see for example \cite{Wasnik2014,bernitt2017fronts}. In what follows, we address the qualitative role of conservation, which is reflected by the existence of a large scale mode, on the dynamics of IAW, using as much as possible generic principles, i.e., extracting conclusions that are qualitatively independent of the specific molecular details that are included in the model.

\section{Actin dynamics as a constrained continuous medium: Implications and applications}\label{sec:physchem}

\begin{figure}[tp]
	\centering
	\includegraphics[width=0.9\textwidth]{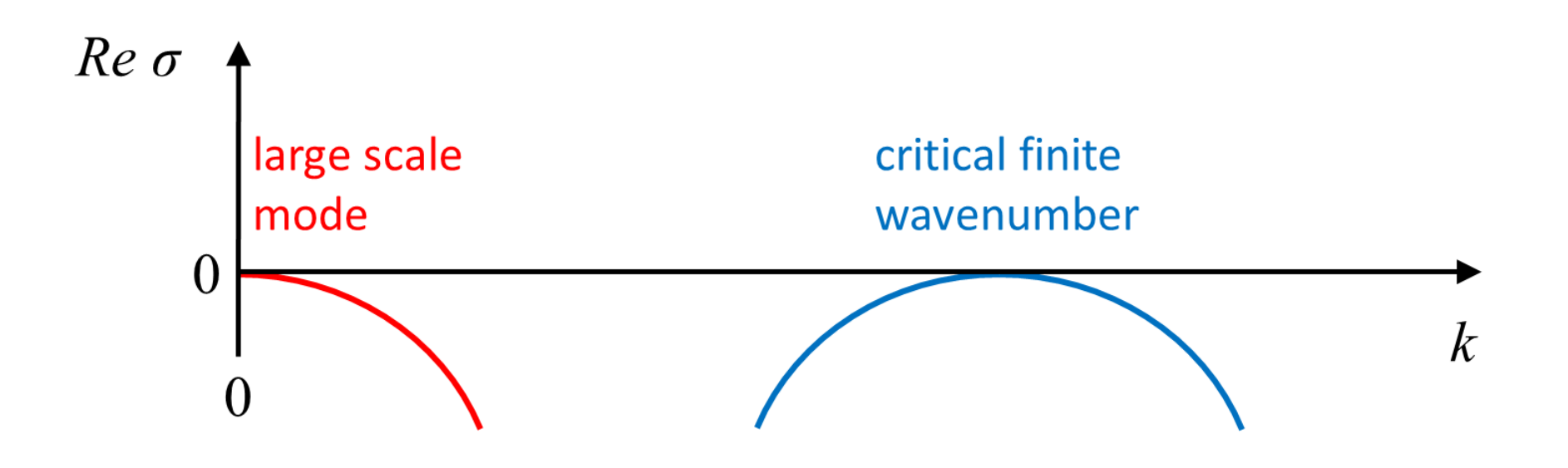}
	\caption{Schematic representation of a dispersion relation obtained from infinitesimal periodic perturbations, proportional to $\exp{(\sigma t+ikx)}$, about a uniform steady state; $Re \, [\sigma]$ is the perturbation growth rate and $k$ its wavenumber. The right-hand part of the dispersion relation represents the onset of an instability of a finite wavenumber type (often also referred to as Turing instability), while the left-hand part reflects a conserved quantity and stays always neutral{; both parts are model independent.} The curves may connect as typically occurs in systems such as~\eqref{eq:consrv} or belong to different curves such as for~\eqref{eq:pRD}. The imaginary part of $\sigma$ corresponds to stationary nonuniform patterns if zero, and otherwise describes time-dependent solutions.} \label{fig:disp}
\end{figure}

The phenomenology of dissipative waves can be demonstrated through a dynamical systems approach via prototypical models, such as FHN. 
{As summarized above, many variants of such activator-inhibitor models have been used to describe different aspects of cytoskeletal dynamics and in particular the formation of actin waves.}
Although these are heuristic models, they are analytically tractable and thus allow for fundamental insights into spatiotemporal behavior, which cannot be obtained through the analysis of more realistic multi-variable equation sets. Propagating waves are traditionally classified into three universality classes~\cite{Cross1993,murray2001mathematical,pismen06,meron2015book}: 
\begin{description}
	\item [Oscillatory dynamics,] which represent traveling waves that develop  via a Hopf instability of a uniform steady state;\\
	\item [Excitability,] corresponding to supra--threshold solitary waves (pulses) that propagate on top of a linearly stable uniform steady state;\\
	\item [Bistability,] which describes traveling domain walls or fronts, i.e., an interface that connects to linearly stable uniform steady states.
\end{description}
While the mathematical mechanisms are distinct, the emerging patterns can show similar characteristics, for example all classes may display the formation of spiral waves~\cite{pismen06,meron2015book}. Consequently, comparisons to experimental observations can often only be qualitative, making insights uncertain. Moreover, it is not always clear whether the simplified models comprise the minimal set of qualitative ingredients, e.g., interactions (local vs. non-local), spatial coupling, essential degrees of freedom and feedback loops, finite domain effects, or existence of conserved observable(s). In a broader context, IAW can be classified as AI type media~\cite{jilkine2011comparison,Allard2013}, although unlike the typical RD media the number of actin monomers is conserved over the time scales of wave dynamics. As such, mass conservation is an inherent constraint of the modeling framework~\cite{jilkine2011comparison,bernitt2017fronts,halatek2018rethinking,halatek2018self}, which is {generically} reflected by coupling to a large scale mode in the dispersion relation, as illustrated in Fig.~\ref{fig:disp}.

\subsection{Conservation in physicochemical systems}

It is convenient to first consider total conservation of an observable, described by the continuity equation
\begin{equation}\label{eq:cont}
\frac{\partial u}{\partial t}=\nabla \cdot \Bra{M(u)\, \nabla \frac{\delta F(u)}{\delta u}},
\end{equation}
where $u$ is a scalar observable, $M$ is a mobility function, and $F$ is a free energy. If the free energy contains an intrinsic length scale, like in the phase field crystal model or wetting, stationary periodic and localized patterns may emerge~\cite{golovin2003self,ziebert2004pattern,golovin2004faceting,weliwita2011skew,robbins2012modeling,thiele2019first,barker2018turing,hilder2018modulating}. The mutual aspect is coupling between the large scale mode ($k=0$), which is {model independent and remains always neutral due to conservation} (also known as the Goldstone mode), and the pattern forming instability of finite wavenumber (Turing) type~\cite{tribelsky1996short}, as shown in Fig.~\ref{fig:disp}. The impact of the conserved quantity has been analyzed mostly via a weakly nonlinear-reduction to a set of two amplitude equations: One is the complex Ginzburg--Landau equation for the finite wave-length mode, while the other is for the neutral large scale mode~\cite{matthews2000pattern,cox2003instability,cox2004envelope,shiwa2005hydrodynamic,dawes2008localized,ohnogi2008instability,golovin2009feedback,huang2010phase,kanevsky2011patterns,thiele2013localized}. Both super-- and sub--critical bifurcation cases have been studied, and showed that indeed inclusion of the large scale mode may qualitatively change the nature of the solution in terms of organization and stability~\cite{knobloch2016localized,schneider2016turing} (and references therein). For example, in the absence of the large scale mode, spatially localized solutions form in coexistence with a periodic (Turing-type) solution and are organized in a vertical homoclinic snaking structure. In the presence of a large scale mode these solutions can also form outside the existence region of periodic solutions and only partially overlap, i.e., the homoclinic snaking structure becomes slanted~\cite{firth2007homoclinic,dawes2008localized}.

In fact, similar asymptotic intuition and analysis methods apply also if the observable has a velocity-like behavior (often Galilean invariance)~\cite{winterbottom2006pattern}, obeying the symmetry $x\to -x$ and $u\to -u$. Such behavior arises in systems that are being driven out of equilibrium, such as convection~\cite{matthews1993travelling,cox2000instability,cox2004swift,jacono2011magnetohydrodynamic}, propagation of flames~\cite{golovin2001traveling,golovin2003complex}, surface waves~\cite{tsimring1997localized,snezhko2006surface,pradenas2017slanted} and electro-diffusion in ion channels~\cite{kramer2002pattern,peter2006traveling}. In such cases leading order approximations show that the dynamics can still be enslaved to an oscillatory (Hopf) finite wavenumber mode and a large scale mode~\cite{coullet1985propagative,riecke1996solitary,ipsen2000finite,winterbottom2005oscillatory,hek2007pulses}. While many fundamental advances have been made in understanding the coupling between the complex Ginzburg--Landau equation and the large scale mode, e.g. in terms of stability of periodic and solitary waves in one space dimension and dynamics of spiral waves in two-dimensional systems, several pattern formation issues remain open~\cite{nepomnyashchy2016longwave}. Consequently, since over the time scales on which IAW occur the system is far from equilibrium, it is natural to assume that a large scale mode due to mass conservation alters the pattern formation mechanism, even without explicit flux conservation.

\subsection{Activator--inhibitor patterns with conservation}

In general, AI systems are modeled in a similar fashion as chemical reactions~\cite{koch1994biological,maini1997spatial,ksp98,ksp08,volpert2009reaction,murray2001mathematical,baker2008partial,deneke2018chemical}, which are not limited by supply of new substrates into the reactor
\begin{subequations}\label{eq:RD}
	\begin{eqnarray}
	\frac{\partial u}{\partial t}&=&f(u,v)+ D_u \nabla^2 u,\\
	\frac{\partial v}{\partial t}&=&g(u,v)+ D_v \nabla^2 v, 
	\end{eqnarray}
\end{subequations}
where $u$ is the activator that typically contains an autocatalytic or enzymatic term and a diffusion constant $D_u$, and $v$ is an inhibitor that diffuses with a diffusion constant $D_v$, where typically $D_v \gg D_u$. 
As intracellular processes often take place on very different time scales, effective mass conservation may arise, for example, in cases where protein synthesis and/or degradation occurs much slower than a particular biochemical reaction of interest.
Conservation in AI models is associated with a local conservation of mass 
\begin{equation}\label{eq:consrv}
\int_\Omega \Bra{u(\mathbf{x},t)+v(\mathbf{x},t)} \text{d}\mathbf{x}=\text{constant},
\end{equation}
where $\Omega$ is the physical domain, or by writing in~\eqref{eq:RD}
\begin{equation}\label{eq:mass}
g(u,v)=-f(u,v).
\end{equation}
{Linear stability analysis about uniform solutions leads to dispersion relations that contain the persistent neutral (large scale) mode, as shown in Fig.~\ref{fig:disp}. As in the case of Eq.~\ref{eq:cont}, also Eq.~\ref{eq:mass} supports multiplicity of uniform solutions since $u$ depends on an arbitrarily chosen value of $v$ (or vise versa), and this degenerate degree of freedom appears as the $k=0$ mode. This constraint plays effectively the role of a chemical potential. However, in the pattern forming case, where an additional bifurcation is present (e.g. a Turing bifurcation), the dispersion relations may contain both the neutral mode at $k=0$ and another at a finite wavenumber.} In this formulation, models for cell polarity~\cite{otsuji2007mass} and molecular motors~\cite{yochelis2015self} had inspired several mathematical works in the context of existence and emergence of stationary~\cite{ishihara2007transient,morita2010stability,chern2018asymptotic,kuwamura2018dynamics,ei2020spike}
and time-dependent~\cite{sakamoto2013hopf,yochelis2016reaction,zmurchok2017application} patterns.

However, as has been described in Section~\ref{sec:modeling}, IAW are multi-component processes and involve a large number activators and inhibitors. Moreover, in such an AI network not all the components obey conservation~\cite{goryachev2008dynamics,jilkine2011comparison}, namely, to Eqs.~\ref{eq:RD} and~\ref{eq:mass} can be added at least one additional non-conserved observable $w$,
\begin{subequations}\label{eq:pRD}
	\begin{eqnarray}
	\frac{\partial u}{\partial t}&=&f(u,v,w)+ D_u \nabla^2 u,\\
	\frac{\partial v}{\partial t}&=&-f(u,v,w)+ D_v \nabla^2 v,\\
	\frac{\partial w}{\partial t}&=&h(u,v,w)+ D_w \nabla^2 w,
	\end{eqnarray}
\end{subequations}
where $h$ can be either a linear or a nonlinear functional and essentially does not have to include transport of $w$ via diffusion; these details are naturally determined by the characteristics of the biological system. Equation~\ref{eq:pRD} thus reflects only a partial conservation and has been employed to study the emergence of IAW in the context of CDR~\cite{bernitt2017fronts}, where a variety of complex behaviors have been observed experimentally, ranging from distinct types of propagating fronts to spatiotemporal chaotic spiral waves.

\section{Discussion and Example}\label{sec:disc}

The complex pattern formation exhibited by CDR raises the question about the modeling strategy, specifically, with respect to the minimal set of equations and the necessity of a conserved quantity. As has already been indicated in Section~\ref{sec:modeling}, there are many ways to model IAW but all of them are prone to subjective interpretations.

In the absence of a clear physical intuition, since IAW are far from equilibrium phenomena, dynamical systems offer
an efficient platform for creating an appropriate qualitative framework. More specifically, the study of bifurcations may provide the minimal qualitative set of constraints, exactly as phase-transitions allow us to classify many types of physical phenomema. On the other hand, bifurcation analysis can also be a tedious task as there may be many local and global bifurcations that coexist in a given parameter range (as an example we refer the reader to a systematic extension of excitable media by Champneys \textit{et al.}~\cite{champneys2007shil}). Nevertheless, utilizing recent advances in nonlinear perturbations~\cite{yochelis2008generation,Mata2013} and numerical path continuation methods~\cite{auto,sherratt2012numerical,uecker2014pde2path,bindel14,coco} it might be possible to navigate between coexisting bifurcations and a multiplicity of emerging stable and unstable solutions~\cite{yochelis2008generation,yochelis2015origin}. Next, we turn to conservation and ask whether it may prescribe a fundamental and robust qualitative change, as compared to typical {local} RD modeling in the absence of conserved quantities. To exemplify this case, we exploit a reduced CDR model (of Eqs.~\ref{eq:pRD} type), which has been used to examine solitary wave collisions in the context of IAW ~\cite{yochelis2020excitable}. In the reduced CDR model the conserved AI system of Eqs.~\ref{eq:pRD} is replaced by the conservation of the actin monomers, as they are converted from the monomeric to the filamentous form (and back) which the IAW propagates.
\begin{figure}[tp]
	\centering
	{\setlength{\unitlength}{0.1\textwidth}
		\begin{picture}(10,3.3)
		\put(0.5,0){\includegraphics[width=0.3\textwidth]{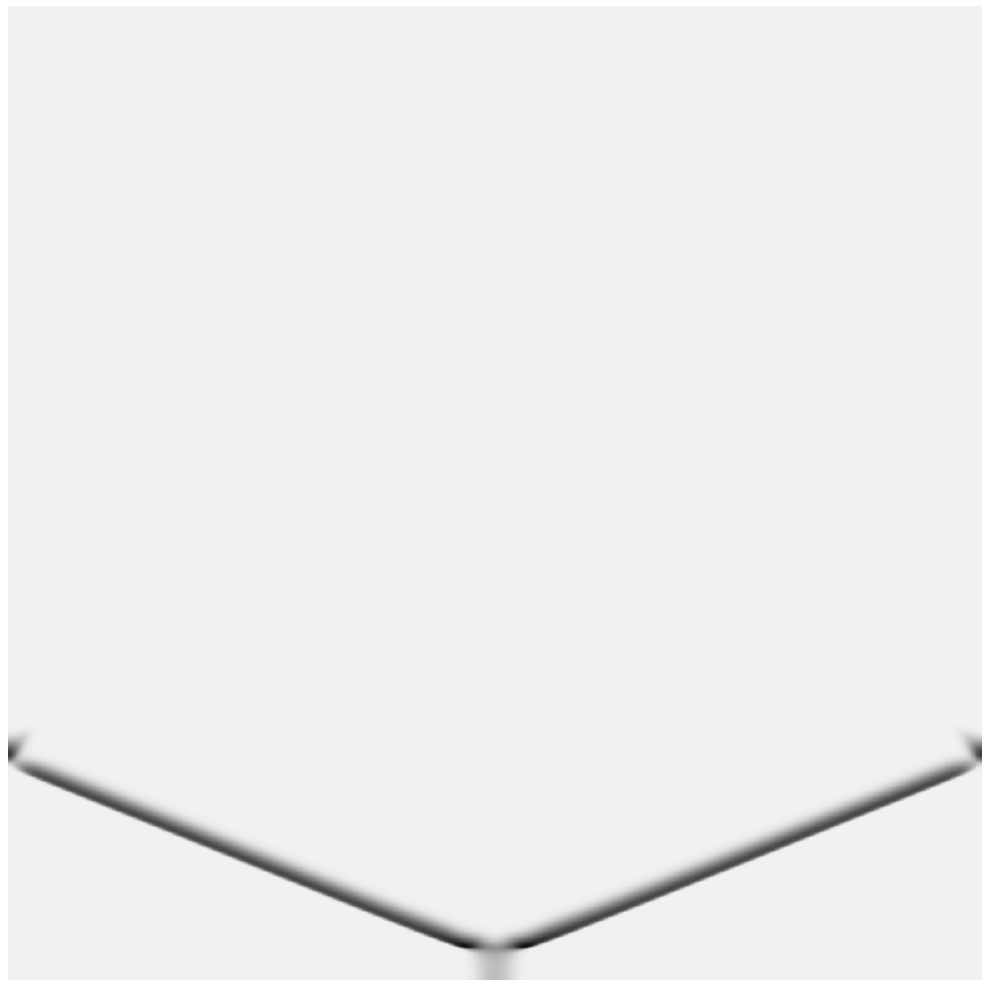}}
		\put(3.55,0){\includegraphics[width=0.3\textwidth]{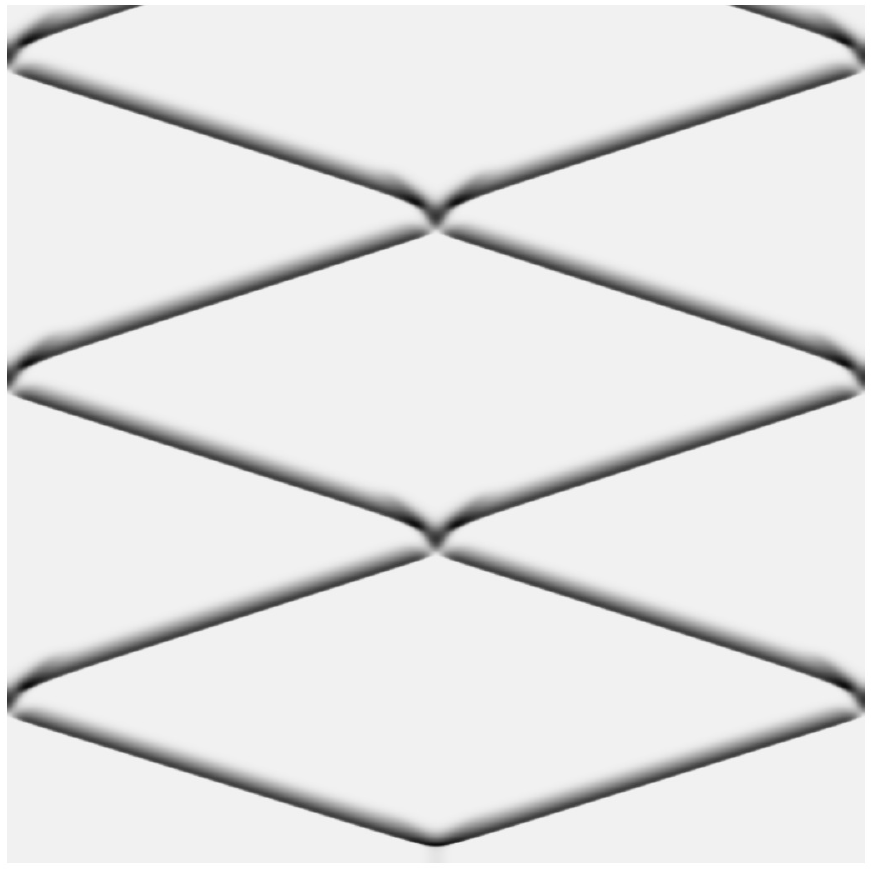}}
		\put(6.6,0){\includegraphics[width=0.3\textwidth]{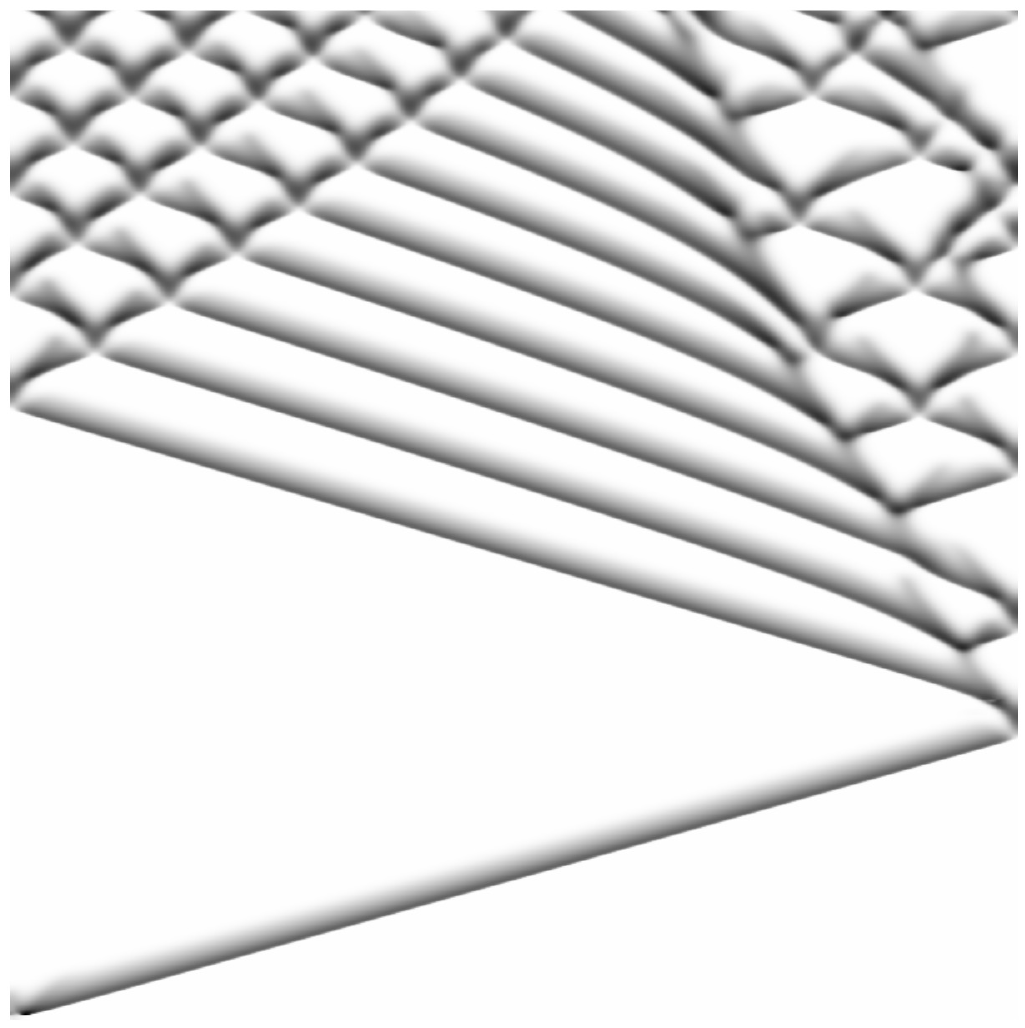}}
		\put(0.6,2.7){$t$}
		\put(0.6,2.5){$\uparrow$}
		\put(0.65,2.4){$\rightarrow$}
		\put(0.9,2.35){$x$}
		\end{picture}}
	\caption{Space--time plots showing (from left to right) annihilation, reflection/crossover, and ``birth'' of new pulses following collision ({a behavior that resembles backfiring}), respectively, as obtained from direct numerical integration of the minimal CDR model equations~\cite{yochelis2020excitable} that have the same structure as Eqs.~\ref{eq:pRD}. No--flux boundary conditions were used. From left to right the amount of actin monomers increases (see details in~\cite{yochelis2020excitable}). The dark shaded color indicates higher values of filamentous actin in the IAW. Reprinted figure with permission from~\cite{yochelis2020excitable} Copyright 2020 by the American Physical Society.} \label{fig:solitons}
\end{figure}
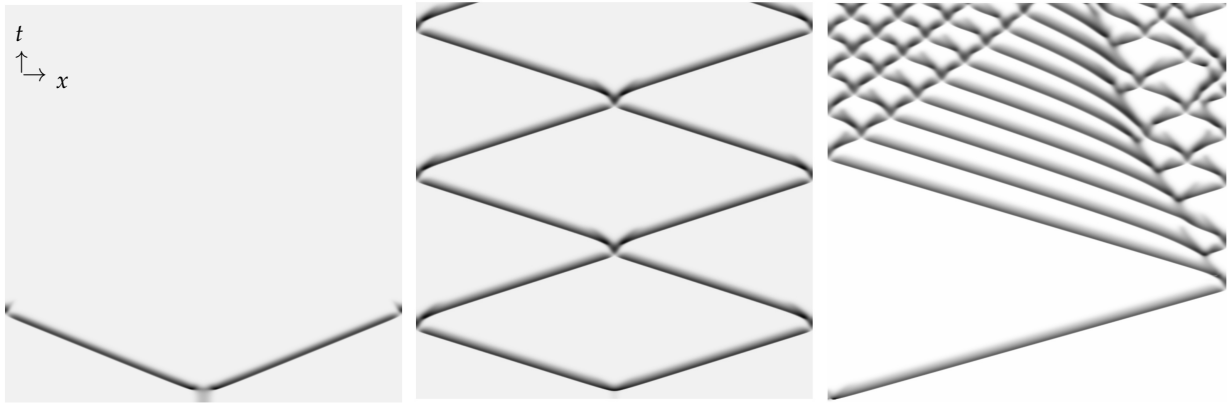

Observation of solitary waves dates back to John S. Russell (1834), yet only after the work of~\citet{zabusky1965interaction} were solitary waves distinguished by their collision properties~\cite{scott1973soliton,scott1975electrophysics}: \emph{solitons} if after collision of two pulses, two pulses emerge (particle-like identity) and \emph{dissipative solitons} or \emph{excitable pulses} if they are annihilated. Solitons are often being discussed in the context of conservative media, which mathematically means exploiting the integrable nature of the governing model equations~\cite{knobloch2015spatial,knobloch2016localized} while, excitable pulses often arise in RD type systems. Although collisions of solitons may involve high spatiotemporal complexity, the outcome of two colliding solitons remains unchanged (i.e., elastic particle-like dynamics)~\cite{ablowitz2012nonlinear,santiago1997dissolution}. On the other hand, the annihilation of excitable pulses after the collision is recognized as paramount for electrophysiological function, i.e., it would be impossible to maintain directionality, and thus rhythmic behavior, under the reflection of action potentials~\cite{alonso2016nonlinear}. 
{Importantly, collision of pulses implies merging of the pulses in space, i.e, through the formation of a \textit{collision zone}. This behaviour is distinct from \textit{interaction} between excitable pulses that is due to repulsion and can exhibit dynamics that may resemble a solitonic behavior~\cite{petrov1994excitability,argentina_colliding_1997}.}{Also more complex scattering scenarios have been observed in generic RD models such as, for example, the Gray-Scott model~\cite{nishiura_scattering_2003,nishiura_dynamic_2003}.}
{Note that there exists a vast literature on the latter topic that we do not intend to review in total here. Taken together,} the distinction between solitons and excitable pulses is important for numerous applications. 

\begin{figure}[tp]
	\centering
	A\includegraphics[width=0.45\textwidth]{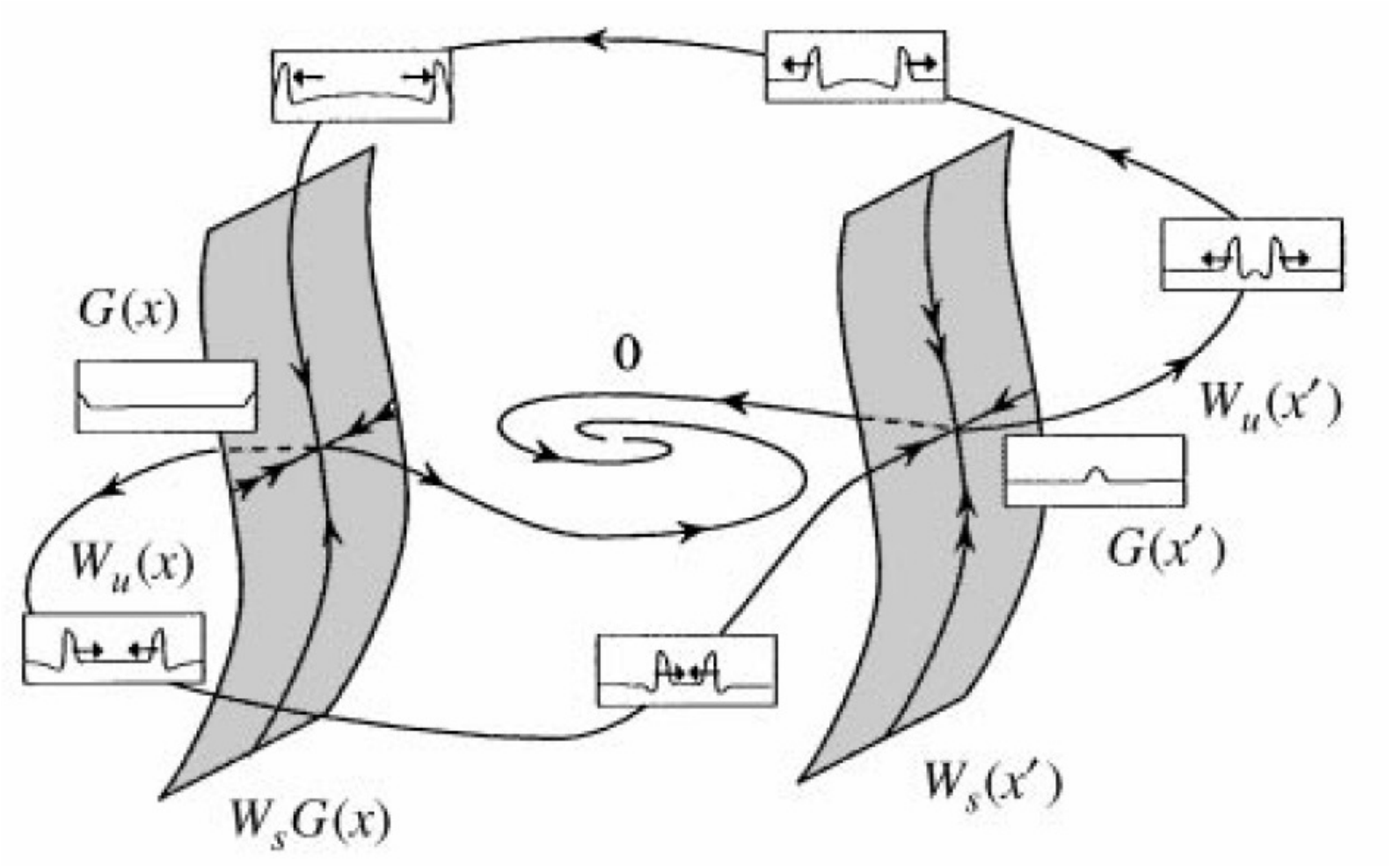}\quad
	B\includegraphics[width=0.38\textwidth]{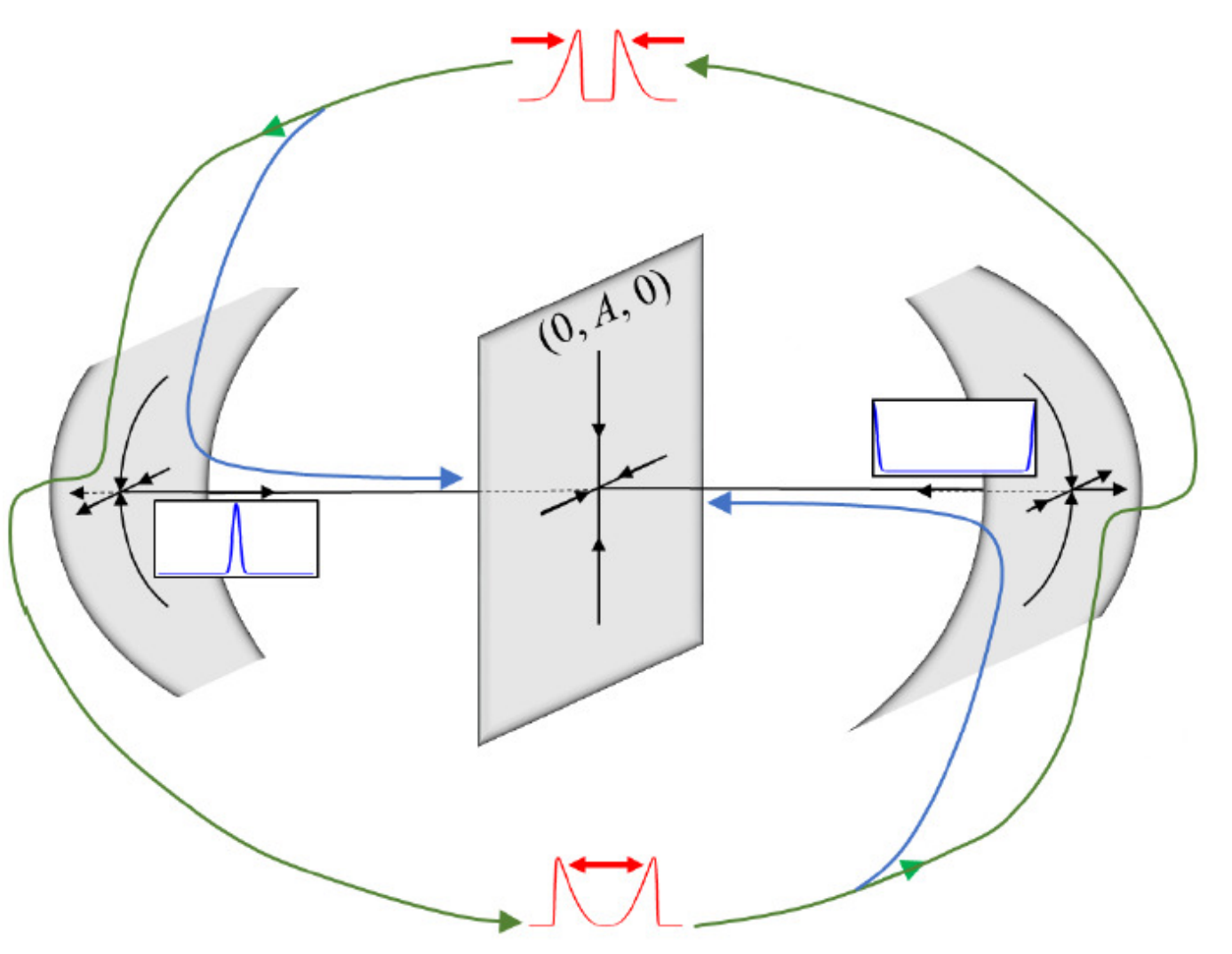}
	\caption{Excitable solitons, geometric analysis of the dynamics during collision of two pulses. (A) FitzHugh--Nagumo model and (B) an reaction--diffusion model with mass-conservation, of Eqs.~\ref{eq:pRD} type. (A) Reprinted from Publication~\cite{argentina2000head}, with permission from Elsevier and (B) from~\cite{yochelis2020excitable}, Copyright 2020 by the American Physical Society.} \label{fig:cAI}
\end{figure}

Yochelis \textit{at el.}~\cite{yochelis2020excitable} showed that the minimal IAW model, in the class of Eqs.~\ref{eq:pRD}, may indeed support rich {and robust} spatiotemporal dynamics following pulse collisions, {in contrast to IAW models which do not contain explicit mass conservation~\cite{whitelam2009transformation,kulawiak2019active,dreher2014spiral,miao2019wave,alonso2018modeling}: annihilation, reflection, and ``birth'' of new pulses after reflection, as shown in Fig.~\ref{fig:solitons}. In a broader RD context, where similar aspects have been also observed,} these dynamics do not require special properties, such as non-locality~\cite{krischer1994bifurcation,bar1994chemical,mimura1998collision,coombes2007exotic}, cross--diffusion~\cite{tsyganov2003quasisoliton}, and heterogeneity~\cite{bar1992solitary,nishiura2007dynamics,yuan2007heterogeneity}. Moreover, the phenomenon is robust and occurs over a wide range of parameter {values}, whereas for a typical RD model without mass conservation, such as FHN, somewhat similar dynamics of propagating pulses are observed only in a narrow range near the onset of an oscillatory Hopf bifurcation about a uniform steady state~\cite{argentina2000head,bordyugov2008anomalous}. The distinction between the FHN model and a system of Eqs.~\ref{eq:pRD} type can be elaborated by geometrical intuition, since pulses are of large amplitude and thus cannot be unfolded using weakly nonlinear analysis such as in Section~\ref{sec:physchem}. Argentina \textit{et al.}~\cite{argentina2000head} showed that in the FHN model a manifold construction about the collision state of two pulses ("collision droplet", Fig.~\ref{fig:cAI}(A)) can explain why a Hopf bifurcation may impact the collision zone and thus generate crossover of pulses (soliton-like behavior). A similar geometric picture shows that mass-conservation in Eqs.~\ref{eq:pRD} changes the nature of the collision zone by addition of a generic two-dimensional neutral manifold (Fig.~\ref{fig:cAI}(B)), relating the pulse crossover behavior to a localized unstable mode and does not require any Hopf bifurcation of the uniform state~\cite{argentina2000head,bordyugov2008anomalous}. {In other words, for the colliding pulses to avoid annihilation, there has to be a mechanism for recovery-- a spontaneous re-growth of the fields after collision. In the FHN model~\cite{argentina2000head}, the proximity to the oscillatory onset can re-initiate the pulses. In the case of actin conservation, the colliding pulses first disintegrate the polymerized actin, thereby releasing a large local pool of monomers. If these monomers do not diffuse too fast, they are available to re-initiate the pulses by polymerization. For more details we refer the reader to~\cite{yochelis2020excitable}}. 

{Note that the nucleation of new pulses after collision should not be confused with the well-known scenario of backfiring, an instability that appears when a localized propagating pulse becomes unstable and splits into two new counterpropagating pulses that, upon collision, annihilate~\cite{argentina_back-firing_2004}.
	Backfiring has been observed in a wide range of model systems~\cite{bar1994chemical,zimmermann_pulse_1997,nishiura_spatio-temporal_2001}, and also in recent experiments of CO electrooxidation on Pt~\cite{bauer_dissipative_2015}.
	However, in contrast to backfiring, the nucleation of new pulses that we addressed here and that is shown in Fig.~\ref{fig:solitons} always requires a preceding collision event and thus has to be distinguished from the classical backfiring scenario.}

\section{Conclusions}\label{sec:concl}

The case of the reduced CDR model discussed above provides a {glimpse} to the profound impact of mass conservation on the dynamics. In conventional FHN-type AI models without mass conservation, colliding pulses typically annihilate upon collision. Here, a soliton-like crossover occurs only under special conditions, e.g. near a Hopf point, and thus requires fine-tuning of the parameters. In contrast, if mass conservation is taken into account, propagating pulses robustly exhibit rich collision scenarios over a wide range of parameters, including crossover and {formation of new pulses following collision}. Even though this has only been demonstrated for a simple toy model, the universal nature of the underlying bifurcations suggests that a similar behavior will be observed also in more detailed, high-dimensional models of IAW, provided that mass conservation is included, {e.g., for mechanochemical waves under conservation of calcium~\cite{radszuweit2013intracellular}}.

{The impact of mass conservation on pattern formation in biological systems has recently attracted increasing attention, in particular in the context of well-controlled, confined systems such as the bacterial Min protein oscillator~\cite{halatek2018rethinking}.
	However, many biological systems involve multiple components not all of which are conserved, so that the consequences of strict mass conservation as implied by Eqs.~\ref{eq:RD} and~\ref{eq:mass} are often relaxed and require a more general view.
	This is provided, in the simplest case, by adding a third dynamical variable to the system that is coupled to the conserved quantities but does not obey mass conservation itself, see Eqs.~\ref{eq:pRD}.
	It demonstrates that a large scale mode is the key feature that mass conservation introduces to the system and that triggers specific dynamical properties, such as soliton-like crossover of pulses and the collision-induced birth of new pulses in a wide range of parameters.}{ Eqs.~\ref{eq:pRD}, and its resulting dynamics, can serve as motivation for further future studies of the synthesis between classical AI models and models with complete mass-conservation (such as those used in the context of the Min and Par systems \cite{gessele2020geometric}).}

Similar to neural systems, where annihilation of colliding pulses is essential to maintain directionality of information transport, we conjecture that also in the case of IAW, the crossover of colliding pulses, which is favored due to the mass conservation constraint, plays an important functional role. This may be particularly true, when sustained wave activity is a key requirement for proper cell functions, as for example in cases where cell locomotion or nutrient uptake depend on IAW (see Section~\ref{sec:exp}).	For traditional excitable pulses that annihilate upon collision, wave activity is likely to get extinguished regularly, thus hampering cellular activities that rely on persistent IAW. In contrast, soliton-like crossover and collision-induced nucleation of new pulses that are robust properties of a mass-conserved system may ensure prolonged wave activity even in the absence of actively triggered pulse nucleation or local heterogeneities that may serve as pacemakers. Moreover, cells may also actively exploit shifts between parameter regimes of pulse annihilation and soliton-like behavior to control their level of IAW activity, as shown in Fig.~\ref{fig:solitons}.

Finally, the study of simplified models to elucidate generic properties of IAW patterns may also prove useful for the future design of synthetic cellular systems. A current focus of bottom-up approaches in synthetic biology is to introduce artificial cytoskeletal structures into membrane vesicles, thus assembling the essential building blocks of a primitive cell~\cite{siton2016toward,schwille_maxsynbio:_2018}. The logical next step along this line of research will be to endow the artificial cytoskeletal components with simple pattern forming properties that may ultimately serve as a basis for essential cellular functions, such as motility and cytokinesis. This requires a thorough understanding of the key properties that are necessary to reconstitute the desired wave patterns in a minimal model system. We thus expect that the understanding of the essential bifurcations and instabilities that govern the dynamics of IAW to provide a useful guideline for the future design of artificial cell cortices.

%


\begin{thebibliography}{182}%
	\makeatletter
	\providecommand \@ifxundefined [1]{%
		\@ifx{#1\undefined}
	}%
	\providecommand \@ifnum [1]{%
		\ifnum #1\expandafter \@firstoftwo
		\else \expandafter \@secondoftwo
		\fi
	}%
	\providecommand \@ifx [1]{%
		\ifx #1\expandafter \@firstoftwo
		\else \expandafter \@secondoftwo
		\fi
	}%
	\providecommand \natexlab [1]{#1}%
	\providecommand \enquote  [1]{``#1''}%
	\providecommand \bibnamefont  [1]{#1}%
	\providecommand \bibfnamefont [1]{#1}%
	\providecommand \citenamefont [1]{#1}%
	\providecommand \href@noop [0]{\@secondoftwo}%
	\providecommand \href [0]{\begingroup \@sanitize@url \@href}%
	\providecommand \@href[1]{\@@startlink{#1}\@@href}%
	\providecommand \@@href[1]{\endgroup#1\@@endlink}%
	\providecommand \@sanitize@url [0]{\catcode `\\12\catcode `\$12\catcode
		`\&12\catcode `\#12\catcode `\^12\catcode `\_12\catcode `\%12\relax}%
	\providecommand \@@startlink[1]{}%
	\providecommand \@@endlink[0]{}%
	\providecommand \url  [0]{\begingroup\@sanitize@url \@url }%
	\providecommand \@url [1]{\endgroup\@href {#1}{\urlprefix }}%
	\providecommand \urlprefix  [0]{URL }%
	\providecommand \Eprint [0]{\href }%
	\providecommand \doibase [0]{http://dx.doi.org/}%
	\providecommand \selectlanguage [0]{\@gobble}%
	\providecommand \bibinfo  [0]{\@secondoftwo}%
	\providecommand \bibfield  [0]{\@secondoftwo}%
	\providecommand \translation [1]{[#1]}%
	\providecommand \BibitemOpen [0]{}%
	\providecommand \bibitemStop [0]{}%
	\providecommand \bibitemNoStop [0]{.\EOS\space}%
	\providecommand \EOS [0]{\spacefactor3000\relax}%
	\providecommand \BibitemShut  [1]{\csname bibitem#1\endcsname}%
	\let\auto@bib@innerbib\@empty
	\bibitem [{\citenamefont {Cross}\ and\ \citenamefont
		{Hohenberg}(1993)}]{Cross1993}%
	\BibitemOpen
	\bibfield  {author} {\bibinfo {author} {\bibfnamefont {M.~C.}\ \bibnamefont
			{Cross}}\ and\ \bibinfo {author} {\bibfnamefont {P.~C.}\ \bibnamefont
			{Hohenberg}},\ }\href@noop {} {\bibfield  {journal} {\bibinfo  {journal}
			{Rev. Mod. Phys.}\ }\textbf {\bibinfo {volume} {65}},\ \bibinfo {pages} {851}
		(\bibinfo {year} {1993})}\BibitemShut {NoStop}%
	\bibitem [{\citenamefont {Pismen}(2006)}]{pismen06}%
	\BibitemOpen
	\bibfield  {author} {\bibinfo {author} {\bibfnamefont {L.}~\bibnamefont
			{Pismen}},\ }\href@noop {} {\emph {\bibinfo {title} {{Patterns and interfaces
					in dissipative dynamics}}}}\ (\bibinfo  {publisher} {{Springer}},\ \bibinfo
	{year} {2006})\BibitemShut {NoStop}%
	\bibitem [{\citenamefont {Cross}\ and\ \citenamefont {Greenside}(2009)}]{CG09}%
	\BibitemOpen
	\bibfield  {author} {\bibinfo {author} {\bibfnamefont {M.}~\bibnamefont
			{Cross}}\ and\ \bibinfo {author} {\bibfnamefont {H.}~\bibnamefont
			{Greenside}},\ }\href@noop {} {\emph {\bibinfo {title} {{Pattern Formation
					and Dynamics in Nonequilibrium Systems}}}}\ (\bibinfo  {publisher}
	{{Cambridge University Press}},\ \bibinfo {year} {2009})\BibitemShut
	{NoStop}%
	\bibitem [{\citenamefont {Turing}(1952)}]{tu52}%
	\BibitemOpen
	\bibfield  {author} {\bibinfo {author} {\bibfnamefont {A.}~\bibnamefont
			{Turing}},\ }\href@noop {} {\bibfield  {journal} {\bibinfo  {journal} {Phil.
				Trans. Roy. Soc. B}\ }\textbf {\bibinfo {volume} {237}},\ \bibinfo {pages}
		{37} (\bibinfo {year} {1952})}\BibitemShut {NoStop}%
	\bibitem [{\citenamefont {Hodgkin}\ and\ \citenamefont
		{Huxley}(1952)}]{hodgkin1952quantitative}%
	\BibitemOpen
	\bibfield  {author} {\bibinfo {author} {\bibfnamefont {A.~L.}\ \bibnamefont
			{Hodgkin}}\ and\ \bibinfo {author} {\bibfnamefont {A.~F.}\ \bibnamefont
			{Huxley}},\ }\href@noop {} {\bibfield  {journal} {\bibinfo  {journal} {The
				Journal of Physiology}\ }\textbf {\bibinfo {volume} {117}},\ \bibinfo {pages}
		{500} (\bibinfo {year} {1952})}\BibitemShut {NoStop}%
	\bibitem [{\citenamefont {Murray}(2001)}]{murray2001mathematical}%
	\BibitemOpen
	\bibfield  {author} {\bibinfo {author} {\bibfnamefont {J.~D.}\ \bibnamefont
			{Murray}},\ }\href@noop {} {\emph {\bibinfo {title} {Mathematical Biology. II
				Spatial Models and Biomedical Applications $\{$Interdisciplinary Applied
				Mathematics V. 18$\}$}}}\ (\bibinfo  {publisher} {Springer-Verlag New York
		Incorporated},\ \bibinfo {year} {2001})\BibitemShut {NoStop}%
	\bibitem [{\citenamefont {FitzHugh}(1961)}]{fitzhugh1961impulses}%
	\BibitemOpen
	\bibfield  {author} {\bibinfo {author} {\bibfnamefont {R.}~\bibnamefont
			{FitzHugh}},\ }\href@noop {} {\bibfield  {journal} {\bibinfo  {journal}
			{Biophysical journal}\ }\textbf {\bibinfo {volume} {1}},\ \bibinfo {pages}
		{445} (\bibinfo {year} {1961})}\BibitemShut {NoStop}%
	\bibitem [{\citenamefont {Nagumo}\ \emph {et~al.}(1962)\citenamefont {Nagumo},
		\citenamefont {Arimoto},\ and\ \citenamefont {Yoshizawa}}]{nagumo1962active}%
	\BibitemOpen
	\bibfield  {author} {\bibinfo {author} {\bibfnamefont {J.}~\bibnamefont
			{Nagumo}}, \bibinfo {author} {\bibfnamefont {S.}~\bibnamefont {Arimoto}}, \
		and\ \bibinfo {author} {\bibfnamefont {S.}~\bibnamefont {Yoshizawa}},\
	}\href@noop {} {\bibfield  {journal} {\bibinfo  {journal} {Proceedings of the
				IRE}\ }\textbf {\bibinfo {volume} {50}},\ \bibinfo {pages} {2061} (\bibinfo
		{year} {1962})}\BibitemShut {NoStop}%
	\bibitem [{\citenamefont {Gierer}\ and\ \citenamefont
		{Meinhardt}(1972)}]{gm72}%
	\BibitemOpen
	\bibfield  {author} {\bibinfo {author} {\bibfnamefont {A.}~\bibnamefont
			{Gierer}}\ and\ \bibinfo {author} {\bibfnamefont {H.}~\bibnamefont
			{Meinhardt}},\ }\href@noop {} {\bibfield  {journal} {\bibinfo  {journal}
			{Kybernetik}\ }\textbf {\bibinfo {volume} {12}},\ \bibinfo {pages} {30}
		(\bibinfo {year} {1972})}\BibitemShut {NoStop}%
	\bibitem [{\citenamefont {Keller}\ and\ \citenamefont
		{Segel}(1971)}]{keller1971model}%
	\BibitemOpen
	\bibfield  {author} {\bibinfo {author} {\bibfnamefont {E.~F.}\ \bibnamefont
			{Keller}}\ and\ \bibinfo {author} {\bibfnamefont {L.~A.}\ \bibnamefont
			{Segel}},\ }\href@noop {} {\bibfield  {journal} {\bibinfo  {journal} {Journal
				of theoretical biology}\ }\textbf {\bibinfo {volume} {30}},\ \bibinfo {pages}
		{225} (\bibinfo {year} {1971})}\BibitemShut {NoStop}%
	\bibitem [{\citenamefont {Swift}\ and\ \citenamefont
		{Hohenberg}(1977)}]{swift1977hydrodynamic}%
	\BibitemOpen
	\bibfield  {author} {\bibinfo {author} {\bibfnamefont {J.}~\bibnamefont
			{Swift}}\ and\ \bibinfo {author} {\bibfnamefont {P.~C.}\ \bibnamefont
			{Hohenberg}},\ }\href@noop {} {\bibfield  {journal} {\bibinfo  {journal}
			{Physical Review A}\ }\textbf {\bibinfo {volume} {15}},\ \bibinfo {pages}
		{319} (\bibinfo {year} {1977})}\BibitemShut {NoStop}%
	\bibitem [{\citenamefont {Gray}\ and\ \citenamefont
		{Scott}(1983)}]{gray1983autocatalytic}%
	\BibitemOpen
	\bibfield  {author} {\bibinfo {author} {\bibfnamefont {P.}~\bibnamefont
			{Gray}}\ and\ \bibinfo {author} {\bibfnamefont {S.}~\bibnamefont {Scott}},\
	}\href@noop {} {\bibfield  {journal} {\bibinfo  {journal} {Chemical
				Engineering Science}\ }\textbf {\bibinfo {volume} {38}},\ \bibinfo {pages}
		{29} (\bibinfo {year} {1983})}\BibitemShut {NoStop}%
	\bibitem [{\citenamefont {Gray}\ and\ \citenamefont
		{Scott}(1984)}]{gray1984autocatalytic}%
	\BibitemOpen
	\bibfield  {author} {\bibinfo {author} {\bibfnamefont {P.}~\bibnamefont
			{Gray}}\ and\ \bibinfo {author} {\bibfnamefont {S.}~\bibnamefont {Scott}},\
	}\href@noop {} {\bibfield  {journal} {\bibinfo  {journal} {Chemical
				Engineering Science}\ }\textbf {\bibinfo {volume} {39}},\ \bibinfo {pages}
		{1087} (\bibinfo {year} {1984})}\BibitemShut {NoStop}%
	\bibitem [{\citenamefont {Pearson}(1993)}]{pearson1993complex}%
	\BibitemOpen
	\bibfield  {author} {\bibinfo {author} {\bibfnamefont {J.~E.}\ \bibnamefont
			{Pearson}},\ }\href@noop {} {\bibfield  {journal} {\bibinfo  {journal}
			{Science}\ }\textbf {\bibinfo {volume} {261}},\ \bibinfo {pages} {189}
		(\bibinfo {year} {1993})}\BibitemShut {NoStop}%
	\bibitem [{\citenamefont {Keener}\ and\ \citenamefont {Sneyd}(1998)}]{ksp98}%
	\BibitemOpen
	\bibfield  {author} {\bibinfo {author} {\bibfnamefont {J.}~\bibnamefont
			{Keener}}\ and\ \bibinfo {author} {\bibfnamefont {J.}~\bibnamefont {Sneyd}},\
	}\href@noop {} {\emph {\bibinfo {title} {Mathematical Physiology. Part I:
				Cellular Physiology}}}\ (\bibinfo  {publisher} {Springer},\ \bibinfo
	{address} {New York},\ \bibinfo {year} {1998})\BibitemShut {NoStop}%
	\bibitem [{\citenamefont {Keener}\ and\ \citenamefont {Sneyd}(2008)}]{ksp08}%
	\BibitemOpen
	\bibfield  {author} {\bibinfo {author} {\bibfnamefont {J.}~\bibnamefont
			{Keener}}\ and\ \bibinfo {author} {\bibfnamefont {J.}~\bibnamefont {Sneyd}},\
	}\href@noop {} {\emph {\bibinfo {title} {Mathematical Physiology. Part II:
				Systems Physiology}}}\ (\bibinfo  {publisher} {Springer Science+Business
		Media},\ \bibinfo {address} {New York},\ \bibinfo {year} {2008})\BibitemShut
	{NoStop}%
	\bibitem [{\citenamefont {Tyson}\ and\ \citenamefont
		{Keener}(1988)}]{tyson1988singular}%
	\BibitemOpen
	\bibfield  {author} {\bibinfo {author} {\bibfnamefont {J.~J.}\ \bibnamefont
			{Tyson}}\ and\ \bibinfo {author} {\bibfnamefont {J.~P.}\ \bibnamefont
			{Keener}},\ }\href@noop {} {\bibfield  {journal} {\bibinfo  {journal}
			{Physica D: Nonlinear Phenomena}\ }\textbf {\bibinfo {volume} {32}},\
		\bibinfo {pages} {327} (\bibinfo {year} {1988})}\BibitemShut {NoStop}%
	\bibitem [{\citenamefont {Golubitsky}\ and\ \citenamefont
		{Stewart}(2002)}]{GoS2002}%
	\BibitemOpen
	\bibfield  {author} {\bibinfo {author} {\bibfnamefont {M.}~\bibnamefont
			{Golubitsky}}\ and\ \bibinfo {author} {\bibfnamefont {I.}~\bibnamefont
			{Stewart}},\ }\href@noop {} {\emph {\bibinfo {title} {The symmetry
				perspective}}}\ (\bibinfo  {publisher} {Birkh\"auser Verlag},\ \bibinfo
	{address} {Basel},\ \bibinfo {year} {2002})\BibitemShut {NoStop}%
	\bibitem [{\citenamefont {Hoyle}(2006)}]{Hoyle}%
	\BibitemOpen
	\bibfield  {author} {\bibinfo {author} {\bibfnamefont {R.}~\bibnamefont
			{Hoyle}},\ }\href@noop {} {\emph {\bibinfo {title} {Pattern formation}}}\
	(\bibinfo  {publisher} {Cambridge University Press.},\ \bibinfo {address}
	{Cambridge, UK},\ \bibinfo {year} {2006})\BibitemShut {NoStop}%
	\bibitem [{\citenamefont {Chow}\ and\ \citenamefont {Hale}(1982)}]{CH82}%
	\BibitemOpen
	\bibfield  {author} {\bibinfo {author} {\bibfnamefont {S.~N.}\ \bibnamefont
			{Chow}}\ and\ \bibinfo {author} {\bibfnamefont {J.~K.}\ \bibnamefont
			{Hale}},\ }\href@noop {} {\emph {\bibinfo {title} {Methods of bifurcation
				theory}}},\ \bibinfo {series} {Grundlehren der Mathematischen Wissenschaften
		[Fundamental Principles of Mathematical Science]}, Vol.\ \bibinfo {volume}
	{251}\ (\bibinfo  {publisher} {Springer-Verlag},\ \bibinfo {address} {New
		York},\ \bibinfo {year} {1982})\BibitemShut {NoStop}%
	\bibitem [{\citenamefont {Collet}\ and\ \citenamefont
		{Eckmann}(1990)}]{Collet90etalbook}%
	\BibitemOpen
	\bibfield  {author} {\bibinfo {author} {\bibfnamefont {P.}~\bibnamefont
			{Collet}}\ and\ \bibinfo {author} {\bibfnamefont {J.-P.}\ \bibnamefont
			{Eckmann}},\ }\href@noop {} {\emph {\bibinfo {title} {Instabilities and
				Fronts in Extended Systems}}}\ (\bibinfo  {publisher} {Princeton University
		Press},\ \bibinfo {year} {1990})\BibitemShut {NoStop}%
	\bibitem [{\citenamefont {Kuznetsov}(2004)}]{kuz04}%
	\BibitemOpen
	\bibfield  {author} {\bibinfo {author} {\bibfnamefont {Y.~A.}\ \bibnamefont
			{Kuznetsov}},\ }\href@noop {} {\emph {\bibinfo {title} {Elements of applied
				bifurcation theory}}},\ \bibinfo {edition} {3rd}\ ed.,\ \bibinfo {series}
	{Applied Mathematical Sciences}, Vol.\ \bibinfo {volume} {112}\ (\bibinfo
	{publisher} {Springer-Verlag, New York},\ \bibinfo {year} {2004})\BibitemShut
	{NoStop}%
	\bibitem [{\citenamefont {Mei}(2000)}]{mei2000}%
	\BibitemOpen
	\bibfield  {author} {\bibinfo {author} {\bibfnamefont {Z.}~\bibnamefont
			{Mei}},\ }\href@noop {} {\emph {\bibinfo {title} {Numerical bifurcation
				analysis for reaction-diffusion equations}}}\ (\bibinfo  {publisher}
	{Springer-Verlag},\ \bibinfo {address} {Berlin},\ \bibinfo {year}
	{2000})\BibitemShut {NoStop}%
	\bibitem [{\citenamefont {Pavliotis}\ and\ \citenamefont
		{Stuart}(2008)}]{ps08}%
	\BibitemOpen
	\bibfield  {author} {\bibinfo {author} {\bibfnamefont {G.}~\bibnamefont
			{Pavliotis}}\ and\ \bibinfo {author} {\bibfnamefont {A.}~\bibnamefont
			{Stuart}},\ }\href@noop {} {\emph {\bibinfo {title} {Multiscale methods:
				Averaging and homogenization}}}\ (\bibinfo  {publisher} {Springer Science \&
		Business Media},\ \bibinfo {year} {2008})\BibitemShut {NoStop}%
	\bibitem [{\citenamefont {Schneider}\ and\ \citenamefont
		{Uecker}(2017)}]{SU17}%
	\BibitemOpen
	\bibfield  {author} {\bibinfo {author} {\bibfnamefont {G.}~\bibnamefont
			{Schneider}}\ and\ \bibinfo {author} {\bibfnamefont {H.}~\bibnamefont
			{Uecker}},\ }\href@noop {} {\emph {\bibinfo {title} {Nonlinear PDE -- a
				dynamical systems approach}}},\ \bibinfo {series} {Graduate Studies
		Mathematics}, Vol.\ \bibinfo {volume} {182}\ (\bibinfo  {publisher} {AMS},\
	\bibinfo {year} {2017})\BibitemShut {NoStop}%
	\bibitem [{\citenamefont {Dupont}(2014)}]{dupont2014modeling}%
	\BibitemOpen
	\bibfield  {author} {\bibinfo {author} {\bibfnamefont {G.}~\bibnamefont
			{Dupont}},\ }\href@noop {} {\bibfield  {journal} {\bibinfo  {journal} {Wiley
				Interdisciplinary Reviews: Systems Biology and Medicine}\ }\textbf {\bibinfo
			{volume} {6}},\ \bibinfo {pages} {227} (\bibinfo {year} {2014})}\BibitemShut
	{NoStop}%
	\bibitem [{\citenamefont {Falcke}(2003)}]{falcke_deterministic_2003}%
	\BibitemOpen
	\bibfield  {author} {\bibinfo {author} {\bibfnamefont {M.}~\bibnamefont
			{Falcke}},\ }\href
	{https://iopscience.iop.org/article/10.1088/1367-2630/5/1/396} {\bibfield
		{journal} {\bibinfo  {journal} {New Journal of Physics}\ }\textbf {\bibinfo
			{volume} {5}},\ \bibinfo {pages} {96} (\bibinfo {year} {2003})}\BibitemShut
	{NoStop}%
	\bibitem [{\citenamefont {Kulawiak}\ \emph {et~al.}(2019)\citenamefont
		{Kulawiak}, \citenamefont {L{\"o}ber}, \citenamefont {B{\"a}r},\ and\
		\citenamefont {Engel}}]{kulawiak2019active}%
	\BibitemOpen
	\bibfield  {author} {\bibinfo {author} {\bibfnamefont {D.~A.}\ \bibnamefont
			{Kulawiak}}, \bibinfo {author} {\bibfnamefont {J.}~\bibnamefont {L{\"o}ber}},
		\bibinfo {author} {\bibfnamefont {M.}~\bibnamefont {B{\"a}r}}, \ and\
		\bibinfo {author} {\bibfnamefont {H.}~\bibnamefont {Engel}},\ }\href@noop {}
	{\bibfield  {journal} {\bibinfo  {journal} {PloS One}\ }\textbf {\bibinfo
			{volume} {14}} (\bibinfo {year} {2019})}\BibitemShut {NoStop}%
	\bibitem [{\citenamefont {Radszuweit}\ \emph {et~al.}(2013)\citenamefont
		{Radszuweit}, \citenamefont {Alonso}, \citenamefont {Engel},\ and\
		\citenamefont {B{\"a}r}}]{radszuweit2013intracellular}%
	\BibitemOpen
	\bibfield  {author} {\bibinfo {author} {\bibfnamefont {M.}~\bibnamefont
			{Radszuweit}}, \bibinfo {author} {\bibfnamefont {S.}~\bibnamefont {Alonso}},
		\bibinfo {author} {\bibfnamefont {H.}~\bibnamefont {Engel}}, \ and\ \bibinfo
		{author} {\bibfnamefont {M.}~\bibnamefont {B{\"a}r}},\ }\href@noop {}
	{\bibfield  {journal} {\bibinfo  {journal} {Physical Review Letters}\
		}\textbf {\bibinfo {volume} {110}},\ \bibinfo {pages} {138102} (\bibinfo
		{year} {2013})}\BibitemShut {NoStop}%
	\bibitem [{\citenamefont {Onsum}\ and\ \citenamefont
		{Rao}(2009)}]{onsum2009calling}%
	\BibitemOpen
	\bibfield  {author} {\bibinfo {author} {\bibfnamefont {M.~D.}\ \bibnamefont
			{Onsum}}\ and\ \bibinfo {author} {\bibfnamefont {C.~V.}\ \bibnamefont
			{Rao}},\ }\href@noop {} {\bibfield  {journal} {\bibinfo  {journal} {Current
				opinion in cell biology}\ }\textbf {\bibinfo {volume} {21}},\ \bibinfo
		{pages} {74} (\bibinfo {year} {2009})}\BibitemShut {NoStop}%
	\bibitem [{\citenamefont {Allard}\ and\ \citenamefont
		{Mogilner}(2013)}]{Allard2013}%
	\BibitemOpen
	\bibfield  {author} {\bibinfo {author} {\bibfnamefont {J.}~\bibnamefont
			{Allard}}\ and\ \bibinfo {author} {\bibfnamefont {A.}~\bibnamefont
			{Mogilner}},\ }\href@noop {} {\bibfield  {journal} {\bibinfo  {journal}
			{Curr. Opin. Cell Biol.}\ }\textbf {\bibinfo {volume} {25}},\ \bibinfo
		{pages} {107} (\bibinfo {year} {2013})}\BibitemShut {NoStop}%
	\bibitem [{\citenamefont {Blanchoin}\ \emph {et~al.}(2014)\citenamefont
		{Blanchoin}, \citenamefont {Boujemaa-Paterski}, \citenamefont {Sykes},\ and\
		\citenamefont {Plastino}}]{Blanchoin2014}%
	\BibitemOpen
	\bibfield  {author} {\bibinfo {author} {\bibfnamefont {L.}~\bibnamefont
			{Blanchoin}}, \bibinfo {author} {\bibfnamefont {R.}~\bibnamefont
			{Boujemaa-Paterski}}, \bibinfo {author} {\bibfnamefont {C.}~\bibnamefont
			{Sykes}}, \ and\ \bibinfo {author} {\bibfnamefont {J.}~\bibnamefont
			{Plastino}},\ }\href@noop {} {\bibfield  {journal} {\bibinfo  {journal}
			{Physiol. Rev.}\ }\textbf {\bibinfo {volume} {94}},\ \bibinfo {pages} {235}
		(\bibinfo {year} {2014})}\BibitemShut {NoStop}%
	\bibitem [{\citenamefont {Inagaki}\ and\ \citenamefont
		{Katsuno}(2017)}]{inagaki2017actin}%
	\BibitemOpen
	\bibfield  {author} {\bibinfo {author} {\bibfnamefont {N.}~\bibnamefont
			{Inagaki}}\ and\ \bibinfo {author} {\bibfnamefont {H.}~\bibnamefont
			{Katsuno}},\ }\href@noop {} {\bibfield  {journal} {\bibinfo  {journal}
			{Trends in Cell Biology}\ }\textbf {\bibinfo {volume} {27}},\ \bibinfo
		{pages} {515} (\bibinfo {year} {2017})}\BibitemShut {NoStop}%
	\bibitem [{\citenamefont {Beta}\ and\ \citenamefont
		{Kruse}(2017)}]{beta2017intracellular}%
	\BibitemOpen
	\bibfield  {author} {\bibinfo {author} {\bibfnamefont {C.}~\bibnamefont
			{Beta}}\ and\ \bibinfo {author} {\bibfnamefont {K.}~\bibnamefont {Kruse}},\
	}\href@noop {} {\bibfield  {journal} {\bibinfo  {journal} {Annual Review of
				Condensed Matter Physics}\ }\textbf {\bibinfo {volume} {8}},\ \bibinfo
		{pages} {239} (\bibinfo {year} {2017})}\BibitemShut {NoStop}%
	\bibitem [{\citenamefont {Halatek}\ \emph {et~al.}(2018)\citenamefont
		{Halatek}, \citenamefont {Brauns},\ and\ \citenamefont
		{Frey}}]{halatek2018self}%
	\BibitemOpen
	\bibfield  {author} {\bibinfo {author} {\bibfnamefont {J.}~\bibnamefont
			{Halatek}}, \bibinfo {author} {\bibfnamefont {F.}~\bibnamefont {Brauns}}, \
		and\ \bibinfo {author} {\bibfnamefont {E.}~\bibnamefont {Frey}},\ }\href@noop
	{} {\bibfield  {journal} {\bibinfo  {journal} {Philosophical Transactions of
				the Royal Society B: Biological Sciences}\ }\textbf {\bibinfo {volume}
			{373}},\ \bibinfo {pages} {20170107} (\bibinfo {year} {2018})}\BibitemShut
	{NoStop}%
	\bibitem [{\citenamefont {Deneke}\ and\ \citenamefont
		{Di~Talia}(2018)}]{deneke2018chemical}%
	\BibitemOpen
	\bibfield  {author} {\bibinfo {author} {\bibfnamefont {V.~E.}\ \bibnamefont
			{Deneke}}\ and\ \bibinfo {author} {\bibfnamefont {S.}~\bibnamefont
			{Di~Talia}},\ }\href@noop {} {\bibfield  {journal} {\bibinfo  {journal}
			{Journal of Cell Biology}\ }\textbf {\bibinfo {volume} {217}},\ \bibinfo
		{pages} {1193} (\bibinfo {year} {2018})}\BibitemShut {NoStop}%
	\bibitem [{\citenamefont {Ruthel}\ and\ \citenamefont
		{Banker}(1998)}]{ruthel_actin-dependent_1998}%
	\BibitemOpen
	\bibfield  {author} {\bibinfo {author} {\bibfnamefont {G.}~\bibnamefont
			{Ruthel}}\ and\ \bibinfo {author} {\bibfnamefont {G.}~\bibnamefont
			{Banker}},\ }\href@noop {} {\bibfield  {journal} {\bibinfo  {journal} {Cell
				Motility and the Cytoskeleton}\ }\textbf {\bibinfo {volume} {40}},\ \bibinfo
		{pages} {160} (\bibinfo {year} {1998})}\BibitemShut {NoStop}%
	\bibitem [{\citenamefont {Toriyama}\ \emph {et~al.}(2006)\citenamefont
		{Toriyama}, \citenamefont {Shimada}, \citenamefont {Kim}, \citenamefont
		{Mitsuba}, \citenamefont {Nomura}, \citenamefont {Katsuta}, \citenamefont
		{Sakumura}, \citenamefont {Roepstorff},\ and\ \citenamefont
		{Inagaki}}]{toriyama_shootin1:_2006}%
	\BibitemOpen
	\bibfield  {author} {\bibinfo {author} {\bibfnamefont {M.}~\bibnamefont
			{Toriyama}}, \bibinfo {author} {\bibfnamefont {T.}~\bibnamefont {Shimada}},
		\bibinfo {author} {\bibfnamefont {K.~B.}\ \bibnamefont {Kim}}, \bibinfo
		{author} {\bibfnamefont {M.}~\bibnamefont {Mitsuba}}, \bibinfo {author}
		{\bibfnamefont {E.}~\bibnamefont {Nomura}}, \bibinfo {author} {\bibfnamefont
			{K.}~\bibnamefont {Katsuta}}, \bibinfo {author} {\bibfnamefont
			{Y.}~\bibnamefont {Sakumura}}, \bibinfo {author} {\bibfnamefont
			{P.}~\bibnamefont {Roepstorff}}, \ and\ \bibinfo {author} {\bibfnamefont
			{N.}~\bibnamefont {Inagaki}},\ }\href
	{http://www.jcb.org/lookup/doi/10.1083/jcb.200604160} {\bibfield  {journal}
		{\bibinfo  {journal} {J Cell Biol}\ }\textbf {\bibinfo {volume} {175}},\
		\bibinfo {pages} {147} (\bibinfo {year} {2006})}\BibitemShut {NoStop}%
	\bibitem [{\citenamefont {Tomba}\ \emph {et~al.}(2017)\citenamefont {Tomba},
		\citenamefont {Bra{\"\i}ni}, \citenamefont {Bugnicourt}, \citenamefont
		{Cohen}, \citenamefont {Friedrich}, \citenamefont {Gov},\ and\ \citenamefont
		{Villard}}]{tomba2017geometrical}%
	\BibitemOpen
	\bibfield  {author} {\bibinfo {author} {\bibfnamefont {C.}~\bibnamefont
			{Tomba}}, \bibinfo {author} {\bibfnamefont {C.}~\bibnamefont {Bra{\"\i}ni}},
		\bibinfo {author} {\bibfnamefont {G.}~\bibnamefont {Bugnicourt}}, \bibinfo
		{author} {\bibfnamefont {F.}~\bibnamefont {Cohen}}, \bibinfo {author}
		{\bibfnamefont {B.~M.}\ \bibnamefont {Friedrich}}, \bibinfo {author}
		{\bibfnamefont {N.~S.}\ \bibnamefont {Gov}}, \ and\ \bibinfo {author}
		{\bibfnamefont {C.}~\bibnamefont {Villard}},\ }\href@noop {} {\bibfield
		{journal} {\bibinfo  {journal} {Frontiers in cellular neuroscience}\ }\textbf
		{\bibinfo {volume} {11}},\ \bibinfo {pages} {86} (\bibinfo {year}
		{2017})}\BibitemShut {NoStop}%
	\bibitem [{\citenamefont {Guetta-Terrier}\ \emph {et~al.}(2015)\citenamefont
		{Guetta-Terrier}, \citenamefont {Monzo}, \citenamefont {Zhu}, \citenamefont
		{Long}, \citenamefont {Venkatraman}, \citenamefont {Zhou}, \citenamefont
		{Wang}, \citenamefont {Chew}, \citenamefont {Mogilner}, \citenamefont
		{Ladoux},\ and\ \citenamefont {Gauthier}}]{guetta-terrier_protrusive_2015}%
	\BibitemOpen
	\bibfield  {author} {\bibinfo {author} {\bibfnamefont {C.}~\bibnamefont
			{Guetta-Terrier}}, \bibinfo {author} {\bibfnamefont {P.}~\bibnamefont
			{Monzo}}, \bibinfo {author} {\bibfnamefont {J.}~\bibnamefont {Zhu}}, \bibinfo
		{author} {\bibfnamefont {H.}~\bibnamefont {Long}}, \bibinfo {author}
		{\bibfnamefont {L.}~\bibnamefont {Venkatraman}}, \bibinfo {author}
		{\bibfnamefont {Y.}~\bibnamefont {Zhou}}, \bibinfo {author} {\bibfnamefont
			{P.}~\bibnamefont {Wang}}, \bibinfo {author} {\bibfnamefont {S.~Y.}\
			\bibnamefont {Chew}}, \bibinfo {author} {\bibfnamefont {A.}~\bibnamefont
			{Mogilner}}, \bibinfo {author} {\bibfnamefont {B.}~\bibnamefont {Ladoux}}, \
		and\ \bibinfo {author} {\bibfnamefont {N.~C.}\ \bibnamefont {Gauthier}},\
	}\href {http://jcb.rupress.org/content/211/3/683} {\bibfield  {journal}
		{\bibinfo  {journal} {J Cell Biol}\ }\textbf {\bibinfo {volume} {211}},\
		\bibinfo {pages} {683} (\bibinfo {year} {2015})}\BibitemShut {NoStop}%
	\bibitem [{\citenamefont {Döbereiner}\ \emph {et~al.}(2006)\citenamefont
		{Döbereiner}, \citenamefont {Dubin-Thaler}, \citenamefont {Hofman},
		\citenamefont {Xenias}, \citenamefont {Sims}, \citenamefont {Giannone},
		\citenamefont {Dustin}, \citenamefont {Wiggins},\ and\ \citenamefont
		{Sheetz}}]{dobereiner_lateral_2006}%
	\BibitemOpen
	\bibfield  {author} {\bibinfo {author} {\bibfnamefont {H.-G.}\ \bibnamefont
			{Döbereiner}}, \bibinfo {author} {\bibfnamefont {B.~J.}\ \bibnamefont
			{Dubin-Thaler}}, \bibinfo {author} {\bibfnamefont {J.~M.}\ \bibnamefont
			{Hofman}}, \bibinfo {author} {\bibfnamefont {H.~S.}\ \bibnamefont {Xenias}},
		\bibinfo {author} {\bibfnamefont {T.~N.}\ \bibnamefont {Sims}}, \bibinfo
		{author} {\bibfnamefont {G.}~\bibnamefont {Giannone}}, \bibinfo {author}
		{\bibfnamefont {M.~L.}\ \bibnamefont {Dustin}}, \bibinfo {author}
		{\bibfnamefont {C.~H.}\ \bibnamefont {Wiggins}}, \ and\ \bibinfo {author}
		{\bibfnamefont {M.~P.}\ \bibnamefont {Sheetz}},\ }\href
	{http://link.aps.org/doi/10.1103/PhysRevLett.97.038102} {\bibfield  {journal}
		{\bibinfo  {journal} {Physical Review Letters}\ }\textbf {\bibinfo {volume}
			{97}},\ \bibinfo {pages} {038102} (\bibinfo {year} {2006})}\BibitemShut
	{NoStop}%
	\bibitem [{\citenamefont {Barnhart}\ \emph {et~al.}(2017)\citenamefont
		{Barnhart}, \citenamefont {Allard}, \citenamefont {Lou}, \citenamefont
		{Theriot},\ and\ \citenamefont {Mogilner}}]{barnhart2017adhesion}%
	\BibitemOpen
	\bibfield  {author} {\bibinfo {author} {\bibfnamefont {E.~L.}\ \bibnamefont
			{Barnhart}}, \bibinfo {author} {\bibfnamefont {J.}~\bibnamefont {Allard}},
		\bibinfo {author} {\bibfnamefont {S.~S.}\ \bibnamefont {Lou}}, \bibinfo
		{author} {\bibfnamefont {J.~A.}\ \bibnamefont {Theriot}}, \ and\ \bibinfo
		{author} {\bibfnamefont {A.}~\bibnamefont {Mogilner}},\ }\href@noop {}
	{\bibfield  {journal} {\bibinfo  {journal} {Current Biology}\ }\textbf
		{\bibinfo {volume} {27}},\ \bibinfo {pages} {27} (\bibinfo {year}
		{2017})}\BibitemShut {NoStop}%
	\bibitem [{\citenamefont {Stankevicins}\ \emph {et~al.}(2020)\citenamefont
		{Stankevicins}, \citenamefont {Ecker}, \citenamefont {Terriac}, \citenamefont
		{Maiuri}, \citenamefont {Schoppmeyer}, \citenamefont {Vargas}, \citenamefont
		{Lennon-Duménil}, \citenamefont {Piel}, \citenamefont {Qu}, \citenamefont
		{Hoth}, \citenamefont {Kruse},\ and\ \citenamefont
		{Lautenschläger}}]{stankevicins_deterministic_2020}%
	\BibitemOpen
	\bibfield  {author} {\bibinfo {author} {\bibfnamefont {L.}~\bibnamefont
			{Stankevicins}}, \bibinfo {author} {\bibfnamefont {N.}~\bibnamefont {Ecker}},
		\bibinfo {author} {\bibfnamefont {E.}~\bibnamefont {Terriac}}, \bibinfo
		{author} {\bibfnamefont {P.}~\bibnamefont {Maiuri}}, \bibinfo {author}
		{\bibfnamefont {R.}~\bibnamefont {Schoppmeyer}}, \bibinfo {author}
		{\bibfnamefont {P.}~\bibnamefont {Vargas}}, \bibinfo {author} {\bibfnamefont
			{A.-M.}\ \bibnamefont {Lennon-Duménil}}, \bibinfo {author} {\bibfnamefont
			{M.}~\bibnamefont {Piel}}, \bibinfo {author} {\bibfnamefont {B.}~\bibnamefont
			{Qu}}, \bibinfo {author} {\bibfnamefont {M.}~\bibnamefont {Hoth}}, \bibinfo
		{author} {\bibfnamefont {K.}~\bibnamefont {Kruse}}, \ and\ \bibinfo {author}
		{\bibfnamefont {F.}~\bibnamefont {Lautenschläger}},\ }\href
	{https://www.pnas.org/content/117/2/826} {\bibfield  {journal} {\bibinfo
			{journal} {PNAS}\ }\textbf {\bibinfo {volume} {117}},\ \bibinfo {pages} {826}
		(\bibinfo {year} {2020})}\BibitemShut {NoStop}%
	\bibitem [{\citenamefont {Flemming}\ \emph {et~al.}(2020)\citenamefont
		{Flemming}, \citenamefont {Font}, \citenamefont {Alonso},\ and\ \citenamefont
		{Beta}}]{flemming_how_2020}%
	\BibitemOpen
	\bibfield  {author} {\bibinfo {author} {\bibfnamefont {S.}~\bibnamefont
			{Flemming}}, \bibinfo {author} {\bibfnamefont {F.}~\bibnamefont {Font}},
		\bibinfo {author} {\bibfnamefont {S.}~\bibnamefont {Alonso}}, \ and\ \bibinfo
		{author} {\bibfnamefont {C.}~\bibnamefont {Beta}},\ }\href@noop {} {\bibfield
		{journal} {\bibinfo  {journal} {Proceedings of the National Academy of
				Sciences}\ }\textbf {\bibinfo {volume} {117}},\ \bibinfo {pages} {6330}
		(\bibinfo {year} {2020})}\BibitemShut {NoStop}%
	\bibitem [{\citenamefont {Weiner}\ \emph {et~al.}(2007)\citenamefont {Weiner},
		\citenamefont {Marganski}, \citenamefont {Wu}, \citenamefont {Altschuler},\
		and\ \citenamefont {Kirschner}}]{weiner_actin-based_2007}%
	\BibitemOpen
	\bibfield  {author} {\bibinfo {author} {\bibfnamefont {O.~D.}\ \bibnamefont
			{Weiner}}, \bibinfo {author} {\bibfnamefont {W.~A.}\ \bibnamefont
			{Marganski}}, \bibinfo {author} {\bibfnamefont {L.~F.}\ \bibnamefont {Wu}},
		\bibinfo {author} {\bibfnamefont {S.~J.}\ \bibnamefont {Altschuler}}, \ and\
		\bibinfo {author} {\bibfnamefont {M.~W.}\ \bibnamefont {Kirschner}},\
	}\href@noop {} {\bibfield  {journal} {\bibinfo  {journal} {PLoS Biol}\
		}\textbf {\bibinfo {volume} {5}},\ \bibinfo {pages} {e221} (\bibinfo {year}
		{2007})}\BibitemShut {NoStop}%
	\bibitem [{\citenamefont {Gerisch}\ \emph {et~al.}(2004)\citenamefont
		{Gerisch}, \citenamefont {Bretschneider}, \citenamefont
		{Müller-Taubenberger}, \citenamefont {Simmeth}, \citenamefont {Ecke},
		\citenamefont {Diez},\ and\ \citenamefont {Anderson}}]{gerisch_mobile_2004}%
	\BibitemOpen
	\bibfield  {author} {\bibinfo {author} {\bibfnamefont {G.}~\bibnamefont
			{Gerisch}}, \bibinfo {author} {\bibfnamefont {T.}~\bibnamefont
			{Bretschneider}}, \bibinfo {author} {\bibfnamefont {A.}~\bibnamefont
			{Müller-Taubenberger}}, \bibinfo {author} {\bibfnamefont {E.}~\bibnamefont
			{Simmeth}}, \bibinfo {author} {\bibfnamefont {M.}~\bibnamefont {Ecke}},
		\bibinfo {author} {\bibfnamefont {S.}~\bibnamefont {Diez}}, \ and\ \bibinfo
		{author} {\bibfnamefont {K.}~\bibnamefont {Anderson}},\ }\href
	{http://www.sciencedirect.com/science/article/pii/S000634950473814X}
	{\bibfield  {journal} {\bibinfo  {journal} {Biophysical Journal}\ }\textbf
		{\bibinfo {volume} {87}},\ \bibinfo {pages} {3493} (\bibinfo {year}
		{2004})}\BibitemShut {NoStop}%
	\bibitem [{\citenamefont {Schroth-Diez}\ \emph {et~al.}(2009)\citenamefont
		{Schroth-Diez}, \citenamefont {Gerwig}, \citenamefont {Ecke}, \citenamefont
		{Hegerl}, \citenamefont {Diez},\ and\ \citenamefont
		{Gerisch}}]{schroth-diez_propagating_2009}%
	\BibitemOpen
	\bibfield  {author} {\bibinfo {author} {\bibfnamefont {B.}~\bibnamefont
			{Schroth-Diez}}, \bibinfo {author} {\bibfnamefont {S.}~\bibnamefont
			{Gerwig}}, \bibinfo {author} {\bibfnamefont {M.}~\bibnamefont {Ecke}},
		\bibinfo {author} {\bibfnamefont {R.}~\bibnamefont {Hegerl}}, \bibinfo
		{author} {\bibfnamefont {S.}~\bibnamefont {Diez}}, \ and\ \bibinfo {author}
		{\bibfnamefont {G.}~\bibnamefont {Gerisch}},\ }\href
	{http://www.ncbi.nlm.nih.gov/pmc/articles/PMC2839813/} {\bibfield  {journal}
		{\bibinfo  {journal} {HFSP J}\ }\textbf {\bibinfo {volume} {3}},\ \bibinfo
		{pages} {412} (\bibinfo {year} {2009})}\BibitemShut {NoStop}%
	\bibitem [{\citenamefont {Jasnin}\ \emph {et~al.}(2019)\citenamefont {Jasnin},
		\citenamefont {Beck}, \citenamefont {Ecke}, \citenamefont {Fukuda},
		\citenamefont {Martinez-Sanchez}, \citenamefont {Baumeister},\ and\
		\citenamefont {Gerisch}}]{jasnin2019architecture}%
	\BibitemOpen
	\bibfield  {author} {\bibinfo {author} {\bibfnamefont {M.}~\bibnamefont
			{Jasnin}}, \bibinfo {author} {\bibfnamefont {F.}~\bibnamefont {Beck}},
		\bibinfo {author} {\bibfnamefont {M.}~\bibnamefont {Ecke}}, \bibinfo {author}
		{\bibfnamefont {Y.}~\bibnamefont {Fukuda}}, \bibinfo {author} {\bibfnamefont
			{A.}~\bibnamefont {Martinez-Sanchez}}, \bibinfo {author} {\bibfnamefont
			{W.}~\bibnamefont {Baumeister}}, \ and\ \bibinfo {author} {\bibfnamefont
			{G.}~\bibnamefont {Gerisch}},\ }\href@noop {} {\bibfield  {journal} {\bibinfo
			{journal} {Structure}\ }\textbf {\bibinfo {volume} {27}},\ \bibinfo {pages}
		{1211} (\bibinfo {year} {2019})}\BibitemShut {NoStop}%
	\bibitem [{\citenamefont {Gerhardt}\ \emph
		{et~al.}(2014{\natexlab{a}})\citenamefont {Gerhardt}, \citenamefont {Ecke},
		\citenamefont {Walz}, \citenamefont {Stengl}, \citenamefont {Beta},\ and\
		\citenamefont {Gerisch}}]{gerhardt_actin_2014}%
	\BibitemOpen
	\bibfield  {author} {\bibinfo {author} {\bibfnamefont {M.}~\bibnamefont
			{Gerhardt}}, \bibinfo {author} {\bibfnamefont {M.}~\bibnamefont {Ecke}},
		\bibinfo {author} {\bibfnamefont {M.}~\bibnamefont {Walz}}, \bibinfo {author}
		{\bibfnamefont {A.}~\bibnamefont {Stengl}}, \bibinfo {author} {\bibfnamefont
			{C.}~\bibnamefont {Beta}}, \ and\ \bibinfo {author} {\bibfnamefont
			{G.}~\bibnamefont {Gerisch}},\ }\href
	{http://jcs.biologists.org/content/127/20/4507} {\bibfield  {journal}
		{\bibinfo  {journal} {J Cell Sci}\ }\textbf {\bibinfo {volume} {127}},\
		\bibinfo {pages} {4507} (\bibinfo {year} {2014}{\natexlab{a}})}\BibitemShut
	{NoStop}%
	\bibitem [{\citenamefont {Miao}\ \emph {et~al.}(2017)\citenamefont {Miao},
		\citenamefont {Bhattacharya}, \citenamefont {Edwards}, \citenamefont {Cai},
		\citenamefont {Inoue}, \citenamefont {Iglesias},\ and\ \citenamefont
		{Devreotes}}]{miao_altering_2017}%
	\BibitemOpen
	\bibfield  {author} {\bibinfo {author} {\bibfnamefont {Y.}~\bibnamefont
			{Miao}}, \bibinfo {author} {\bibfnamefont {S.}~\bibnamefont {Bhattacharya}},
		\bibinfo {author} {\bibfnamefont {M.}~\bibnamefont {Edwards}}, \bibinfo
		{author} {\bibfnamefont {H.}~\bibnamefont {Cai}}, \bibinfo {author}
		{\bibfnamefont {T.}~\bibnamefont {Inoue}}, \bibinfo {author} {\bibfnamefont
			{P.~A.}\ \bibnamefont {Iglesias}}, \ and\ \bibinfo {author} {\bibfnamefont
			{P.~N.}\ \bibnamefont {Devreotes}},\ }\href
	{http://www.nature.com/doifinder/10.1038/ncb3495} {\bibfield  {journal}
		{\bibinfo  {journal} {Nature Cell Biology}\ }\textbf {\bibinfo {volume}
			{19}},\ \bibinfo {pages} {329} (\bibinfo {year} {2017})}\BibitemShut
	{NoStop}%
	\bibitem [{\citenamefont {Gerhardt}\ \emph
		{et~al.}(2014{\natexlab{b}})\citenamefont {Gerhardt}, \citenamefont {Walz},\
		and\ \citenamefont {Beta}}]{gerhardt_signaling_2014}%
	\BibitemOpen
	\bibfield  {author} {\bibinfo {author} {\bibfnamefont {M.}~\bibnamefont
			{Gerhardt}}, \bibinfo {author} {\bibfnamefont {M.}~\bibnamefont {Walz}}, \
		and\ \bibinfo {author} {\bibfnamefont {C.}~\bibnamefont {Beta}},\ }\href
	{http://jcs.biologists.org/content/127/23/5115} {\bibfield  {journal}
		{\bibinfo  {journal} {J Cell Sci}\ }\textbf {\bibinfo {volume} {127}},\
		\bibinfo {pages} {5115} (\bibinfo {year} {2014}{\natexlab{b}})}\BibitemShut
	{NoStop}%
	\bibitem [{\citenamefont {Bement}\ \emph {et~al.}(2015)\citenamefont {Bement},
		\citenamefont {Leda}, \citenamefont {Moe}, \citenamefont {Kita},
		\citenamefont {Larson}, \citenamefont {Golding}, \citenamefont {Pfeuti},
		\citenamefont {Su}, \citenamefont {Miller}, \citenamefont {Goryachev},\ and\
		\citenamefont {von Dassow}}]{bement_activator-inhibitor_2015}%
	\BibitemOpen
	\bibfield  {author} {\bibinfo {author} {\bibfnamefont {W.~M.}\ \bibnamefont
			{Bement}}, \bibinfo {author} {\bibfnamefont {M.}~\bibnamefont {Leda}},
		\bibinfo {author} {\bibfnamefont {A.~M.}\ \bibnamefont {Moe}}, \bibinfo
		{author} {\bibfnamefont {A.~M.}\ \bibnamefont {Kita}}, \bibinfo {author}
		{\bibfnamefont {M.~E.}\ \bibnamefont {Larson}}, \bibinfo {author}
		{\bibfnamefont {A.~E.}\ \bibnamefont {Golding}}, \bibinfo {author}
		{\bibfnamefont {C.}~\bibnamefont {Pfeuti}}, \bibinfo {author} {\bibfnamefont
			{K.-C.}\ \bibnamefont {Su}}, \bibinfo {author} {\bibfnamefont {A.~L.}\
			\bibnamefont {Miller}}, \bibinfo {author} {\bibfnamefont {A.~B.}\
			\bibnamefont {Goryachev}}, \ and\ \bibinfo {author} {\bibfnamefont
			{G.}~\bibnamefont {von Dassow}},\ }\href
	{http://www.nature.com/ncb/journal/v17/n11/full/ncb3251.html} {\bibfield
		{journal} {\bibinfo  {journal} {Nat Cell Biol}\ }\textbf {\bibinfo {volume}
			{17}},\ \bibinfo {pages} {1471} (\bibinfo {year} {2015})}\BibitemShut
	{NoStop}%
	\bibitem [{\citenamefont {Xiao}\ \emph {et~al.}(2017)\citenamefont {Xiao},
		\citenamefont {Tong}, \citenamefont {Yang},\ and\ \citenamefont
		{Wu}}]{xiao_mitotic_2017}%
	\BibitemOpen
	\bibfield  {author} {\bibinfo {author} {\bibfnamefont {S.}~\bibnamefont
			{Xiao}}, \bibinfo {author} {\bibfnamefont {C.}~\bibnamefont {Tong}}, \bibinfo
		{author} {\bibfnamefont {Y.}~\bibnamefont {Yang}}, \ and\ \bibinfo {author}
		{\bibfnamefont {M.}~\bibnamefont {Wu}},\ }\href
	{http://www.sciencedirect.com/science/article/pii/S1534580717308663}
	{\bibfield  {journal} {\bibinfo  {journal} {Developmental Cell}\ }\textbf
		{\bibinfo {volume} {43}},\ \bibinfo {pages} {493} (\bibinfo {year}
		{2017})}\BibitemShut {NoStop}%
	\bibitem [{\citenamefont {Gerisch}\ \emph {et~al.}(2009)\citenamefont
		{Gerisch}, \citenamefont {Ecke}, \citenamefont {Schroth-Diez}, \citenamefont
		{Gerwig}, \citenamefont {Engel}, \citenamefont {Maddera},\ and\ \citenamefont
		{Clarke}}]{gerisch_self-organizing_2009}%
	\BibitemOpen
	\bibfield  {author} {\bibinfo {author} {\bibfnamefont {G.}~\bibnamefont
			{Gerisch}}, \bibinfo {author} {\bibfnamefont {M.}~\bibnamefont {Ecke}},
		\bibinfo {author} {\bibfnamefont {B.}~\bibnamefont {Schroth-Diez}}, \bibinfo
		{author} {\bibfnamefont {S.}~\bibnamefont {Gerwig}}, \bibinfo {author}
		{\bibfnamefont {U.}~\bibnamefont {Engel}}, \bibinfo {author} {\bibfnamefont
			{L.}~\bibnamefont {Maddera}}, \ and\ \bibinfo {author} {\bibfnamefont
			{M.}~\bibnamefont {Clarke}},\ }\href
	{http://www.landesbioscience.com/journals/celladhesion/article/9708/}
	{\bibfield  {journal} {\bibinfo  {journal} {Cell Adhesion \& Migration}\
		}\textbf {\bibinfo {volume} {3}},\ \bibinfo {pages} {373} (\bibinfo {year}
		{2009})}\BibitemShut {NoStop}%
	\bibitem [{\citenamefont {Miao}\ \emph {et~al.}(2019)\citenamefont {Miao},
		\citenamefont {Bhattacharya}, \citenamefont {Banerjee}, \citenamefont
		{Abubaker-Sharif}, \citenamefont {Long}, \citenamefont {Inoue}, \citenamefont
		{Iglesias},\ and\ \citenamefont {Devreotes}}]{miao2019wave}%
	\BibitemOpen
	\bibfield  {author} {\bibinfo {author} {\bibfnamefont {Y.}~\bibnamefont
			{Miao}}, \bibinfo {author} {\bibfnamefont {S.}~\bibnamefont {Bhattacharya}},
		\bibinfo {author} {\bibfnamefont {T.}~\bibnamefont {Banerjee}}, \bibinfo
		{author} {\bibfnamefont {B.}~\bibnamefont {Abubaker-Sharif}}, \bibinfo
		{author} {\bibfnamefont {Y.}~\bibnamefont {Long}}, \bibinfo {author}
		{\bibfnamefont {T.}~\bibnamefont {Inoue}}, \bibinfo {author} {\bibfnamefont
			{P.~A.}\ \bibnamefont {Iglesias}}, \ and\ \bibinfo {author} {\bibfnamefont
			{P.~N.}\ \bibnamefont {Devreotes}},\ }\href@noop {} {\bibfield  {journal}
		{\bibinfo  {journal} {Molecular Systems Biology}\ }\textbf {\bibinfo {volume}
			{15}} (\bibinfo {year} {2019})}\BibitemShut {NoStop}%
	\bibitem [{\citenamefont {Veltman}\ \emph {et~al.}(2016)\citenamefont
		{Veltman}, \citenamefont {Williams}, \citenamefont {Bloomfield},
		\citenamefont {Chen}, \citenamefont {Betzig}, \citenamefont {Insall},\ and\
		\citenamefont {Kay}}]{veltman_plasma_2016}%
	\BibitemOpen
	\bibfield  {author} {\bibinfo {author} {\bibfnamefont {D.~M.}\ \bibnamefont
			{Veltman}}, \bibinfo {author} {\bibfnamefont {T.~D.}\ \bibnamefont
			{Williams}}, \bibinfo {author} {\bibfnamefont {G.}~\bibnamefont
			{Bloomfield}}, \bibinfo {author} {\bibfnamefont {B.-C.}\ \bibnamefont
			{Chen}}, \bibinfo {author} {\bibfnamefont {E.}~\bibnamefont {Betzig}},
		\bibinfo {author} {\bibfnamefont {R.~H.}\ \bibnamefont {Insall}}, \ and\
		\bibinfo {author} {\bibfnamefont {R.~R.}\ \bibnamefont {Kay}},\ }\href@noop
	{} {\bibfield  {journal} {\bibinfo  {journal} {Elife}\ }\textbf {\bibinfo
			{volume} {5}},\ \bibinfo {pages} {e20085} (\bibinfo {year}
		{2016})}\BibitemShut {NoStop}%
	\bibitem [{\citenamefont {Bernitt}\ \emph {et~al.}(2015)\citenamefont
		{Bernitt}, \citenamefont {Koh}, \citenamefont {Gov},\ and\ \citenamefont
		{Döbereiner}}]{bernitt_dynamics_2015}%
	\BibitemOpen
	\bibfield  {author} {\bibinfo {author} {\bibfnamefont {E.}~\bibnamefont
			{Bernitt}}, \bibinfo {author} {\bibfnamefont {C.~G.}\ \bibnamefont {Koh}},
		\bibinfo {author} {\bibfnamefont {N.}~\bibnamefont {Gov}}, \ and\ \bibinfo
		{author} {\bibfnamefont {H.-G.}\ \bibnamefont {Döbereiner}},\ }\href
	{http://journals.plos.org/plosone/article?id=10.1371/journal.pone.0115857}
	{\bibfield  {journal} {\bibinfo  {journal} {PLOS ONE}\ }\textbf {\bibinfo
			{volume} {10}},\ \bibinfo {pages} {e0115857} (\bibinfo {year}
		{2015})}\BibitemShut {NoStop}%
	\bibitem [{\citenamefont {Bernitt}\ \emph {et~al.}(2017)\citenamefont
		{Bernitt}, \citenamefont {D{\"o}bereiner}, \citenamefont {Gov},\ and\
		\citenamefont {Yochelis}}]{bernitt2017fronts}%
	\BibitemOpen
	\bibfield  {author} {\bibinfo {author} {\bibfnamefont {E.}~\bibnamefont
			{Bernitt}}, \bibinfo {author} {\bibfnamefont {H.-G.}\ \bibnamefont
			{D{\"o}bereiner}}, \bibinfo {author} {\bibfnamefont {N.~S.}\ \bibnamefont
			{Gov}}, \ and\ \bibinfo {author} {\bibfnamefont {A.}~\bibnamefont
			{Yochelis}},\ }\href@noop {} {\bibfield  {journal} {\bibinfo  {journal}
			{Nature Communications}\ }\textbf {\bibinfo {volume} {8}},\ \bibinfo {pages}
		{15863} (\bibinfo {year} {2017})}\BibitemShut {NoStop}%
	\bibitem [{\citenamefont {Buccione}\ \emph {et~al.}(2004)\citenamefont
		{Buccione}, \citenamefont {Orth},\ and\ \citenamefont
		{McNiven}}]{buccione_foot_2004}%
	\BibitemOpen
	\bibfield  {author} {\bibinfo {author} {\bibfnamefont {R.}~\bibnamefont
			{Buccione}}, \bibinfo {author} {\bibfnamefont {J.~D.}\ \bibnamefont {Orth}},
		\ and\ \bibinfo {author} {\bibfnamefont {M.~A.}\ \bibnamefont {McNiven}},\
	}\href {http://www.nature.com/nrm/journal/v5/n8/full/nrm1436.html} {\bibfield
		{journal} {\bibinfo  {journal} {Nat Rev Mol Cell Biol}\ }\textbf {\bibinfo
			{volume} {5}},\ \bibinfo {pages} {647} (\bibinfo {year} {2004})}\BibitemShut
	{NoStop}%
	\bibitem [{\citenamefont {Commisso}\ \emph {et~al.}(2013)\citenamefont
		{Commisso}, \citenamefont {Davidson}, \citenamefont {Soydaner-Azeloglu},
		\citenamefont {Parker}, \citenamefont {Kamphorst}, \citenamefont {Hackett},
		\citenamefont {Grabocka}, \citenamefont {Nofal}, \citenamefont {Drebin},
		\citenamefont {Thompson},\ and\ \citenamefont {et~al.}}]{Commisso2013}%
	\BibitemOpen
	\bibfield  {author} {\bibinfo {author} {\bibfnamefont {C.}~\bibnamefont
			{Commisso}}, \bibinfo {author} {\bibfnamefont {S.~M.}\ \bibnamefont
			{Davidson}}, \bibinfo {author} {\bibfnamefont {R.~G.}\ \bibnamefont
			{Soydaner-Azeloglu}}, \bibinfo {author} {\bibfnamefont {S.~J.}\ \bibnamefont
			{Parker}}, \bibinfo {author} {\bibfnamefont {J.~J.}\ \bibnamefont
			{Kamphorst}}, \bibinfo {author} {\bibfnamefont {S.}~\bibnamefont {Hackett}},
		\bibinfo {author} {\bibfnamefont {E.}~\bibnamefont {Grabocka}}, \bibinfo
		{author} {\bibfnamefont {M.}~\bibnamefont {Nofal}}, \bibinfo {author}
		{\bibfnamefont {J.~A.}\ \bibnamefont {Drebin}}, \bibinfo {author}
		{\bibfnamefont {C.~B.}\ \bibnamefont {Thompson}}, \ and\ \bibinfo {author}
		{\bibnamefont {et~al.}},\ }\href@noop {} {\bibfield  {journal} {\bibinfo
			{journal} {Nature}\ }\textbf {\bibinfo {volume} {497}},\ \bibinfo {pages}
		{633} (\bibinfo {year} {2013})}\BibitemShut {NoStop}%
	\bibitem [{\citenamefont {Itoh}\ and\ \citenamefont
		{Hasegawa}(2012)}]{Itoh2012}%
	\BibitemOpen
	\bibfield  {author} {\bibinfo {author} {\bibfnamefont {T.}~\bibnamefont
			{Itoh}}\ and\ \bibinfo {author} {\bibfnamefont {J.}~\bibnamefont
			{Hasegawa}},\ }\href@noop {} {\bibfield  {journal} {\bibinfo  {journal} {J.
				Biochem}\ }\textbf {\bibinfo {volume} {153}},\ \bibinfo {pages} {21}
		(\bibinfo {year} {2012})}\BibitemShut {NoStop}%
	\bibitem [{\citenamefont {Hoon}\ \emph {et~al.}(2012)\citenamefont {Hoon},
		\citenamefont {Wong},\ and\ \citenamefont {Koh}}]{Hoon2012}%
	\BibitemOpen
	\bibfield  {author} {\bibinfo {author} {\bibfnamefont {J.-L.}\ \bibnamefont
			{Hoon}}, \bibinfo {author} {\bibfnamefont {W.-K.}\ \bibnamefont {Wong}}, \
		and\ \bibinfo {author} {\bibfnamefont {C.-G.}\ \bibnamefont {Koh}},\
	}\href@noop {} {\bibfield  {journal} {\bibinfo  {journal} {Mol. Cell. Biol.}\
		}\textbf {\bibinfo {volume} {32}},\ \bibinfo {pages} {4246} (\bibinfo {year}
		{2012})}\BibitemShut {NoStop}%
	\bibitem [{\citenamefont {Ryan}\ \emph {et~al.}(2012)\citenamefont {Ryan},
		\citenamefont {Watanabe},\ and\ \citenamefont {Vavylonis}}]{ryan2012review}%
	\BibitemOpen
	\bibfield  {author} {\bibinfo {author} {\bibfnamefont {G.~L.}\ \bibnamefont
			{Ryan}}, \bibinfo {author} {\bibfnamefont {N.}~\bibnamefont {Watanabe}}, \
		and\ \bibinfo {author} {\bibfnamefont {D.}~\bibnamefont {Vavylonis}},\
	}\href@noop {} {\bibfield  {journal} {\bibinfo  {journal} {Cytoskeleton}\
		}\textbf {\bibinfo {volume} {69}},\ \bibinfo {pages} {195} (\bibinfo {year}
		{2012})}\BibitemShut {NoStop}%
	\bibitem [{\citenamefont {Sept}\ and\ \citenamefont
		{Carlsson}(2014)}]{sept2014modeling}%
	\BibitemOpen
	\bibfield  {author} {\bibinfo {author} {\bibfnamefont {D.}~\bibnamefont
			{Sept}}\ and\ \bibinfo {author} {\bibfnamefont {A.~E.}\ \bibnamefont
			{Carlsson}},\ }\href@noop {} {\bibfield  {journal} {\bibinfo  {journal}
			{Quarterly reviews of biophysics}\ }\textbf {\bibinfo {volume} {47}},\
		\bibinfo {pages} {221} (\bibinfo {year} {2014})}\BibitemShut {NoStop}%
	\bibitem [{\citenamefont {Carlsson}(2010)}]{carlsson2010dendritic}%
	\BibitemOpen
	\bibfield  {author} {\bibinfo {author} {\bibfnamefont {A.~E.}\ \bibnamefont
			{Carlsson}},\ }\href@noop {} {\bibfield  {journal} {\bibinfo  {journal}
			{Physical review letters}\ }\textbf {\bibinfo {volume} {104}},\ \bibinfo
		{pages} {228102} (\bibinfo {year} {2010})}\BibitemShut {NoStop}%
	\bibitem [{\citenamefont {Huber}\ \emph {et~al.}(2008)\citenamefont {Huber},
		\citenamefont {K{\"a}s},\ and\ \citenamefont {Stuhrmann}}]{huber2008growing}%
	\BibitemOpen
	\bibfield  {author} {\bibinfo {author} {\bibfnamefont {F.}~\bibnamefont
			{Huber}}, \bibinfo {author} {\bibfnamefont {J.}~\bibnamefont {K{\"a}s}}, \
		and\ \bibinfo {author} {\bibfnamefont {B.}~\bibnamefont {Stuhrmann}},\
	}\href@noop {} {\bibfield  {journal} {\bibinfo  {journal} {Biophysical
				journal}\ }\textbf {\bibinfo {volume} {95}},\ \bibinfo {pages} {5508}
		(\bibinfo {year} {2008})}\BibitemShut {NoStop}%
	\bibitem [{\citenamefont {Khamviwath}\ \emph {et~al.}(2013)\citenamefont
		{Khamviwath}, \citenamefont {Hu},\ and\ \citenamefont
		{Othmer}}]{khamviwath2013continuum}%
	\BibitemOpen
	\bibfield  {author} {\bibinfo {author} {\bibfnamefont {V.}~\bibnamefont
			{Khamviwath}}, \bibinfo {author} {\bibfnamefont {J.}~\bibnamefont {Hu}}, \
		and\ \bibinfo {author} {\bibfnamefont {H.~G.}\ \bibnamefont {Othmer}},\
	}\href@noop {} {\bibfield  {journal} {\bibinfo  {journal} {PloS one}\
		}\textbf {\bibinfo {volume} {8}},\ \bibinfo {pages} {e64272} (\bibinfo {year}
		{2013})}\BibitemShut {NoStop}%
	\bibitem [{\citenamefont {Gholami}\ \emph {et~al.}(2012)\citenamefont
		{Gholami}, \citenamefont {Enculescu},\ and\ \citenamefont
		{Falcke}}]{gholami2012membrane}%
	\BibitemOpen
	\bibfield  {author} {\bibinfo {author} {\bibfnamefont {A.}~\bibnamefont
			{Gholami}}, \bibinfo {author} {\bibfnamefont {M.}~\bibnamefont {Enculescu}},
		\ and\ \bibinfo {author} {\bibfnamefont {M.}~\bibnamefont {Falcke}},\
	}\href@noop {} {\bibfield  {journal} {\bibinfo  {journal} {New Journal of
				Physics}\ }\textbf {\bibinfo {volume} {14}},\ \bibinfo {pages} {115002}
		(\bibinfo {year} {2012})}\BibitemShut {NoStop}%
	\bibitem [{\citenamefont {Gov}\ and\ \citenamefont
		{Gopinathan}(2006)}]{gov2006dynamics}%
	\BibitemOpen
	\bibfield  {author} {\bibinfo {author} {\bibfnamefont {N.~S.}\ \bibnamefont
			{Gov}}\ and\ \bibinfo {author} {\bibfnamefont {A.}~\bibnamefont
			{Gopinathan}},\ }\href@noop {} {\bibfield  {journal} {\bibinfo  {journal}
			{Biophysical journal}\ }\textbf {\bibinfo {volume} {90}},\ \bibinfo {pages}
		{454} (\bibinfo {year} {2006})}\BibitemShut {NoStop}%
	\bibitem [{\citenamefont {Shlomovitz}\ and\ \citenamefont
		{Gov}(2007)}]{shlomovitz2007membrane}%
	\BibitemOpen
	\bibfield  {author} {\bibinfo {author} {\bibfnamefont {R.}~\bibnamefont
			{Shlomovitz}}\ and\ \bibinfo {author} {\bibfnamefont {N.}~\bibnamefont
			{Gov}},\ }\href@noop {} {\bibfield  {journal} {\bibinfo  {journal} {Physical
				review letters}\ }\textbf {\bibinfo {volume} {98}},\ \bibinfo {pages}
		{168103} (\bibinfo {year} {2007})}\BibitemShut {NoStop}%
	\bibitem [{\citenamefont {Veksler}\ and\ \citenamefont
		{Gov}(2009)}]{veksler2009calcium}%
	\BibitemOpen
	\bibfield  {author} {\bibinfo {author} {\bibfnamefont {A.}~\bibnamefont
			{Veksler}}\ and\ \bibinfo {author} {\bibfnamefont {N.~S.}\ \bibnamefont
			{Gov}},\ }\href@noop {} {\bibfield  {journal} {\bibinfo  {journal}
			{Biophysical journal}\ }\textbf {\bibinfo {volume} {97}},\ \bibinfo {pages}
		{1558} (\bibinfo {year} {2009})}\BibitemShut {NoStop}%
	\bibitem [{\citenamefont {Peleg}\ \emph {et~al.}(2011)\citenamefont {Peleg},
		\citenamefont {Disanza}, \citenamefont {Scita},\ and\ \citenamefont
		{Gov}}]{peleg2011propagating}%
	\BibitemOpen
	\bibfield  {author} {\bibinfo {author} {\bibfnamefont {B.}~\bibnamefont
			{Peleg}}, \bibinfo {author} {\bibfnamefont {A.}~\bibnamefont {Disanza}},
		\bibinfo {author} {\bibfnamefont {G.}~\bibnamefont {Scita}}, \ and\ \bibinfo
		{author} {\bibfnamefont {N.}~\bibnamefont {Gov}},\ }\href@noop {} {\bibfield
		{journal} {\bibinfo  {journal} {PloS one}\ }\textbf {\bibinfo {volume} {6}}
		(\bibinfo {year} {2011})}\BibitemShut {NoStop}%
	\bibitem [{\citenamefont {Naoz}\ and\ \citenamefont
		{Gov}(2020)}]{naoz2020cell}%
	\BibitemOpen
	\bibfield  {author} {\bibinfo {author} {\bibfnamefont {M.}~\bibnamefont
			{Naoz}}\ and\ \bibinfo {author} {\bibfnamefont {N.~S.}\ \bibnamefont {Gov}},\
	}\href@noop {} {\bibfield  {journal} {\bibinfo  {journal} {Cells}\ }\textbf
		{\bibinfo {volume} {9}},\ \bibinfo {pages} {782} (\bibinfo {year}
		{2020})}\BibitemShut {NoStop}%
	\bibitem [{\citenamefont {Wu}\ \emph {et~al.}(2018)\citenamefont {Wu},
		\citenamefont {Su}, \citenamefont {Tong}, \citenamefont {Wu},\ and\
		\citenamefont {Liu}}]{wu2018membrane}%
	\BibitemOpen
	\bibfield  {author} {\bibinfo {author} {\bibfnamefont {Z.}~\bibnamefont
			{Wu}}, \bibinfo {author} {\bibfnamefont {M.}~\bibnamefont {Su}}, \bibinfo
		{author} {\bibfnamefont {C.}~\bibnamefont {Tong}}, \bibinfo {author}
		{\bibfnamefont {M.}~\bibnamefont {Wu}}, \ and\ \bibinfo {author}
		{\bibfnamefont {J.}~\bibnamefont {Liu}},\ }\href@noop {} {\bibfield
		{journal} {\bibinfo  {journal} {Nature communications}\ }\textbf {\bibinfo
			{volume} {9}},\ \bibinfo {pages} {1} (\bibinfo {year} {2018})}\BibitemShut
	{NoStop}%
	\bibitem [{\citenamefont {Katsuno}\ \emph {et~al.}(2015)\citenamefont
		{Katsuno}, \citenamefont {Toriyama}, \citenamefont {Hosokawa}, \citenamefont
		{Mizuno}, \citenamefont {Ikeda}, \citenamefont {Sakumura},\ and\
		\citenamefont {Inagaki}}]{katsuno2015actin}%
	\BibitemOpen
	\bibfield  {author} {\bibinfo {author} {\bibfnamefont {H.}~\bibnamefont
			{Katsuno}}, \bibinfo {author} {\bibfnamefont {M.}~\bibnamefont {Toriyama}},
		\bibinfo {author} {\bibfnamefont {Y.}~\bibnamefont {Hosokawa}}, \bibinfo
		{author} {\bibfnamefont {K.}~\bibnamefont {Mizuno}}, \bibinfo {author}
		{\bibfnamefont {K.}~\bibnamefont {Ikeda}}, \bibinfo {author} {\bibfnamefont
			{Y.}~\bibnamefont {Sakumura}}, \ and\ \bibinfo {author} {\bibfnamefont
			{N.}~\bibnamefont {Inagaki}},\ }\href@noop {} {\bibfield  {journal} {\bibinfo
			{journal} {Cell reports}\ }\textbf {\bibinfo {volume} {12}},\ \bibinfo
		{pages} {648} (\bibinfo {year} {2015})}\BibitemShut {NoStop}%
	\bibitem [{\citenamefont {Chen}\ \emph {et~al.}(2009)\citenamefont {Chen},
		\citenamefont {Tsai}, \citenamefont {Wang},\ and\ \citenamefont
		{Lee}}]{chen2009three}%
	\BibitemOpen
	\bibfield  {author} {\bibinfo {author} {\bibfnamefont {C.-H.}\ \bibnamefont
			{Chen}}, \bibinfo {author} {\bibfnamefont {F.-C.}\ \bibnamefont {Tsai}},
		\bibinfo {author} {\bibfnamefont {C.-C.}\ \bibnamefont {Wang}}, \ and\
		\bibinfo {author} {\bibfnamefont {C.-H.}\ \bibnamefont {Lee}},\ }\href@noop
	{} {\bibfield  {journal} {\bibinfo  {journal} {Physical review letters}\
		}\textbf {\bibinfo {volume} {103}},\ \bibinfo {pages} {238101} (\bibinfo
		{year} {2009})}\BibitemShut {NoStop}%
	\bibitem [{\citenamefont {Beta}\ \emph {et~al.}(2008)\citenamefont {Beta},
		\citenamefont {Amselem},\ and\ \citenamefont
		{Bodenschatz}}]{beta_bistable_2008}%
	\BibitemOpen
	\bibfield  {author} {\bibinfo {author} {\bibfnamefont {C.}~\bibnamefont
			{Beta}}, \bibinfo {author} {\bibfnamefont {G.}~\bibnamefont {Amselem}}, \
		and\ \bibinfo {author} {\bibfnamefont {E.}~\bibnamefont {Bodenschatz}},\
	}\href {http://iopscience.iop.org/1367-2630/10/8/083015} {\bibfield
		{journal} {\bibinfo  {journal} {New Journal of Physics}\ }\textbf {\bibinfo
			{volume} {10}},\ \bibinfo {pages} {083015} (\bibinfo {year}
		{2008})}\BibitemShut {NoStop}%
	\bibitem [{\citenamefont {Xiong}\ \emph {et~al.}(2010)\citenamefont {Xiong},
		\citenamefont {Huang}, \citenamefont {Iglesias},\ and\ \citenamefont
		{Devreotes}}]{xiong_cells_2010}%
	\BibitemOpen
	\bibfield  {author} {\bibinfo {author} {\bibfnamefont {Y.}~\bibnamefont
			{Xiong}}, \bibinfo {author} {\bibfnamefont {C.-H.}\ \bibnamefont {Huang}},
		\bibinfo {author} {\bibfnamefont {P.~A.}\ \bibnamefont {Iglesias}}, \ and\
		\bibinfo {author} {\bibfnamefont {P.~N.}\ \bibnamefont {Devreotes}},\ }\href
	{http://www.pnas.org/content/107/40/17079} {\bibfield  {journal} {\bibinfo
			{journal} {Proceedings of the National Academy of Sciences}\ }\textbf
		{\bibinfo {volume} {107}},\ \bibinfo {pages} {17079} (\bibinfo {year}
		{2010})}\BibitemShut {NoStop}%
	\bibitem [{\citenamefont {Devreotes}\ \emph {et~al.}(2017)\citenamefont
		{Devreotes}, \citenamefont {Bhattacharya}, \citenamefont {Edwards},
		\citenamefont {Iglesias}, \citenamefont {Lampert},\ and\ \citenamefont
		{Miao}}]{devreotes_excitable_2017}%
	\BibitemOpen
	\bibfield  {author} {\bibinfo {author} {\bibfnamefont {P.~N.}\ \bibnamefont
			{Devreotes}}, \bibinfo {author} {\bibfnamefont {S.}~\bibnamefont
			{Bhattacharya}}, \bibinfo {author} {\bibfnamefont {M.}~\bibnamefont
			{Edwards}}, \bibinfo {author} {\bibfnamefont {P.~A.}\ \bibnamefont
			{Iglesias}}, \bibinfo {author} {\bibfnamefont {T.}~\bibnamefont {Lampert}}, \
		and\ \bibinfo {author} {\bibfnamefont {Y.}~\bibnamefont {Miao}},\ }\href
	{https://doi.org/10.1146/annurev-cellbio-100616-060739} {\bibfield  {journal}
		{\bibinfo  {journal} {Annual Review of Cell and Developmental Biology}\
		}\textbf {\bibinfo {volume} {33}},\ \bibinfo {pages} {103} (\bibinfo {year}
		{2017})}\BibitemShut {NoStop}%
	\bibitem [{\citenamefont {Sambeth}\ and\ \citenamefont
		{Baumgaertner}(2001)}]{sambeth2001autocatalytic}%
	\BibitemOpen
	\bibfield  {author} {\bibinfo {author} {\bibfnamefont {R.}~\bibnamefont
			{Sambeth}}\ and\ \bibinfo {author} {\bibfnamefont {A.}~\bibnamefont
			{Baumgaertner}},\ }\href@noop {} {\bibfield  {journal} {\bibinfo  {journal}
			{Physical review letters}\ }\textbf {\bibinfo {volume} {86}},\ \bibinfo
		{pages} {5196} (\bibinfo {year} {2001})}\BibitemShut {NoStop}%
	\bibitem [{\citenamefont {Dreher}\ \emph
		{et~al.}(2014{\natexlab{a}})\citenamefont {Dreher}, \citenamefont {Aranson},\
		and\ \citenamefont {Kruse}}]{Dreher2014}%
	\BibitemOpen
	\bibfield  {author} {\bibinfo {author} {\bibfnamefont {A.}~\bibnamefont
			{Dreher}}, \bibinfo {author} {\bibfnamefont {I.~S.}\ \bibnamefont {Aranson}},
		\ and\ \bibinfo {author} {\bibfnamefont {K.}~\bibnamefont {Kruse}},\
	}\href@noop {} {\bibfield  {journal} {\bibinfo  {journal} {New J. Phys.}\
		}\textbf {\bibinfo {volume} {16}},\ \bibinfo {pages} {055007} (\bibinfo
		{year} {2014}{\natexlab{a}})}\BibitemShut {NoStop}%
	\bibitem [{\citenamefont {Cao}\ \emph {et~al.}(2019)\citenamefont {Cao},
		\citenamefont {Ghabache},\ and\ \citenamefont {Rappel}}]{cao2019plasticity}%
	\BibitemOpen
	\bibfield  {author} {\bibinfo {author} {\bibfnamefont {Y.}~\bibnamefont
			{Cao}}, \bibinfo {author} {\bibfnamefont {E.}~\bibnamefont {Ghabache}}, \
		and\ \bibinfo {author} {\bibfnamefont {W.-J.}\ \bibnamefont {Rappel}},\
	}\href@noop {} {\bibfield  {journal} {\bibinfo  {journal} {Elife}\ }\textbf
		{\bibinfo {volume} {8}} (\bibinfo {year} {2019})}\BibitemShut {NoStop}%
	\bibitem [{\citenamefont {Mata}\ \emph {et~al.}(2013)\citenamefont {Mata},
		\citenamefont {Dutot}, \citenamefont {Edelstein-Keshet},\ and\ \citenamefont
		{Holmes}}]{Mata2013}%
	\BibitemOpen
	\bibfield  {author} {\bibinfo {author} {\bibfnamefont {M.~A.}\ \bibnamefont
			{Mata}}, \bibinfo {author} {\bibfnamefont {M.}~\bibnamefont {Dutot}},
		\bibinfo {author} {\bibfnamefont {L.}~\bibnamefont {Edelstein-Keshet}}, \
		and\ \bibinfo {author} {\bibfnamefont {W.~R.}\ \bibnamefont {Holmes}},\
	}\href@noop {} {\bibfield  {journal} {\bibinfo  {journal} {J. Theo. Biol.}\
		}\textbf {\bibinfo {volume} {334}},\ \bibinfo {pages} {149} (\bibinfo {year}
		{2013})}\BibitemShut {NoStop}%
	\bibitem [{\citenamefont {Wasnik}\ and\ \citenamefont
		{Mukhopadhyay}(2014)}]{Wasnik2014}%
	\BibitemOpen
	\bibfield  {author} {\bibinfo {author} {\bibfnamefont {V.}~\bibnamefont
			{Wasnik}}\ and\ \bibinfo {author} {\bibfnamefont {R.}~\bibnamefont
			{Mukhopadhyay}},\ }\href@noop {} {\bibfield  {journal} {\bibinfo  {journal}
			{Phys. Rev. E}\ }\textbf {\bibinfo {volume} {90}},\ \bibinfo {pages} {052707}
		(\bibinfo {year} {2014})}\BibitemShut {NoStop}%
	\bibitem [{\citenamefont {Meron}(2015)}]{meron2015book}%
	\BibitemOpen
	\bibfield  {author} {\bibinfo {author} {\bibfnamefont {E.}~\bibnamefont
			{Meron}},\ }\href@noop {} {\emph {\bibinfo {title} {Nonlinear physics of
				ecosystems}}}\ (\bibinfo  {publisher} {CRC Press},\ \bibinfo {year}
	{2015})\BibitemShut {NoStop}%
	\bibitem [{\citenamefont {Jilkine}\ and\ \citenamefont
		{Edelstein-Keshet}(2011)}]{jilkine2011comparison}%
	\BibitemOpen
	\bibfield  {author} {\bibinfo {author} {\bibfnamefont {A.}~\bibnamefont
			{Jilkine}}\ and\ \bibinfo {author} {\bibfnamefont {L.}~\bibnamefont
			{Edelstein-Keshet}},\ }\href@noop {} {\bibfield  {journal} {\bibinfo
			{journal} {PLoS computational biology}\ }\textbf {\bibinfo {volume} {7}}
		(\bibinfo {year} {2011})}\BibitemShut {NoStop}%
	\bibitem [{\citenamefont {Halatek}\ and\ \citenamefont
		{Frey}(2018)}]{halatek2018rethinking}%
	\BibitemOpen
	\bibfield  {author} {\bibinfo {author} {\bibfnamefont {J.}~\bibnamefont
			{Halatek}}\ and\ \bibinfo {author} {\bibfnamefont {E.}~\bibnamefont {Frey}},\
	}\href@noop {} {\bibfield  {journal} {\bibinfo  {journal} {Nature Physics}\
		}\textbf {\bibinfo {volume} {14}},\ \bibinfo {pages} {507} (\bibinfo {year}
		{2018})}\BibitemShut {NoStop}%
	\bibitem [{\citenamefont {Golovin}\ \emph
		{et~al.}(2003{\natexlab{a}})\citenamefont {Golovin}, \citenamefont {Davis},\
		and\ \citenamefont {Voorhees}}]{golovin2003self}%
	\BibitemOpen
	\bibfield  {author} {\bibinfo {author} {\bibfnamefont {A.}~\bibnamefont
			{Golovin}}, \bibinfo {author} {\bibfnamefont {S.~H.}\ \bibnamefont {Davis}},
		\ and\ \bibinfo {author} {\bibfnamefont {P.~W.}\ \bibnamefont {Voorhees}},\
	}\href@noop {} {\bibfield  {journal} {\bibinfo  {journal} {Physical Review
				E}\ }\textbf {\bibinfo {volume} {68}},\ \bibinfo {pages} {056203} (\bibinfo
		{year} {2003}{\natexlab{a}})}\BibitemShut {NoStop}%
	\bibitem [{\citenamefont {Ziebert}\ and\ \citenamefont
		{Zimmermann}(2004)}]{ziebert2004pattern}%
	\BibitemOpen
	\bibfield  {author} {\bibinfo {author} {\bibfnamefont {F.}~\bibnamefont
			{Ziebert}}\ and\ \bibinfo {author} {\bibfnamefont {W.}~\bibnamefont
			{Zimmermann}},\ }\href@noop {} {\bibfield  {journal} {\bibinfo  {journal}
			{Physical Review E}\ }\textbf {\bibinfo {volume} {70}},\ \bibinfo {pages}
		{022902} (\bibinfo {year} {2004})}\BibitemShut {NoStop}%
	\bibitem [{\citenamefont {Golovin}\ \emph {et~al.}(2004)\citenamefont
		{Golovin}, \citenamefont {Levine}, \citenamefont {Savina},\ and\
		\citenamefont {Davis}}]{golovin2004faceting}%
	\BibitemOpen
	\bibfield  {author} {\bibinfo {author} {\bibfnamefont {A.}~\bibnamefont
			{Golovin}}, \bibinfo {author} {\bibfnamefont {M.}~\bibnamefont {Levine}},
		\bibinfo {author} {\bibfnamefont {T.}~\bibnamefont {Savina}}, \ and\ \bibinfo
		{author} {\bibfnamefont {S.~H.}\ \bibnamefont {Davis}},\ }\href@noop {}
	{\bibfield  {journal} {\bibinfo  {journal} {Physical Review B}\ }\textbf
		{\bibinfo {volume} {70}},\ \bibinfo {pages} {235342} (\bibinfo {year}
		{2004})}\BibitemShut {NoStop}%
	\bibitem [{\citenamefont {Weliwita}\ \emph {et~al.}(2011)\citenamefont
		{Weliwita}, \citenamefont {Rucklidge},\ and\ \citenamefont
		{Tobias}}]{weliwita2011skew}%
	\BibitemOpen
	\bibfield  {author} {\bibinfo {author} {\bibfnamefont {J.}~\bibnamefont
			{Weliwita}}, \bibinfo {author} {\bibfnamefont {A.}~\bibnamefont {Rucklidge}},
		\ and\ \bibinfo {author} {\bibfnamefont {S.}~\bibnamefont {Tobias}},\
	}\href@noop {} {\bibfield  {journal} {\bibinfo  {journal} {Physical Review
				E}\ }\textbf {\bibinfo {volume} {84}},\ \bibinfo {pages} {036201} (\bibinfo
		{year} {2011})}\BibitemShut {NoStop}%
	\bibitem [{\citenamefont {Robbins}\ \emph {et~al.}(2012)\citenamefont
		{Robbins}, \citenamefont {Archer}, \citenamefont {Thiele},\ and\
		\citenamefont {Knobloch}}]{robbins2012modeling}%
	\BibitemOpen
	\bibfield  {author} {\bibinfo {author} {\bibfnamefont {M.~J.}\ \bibnamefont
			{Robbins}}, \bibinfo {author} {\bibfnamefont {A.~J.}\ \bibnamefont {Archer}},
		\bibinfo {author} {\bibfnamefont {U.}~\bibnamefont {Thiele}}, \ and\ \bibinfo
		{author} {\bibfnamefont {E.}~\bibnamefont {Knobloch}},\ }\href@noop {}
	{\bibfield  {journal} {\bibinfo  {journal} {Physical Review E}\ }\textbf
		{\bibinfo {volume} {85}},\ \bibinfo {pages} {061408} (\bibinfo {year}
		{2012})}\BibitemShut {NoStop}%
	\bibitem [{\citenamefont {Thiele}\ \emph {et~al.}(2019)\citenamefont {Thiele},
		\citenamefont {Frohoff-H{\"u}lsmann}, \citenamefont {Engelnkemper},
		\citenamefont {Knobloch},\ and\ \citenamefont {Archer}}]{thiele2019first}%
	\BibitemOpen
	\bibfield  {author} {\bibinfo {author} {\bibfnamefont {U.}~\bibnamefont
			{Thiele}}, \bibinfo {author} {\bibfnamefont {T.}~\bibnamefont
			{Frohoff-H{\"u}lsmann}}, \bibinfo {author} {\bibfnamefont {S.}~\bibnamefont
			{Engelnkemper}}, \bibinfo {author} {\bibfnamefont {E.}~\bibnamefont
			{Knobloch}}, \ and\ \bibinfo {author} {\bibfnamefont {A.~J.}\ \bibnamefont
			{Archer}},\ }\href@noop {} {\bibfield  {journal} {\bibinfo  {journal} {New
				Journal of Physics}\ }\textbf {\bibinfo {volume} {21}},\ \bibinfo {pages}
		{123021} (\bibinfo {year} {2019})}\BibitemShut {NoStop}%
	\bibitem [{\citenamefont {Barker}\ \emph {et~al.}(2018)\citenamefont {Barker},
		\citenamefont {Jung},\ and\ \citenamefont {Zumbrun}}]{barker2018turing}%
	\BibitemOpen
	\bibfield  {author} {\bibinfo {author} {\bibfnamefont {B.}~\bibnamefont
			{Barker}}, \bibinfo {author} {\bibfnamefont {S.}~\bibnamefont {Jung}}, \ and\
		\bibinfo {author} {\bibfnamefont {K.}~\bibnamefont {Zumbrun}},\ }\href@noop
	{} {\bibfield  {journal} {\bibinfo  {journal} {Physica D: Nonlinear
				Phenomena}\ }\textbf {\bibinfo {volume} {367}},\ \bibinfo {pages} {11}
		(\bibinfo {year} {2018})}\BibitemShut {NoStop}%
	\bibitem [{\citenamefont {Hilder}(2018)}]{hilder2018modulating}%
	\BibitemOpen
	\bibfield  {author} {\bibinfo {author} {\bibfnamefont {B.}~\bibnamefont
			{Hilder}},\ }\href@noop {} {\bibfield  {journal} {\bibinfo  {journal} {arXiv
				preprint arXiv:1811.12178}\ } (\bibinfo {year} {2018})}\BibitemShut {NoStop}%
	\bibitem [{\citenamefont {Tribelsky}\ and\ \citenamefont
		{Velarde}(1996)}]{tribelsky1996short}%
	\BibitemOpen
	\bibfield  {author} {\bibinfo {author} {\bibfnamefont {M.~I.}\ \bibnamefont
			{Tribelsky}}\ and\ \bibinfo {author} {\bibfnamefont {M.~G.}\ \bibnamefont
			{Velarde}},\ }\href@noop {} {\bibfield  {journal} {\bibinfo  {journal}
			{Physical Review E}\ }\textbf {\bibinfo {volume} {54}},\ \bibinfo {pages}
		{4973} (\bibinfo {year} {1996})}\BibitemShut {NoStop}%
	\bibitem [{\citenamefont {Matthews}\ and\ \citenamefont
		{Cox}(2000)}]{matthews2000pattern}%
	\BibitemOpen
	\bibfield  {author} {\bibinfo {author} {\bibfnamefont {P.}~\bibnamefont
			{Matthews}}\ and\ \bibinfo {author} {\bibfnamefont {S.~M.}\ \bibnamefont
			{Cox}},\ }\href@noop {} {\bibfield  {journal} {\bibinfo  {journal}
			{Nonlinearity}\ }\textbf {\bibinfo {volume} {13}},\ \bibinfo {pages} {1293}
		(\bibinfo {year} {2000})}\BibitemShut {NoStop}%
	\bibitem [{\citenamefont {Cox}\ and\ \citenamefont
		{Matthews}(2003)}]{cox2003instability}%
	\BibitemOpen
	\bibfield  {author} {\bibinfo {author} {\bibfnamefont {S.}~\bibnamefont
			{Cox}}\ and\ \bibinfo {author} {\bibfnamefont {P.}~\bibnamefont {Matthews}},\
	}\href@noop {} {\bibfield  {journal} {\bibinfo  {journal} {Physica D:
				Nonlinear Phenomena}\ }\textbf {\bibinfo {volume} {175}},\ \bibinfo {pages}
		{196} (\bibinfo {year} {2003})}\BibitemShut {NoStop}%
	\bibitem [{\citenamefont {Cox}(2004)}]{cox2004envelope}%
	\BibitemOpen
	\bibfield  {author} {\bibinfo {author} {\bibfnamefont {S.~M.}\ \bibnamefont
			{Cox}},\ }\href@noop {} {\bibfield  {journal} {\bibinfo  {journal} {Physics
				Letters A}\ }\textbf {\bibinfo {volume} {333}},\ \bibinfo {pages} {91}
		(\bibinfo {year} {2004})}\BibitemShut {NoStop}%
	\bibitem [{\citenamefont {Shiwa}(2005)}]{shiwa2005hydrodynamic}%
	\BibitemOpen
	\bibfield  {author} {\bibinfo {author} {\bibfnamefont {Y.}~\bibnamefont
			{Shiwa}},\ }\href@noop {} {\bibfield  {journal} {\bibinfo  {journal}
			{Physical Review E}\ }\textbf {\bibinfo {volume} {72}},\ \bibinfo {pages}
		{016204} (\bibinfo {year} {2005})}\BibitemShut {NoStop}%
	\bibitem [{\citenamefont {Dawes}(2008)}]{dawes2008localized}%
	\BibitemOpen
	\bibfield  {author} {\bibinfo {author} {\bibfnamefont {J.~H.}\ \bibnamefont
			{Dawes}},\ }\href@noop {} {\bibfield  {journal} {\bibinfo  {journal} {SIAM
				Journal on Applied Dynamical Systems}\ }\textbf {\bibinfo {volume} {7}},\
		\bibinfo {pages} {186} (\bibinfo {year} {2008})}\BibitemShut {NoStop}%
	\bibitem [{\citenamefont {Ohnogi}\ and\ \citenamefont
		{Shiwa}(2008)}]{ohnogi2008instability}%
	\BibitemOpen
	\bibfield  {author} {\bibinfo {author} {\bibfnamefont {H.}~\bibnamefont
			{Ohnogi}}\ and\ \bibinfo {author} {\bibfnamefont {Y.}~\bibnamefont {Shiwa}},\
	}\href@noop {} {\bibfield  {journal} {\bibinfo  {journal} {Physica D:
				Nonlinear Phenomena}\ }\textbf {\bibinfo {volume} {237}},\ \bibinfo {pages}
		{3046} (\bibinfo {year} {2008})}\BibitemShut {NoStop}%
	\bibitem [{\citenamefont {Golovin}\ \emph {et~al.}(2009)\citenamefont
		{Golovin}, \citenamefont {Kanevsky},\ and\ \citenamefont
		{Nepomnyashchy}}]{golovin2009feedback}%
	\BibitemOpen
	\bibfield  {author} {\bibinfo {author} {\bibfnamefont {A.}~\bibnamefont
			{Golovin}}, \bibinfo {author} {\bibfnamefont {Y.}~\bibnamefont {Kanevsky}}, \
		and\ \bibinfo {author} {\bibfnamefont {A.}~\bibnamefont {Nepomnyashchy}},\
	}\href@noop {} {\bibfield  {journal} {\bibinfo  {journal} {Physical Review
				E}\ }\textbf {\bibinfo {volume} {79}},\ \bibinfo {pages} {046218} (\bibinfo
		{year} {2009})}\BibitemShut {NoStop}%
	\bibitem [{\citenamefont {Huang}\ \emph {et~al.}(2010)\citenamefont {Huang},
		\citenamefont {Elder},\ and\ \citenamefont {Provatas}}]{huang2010phase}%
	\BibitemOpen
	\bibfield  {author} {\bibinfo {author} {\bibfnamefont {Z.-F.}\ \bibnamefont
			{Huang}}, \bibinfo {author} {\bibfnamefont {K.}~\bibnamefont {Elder}}, \ and\
		\bibinfo {author} {\bibfnamefont {N.}~\bibnamefont {Provatas}},\ }\href@noop
	{} {\bibfield  {journal} {\bibinfo  {journal} {Physical Review E}\ }\textbf
		{\bibinfo {volume} {82}},\ \bibinfo {pages} {021605} (\bibinfo {year}
		{2010})}\BibitemShut {NoStop}%
	\bibitem [{\citenamefont {Kanevsky}\ and\ \citenamefont
		{Nepomnyashchy}(2011)}]{kanevsky2011patterns}%
	\BibitemOpen
	\bibfield  {author} {\bibinfo {author} {\bibfnamefont {Y.}~\bibnamefont
			{Kanevsky}}\ and\ \bibinfo {author} {\bibfnamefont {A.}~\bibnamefont
			{Nepomnyashchy}},\ }\href@noop {} {\bibfield  {journal} {\bibinfo  {journal}
			{Mathematical Modelling of Natural Phenomena}\ }\textbf {\bibinfo {volume}
			{6}},\ \bibinfo {pages} {188} (\bibinfo {year} {2011})}\BibitemShut {NoStop}%
	\bibitem [{\citenamefont {Thiele}\ \emph {et~al.}(2013)\citenamefont {Thiele},
		\citenamefont {Archer}, \citenamefont {Robbins}, \citenamefont {Gomez},\ and\
		\citenamefont {Knobloch}}]{thiele2013localized}%
	\BibitemOpen
	\bibfield  {author} {\bibinfo {author} {\bibfnamefont {U.}~\bibnamefont
			{Thiele}}, \bibinfo {author} {\bibfnamefont {A.~J.}\ \bibnamefont {Archer}},
		\bibinfo {author} {\bibfnamefont {M.~J.}\ \bibnamefont {Robbins}}, \bibinfo
		{author} {\bibfnamefont {H.}~\bibnamefont {Gomez}}, \ and\ \bibinfo {author}
		{\bibfnamefont {E.}~\bibnamefont {Knobloch}},\ }\href@noop {} {\bibfield
		{journal} {\bibinfo  {journal} {Physical Review E}\ }\textbf {\bibinfo
			{volume} {87}},\ \bibinfo {pages} {042915} (\bibinfo {year}
		{2013})}\BibitemShut {NoStop}%
	\bibitem [{\citenamefont {Knobloch}(2016)}]{knobloch2016localized}%
	\BibitemOpen
	\bibfield  {author} {\bibinfo {author} {\bibfnamefont {E.}~\bibnamefont
			{Knobloch}},\ }\href@noop {} {\bibfield  {journal} {\bibinfo  {journal} {IMA
				Journal of Applied Mathematics}\ }\textbf {\bibinfo {volume} {81}},\ \bibinfo
		{pages} {457} (\bibinfo {year} {2016})}\BibitemShut {NoStop}%
	\bibitem [{\citenamefont {Schneider}\ and\ \citenamefont
		{Zimmermann}(2016)}]{schneider2016turing}%
	\BibitemOpen
	\bibfield  {author} {\bibinfo {author} {\bibfnamefont {G.}~\bibnamefont
			{Schneider}}\ and\ \bibinfo {author} {\bibfnamefont {D.}~\bibnamefont
			{Zimmermann}},\ }in\ \href@noop {} {\emph {\bibinfo {booktitle}
			{International Conference on Patterns of Dynamics}}}\ (\bibinfo
	{organization} {Springer},\ \bibinfo {year} {2016})\ pp.\ \bibinfo {pages}
	{28--43}\BibitemShut {NoStop}%
	\bibitem [{\citenamefont {Firth}\ \emph {et~al.}(2007)\citenamefont {Firth},
		\citenamefont {Columbo},\ and\ \citenamefont
		{Maggipinto}}]{firth2007homoclinic}%
	\BibitemOpen
	\bibfield  {author} {\bibinfo {author} {\bibfnamefont {W.}~\bibnamefont
			{Firth}}, \bibinfo {author} {\bibfnamefont {L.}~\bibnamefont {Columbo}}, \
		and\ \bibinfo {author} {\bibfnamefont {T.}~\bibnamefont {Maggipinto}},\
	}\href@noop {} {\bibfield  {journal} {\bibinfo  {journal} {Chaos: An
				Interdisciplinary Journal of Nonlinear Science}\ }\textbf {\bibinfo {volume}
			{17}},\ \bibinfo {pages} {037115} (\bibinfo {year} {2007})}\BibitemShut
	{NoStop}%
	\bibitem [{\citenamefont {Winterbottom}(2006)}]{winterbottom2006pattern}%
	\BibitemOpen
	\bibfield  {author} {\bibinfo {author} {\bibfnamefont {D.~M.}\ \bibnamefont
			{Winterbottom}},\ }\emph {\bibinfo {title} {Pattern formation with a
			conservation law}},\ \href@noop {} {Ph.D. thesis},\ \bibinfo  {school}
	{University of Nottingham} (\bibinfo {year} {2006})\BibitemShut {NoStop}%
	\bibitem [{\citenamefont {Matthews}\ and\ \citenamefont
		{Ruckllidge}(1993)}]{matthews1993travelling}%
	\BibitemOpen
	\bibfield  {author} {\bibinfo {author} {\bibfnamefont {P.}~\bibnamefont
			{Matthews}}\ and\ \bibinfo {author} {\bibfnamefont {A.}~\bibnamefont
			{Ruckllidge}},\ }\href@noop {} {\bibfield  {journal} {\bibinfo  {journal}
			{Proceedings of the Royal Society of London. Series A: Mathematical and
				Physical Sciences}\ }\textbf {\bibinfo {volume} {441}},\ \bibinfo {pages}
		{649} (\bibinfo {year} {1993})}\BibitemShut {NoStop}%
	\bibitem [{\citenamefont {Cox}\ and\ \citenamefont
		{Matthews}(2000)}]{cox2000instability}%
	\BibitemOpen
	\bibfield  {author} {\bibinfo {author} {\bibfnamefont {S.}~\bibnamefont
			{Cox}}\ and\ \bibinfo {author} {\bibfnamefont {P.}~\bibnamefont {Matthews}},\
	}\href@noop {} {\bibfield  {journal} {\bibinfo  {journal} {Journal of Fluid
				Mechanics}\ }\textbf {\bibinfo {volume} {403}},\ \bibinfo {pages} {153}
		(\bibinfo {year} {2000})}\BibitemShut {NoStop}%
	\bibitem [{\citenamefont {Cox}\ \emph {et~al.}(2004)\citenamefont {Cox},
		\citenamefont {Matthews},\ and\ \citenamefont {Pollicott}}]{cox2004swift}%
	\BibitemOpen
	\bibfield  {author} {\bibinfo {author} {\bibfnamefont {S.~M.}\ \bibnamefont
			{Cox}}, \bibinfo {author} {\bibfnamefont {P.}~\bibnamefont {Matthews}}, \
		and\ \bibinfo {author} {\bibfnamefont {S.}~\bibnamefont {Pollicott}},\
	}\href@noop {} {\bibfield  {journal} {\bibinfo  {journal} {Physical Review
				E}\ }\textbf {\bibinfo {volume} {69}},\ \bibinfo {pages} {066314} (\bibinfo
		{year} {2004})}\BibitemShut {NoStop}%
	\bibitem [{\citenamefont {Jacono}\ \emph {et~al.}(2011)\citenamefont {Jacono},
		\citenamefont {Bergeon},\ and\ \citenamefont
		{Knobloch}}]{jacono2011magnetohydrodynamic}%
	\BibitemOpen
	\bibfield  {author} {\bibinfo {author} {\bibfnamefont {D.~L.}\ \bibnamefont
			{Jacono}}, \bibinfo {author} {\bibfnamefont {A.}~\bibnamefont {Bergeon}}, \
		and\ \bibinfo {author} {\bibfnamefont {E.}~\bibnamefont {Knobloch}},\
	}\href@noop {} {\bibfield  {journal} {\bibinfo  {journal} {Journal of Fluid
				Mechanics}\ }\textbf {\bibinfo {volume} {687}},\ \bibinfo {pages} {595}
		(\bibinfo {year} {2011})}\BibitemShut {NoStop}%
	\bibitem [{\citenamefont {Golovin}\ \emph {et~al.}(2001)\citenamefont
		{Golovin}, \citenamefont {Nepomnyashchy},\ and\ \citenamefont
		{Matkowsky}}]{golovin2001traveling}%
	\BibitemOpen
	\bibfield  {author} {\bibinfo {author} {\bibfnamefont {A.}~\bibnamefont
			{Golovin}}, \bibinfo {author} {\bibfnamefont {A.}~\bibnamefont
			{Nepomnyashchy}}, \ and\ \bibinfo {author} {\bibfnamefont {B.~J.}\
			\bibnamefont {Matkowsky}},\ }\href@noop {} {\bibfield  {journal} {\bibinfo
			{journal} {Physica D: Nonlinear Phenomena}\ }\textbf {\bibinfo {volume}
			{160}},\ \bibinfo {pages} {1} (\bibinfo {year} {2001})}\BibitemShut {NoStop}%
	\bibitem [{\citenamefont {Golovin}\ \emph
		{et~al.}(2003{\natexlab{b}})\citenamefont {Golovin}, \citenamefont
		{Matkowsky},\ and\ \citenamefont {Nepomnyashchy}}]{golovin2003complex}%
	\BibitemOpen
	\bibfield  {author} {\bibinfo {author} {\bibfnamefont {A.}~\bibnamefont
			{Golovin}}, \bibinfo {author} {\bibfnamefont {B.~J.}\ \bibnamefont
			{Matkowsky}}, \ and\ \bibinfo {author} {\bibfnamefont {A.}~\bibnamefont
			{Nepomnyashchy}},\ }\href@noop {} {\bibfield  {journal} {\bibinfo  {journal}
			{Physica D: Nonlinear Phenomena}\ }\textbf {\bibinfo {volume} {179}},\
		\bibinfo {pages} {183} (\bibinfo {year} {2003}{\natexlab{b}})}\BibitemShut
	{NoStop}%
	\bibitem [{\citenamefont {Tsimring}\ and\ \citenamefont
		{Aranson}(1997)}]{tsimring1997localized}%
	\BibitemOpen
	\bibfield  {author} {\bibinfo {author} {\bibfnamefont {L.~S.}\ \bibnamefont
			{Tsimring}}\ and\ \bibinfo {author} {\bibfnamefont {I.~S.}\ \bibnamefont
			{Aranson}},\ }\href@noop {} {\bibfield  {journal} {\bibinfo  {journal}
			{Physical review letters}\ }\textbf {\bibinfo {volume} {79}},\ \bibinfo
		{pages} {213} (\bibinfo {year} {1997})}\BibitemShut {NoStop}%
	\bibitem [{\citenamefont {Snezhko}\ \emph {et~al.}(2006)\citenamefont
		{Snezhko}, \citenamefont {Aranson},\ and\ \citenamefont
		{Kwok}}]{snezhko2006surface}%
	\BibitemOpen
	\bibfield  {author} {\bibinfo {author} {\bibfnamefont {A.}~\bibnamefont
			{Snezhko}}, \bibinfo {author} {\bibfnamefont {I.}~\bibnamefont {Aranson}}, \
		and\ \bibinfo {author} {\bibfnamefont {W.-K.}\ \bibnamefont {Kwok}},\
	}\href@noop {} {\bibfield  {journal} {\bibinfo  {journal} {Physical Review
				Letters}\ }\textbf {\bibinfo {volume} {96}},\ \bibinfo {pages} {078701}
		(\bibinfo {year} {2006})}\BibitemShut {NoStop}%
	\bibitem [{\citenamefont {Pradenas}\ \emph {et~al.}(2017)\citenamefont
		{Pradenas}, \citenamefont {Araya}, \citenamefont {Clerc}, \citenamefont
		{Falc{\'o}n}, \citenamefont {Gandhi},\ and\ \citenamefont
		{Knobloch}}]{pradenas2017slanted}%
	\BibitemOpen
	\bibfield  {author} {\bibinfo {author} {\bibfnamefont {B.}~\bibnamefont
			{Pradenas}}, \bibinfo {author} {\bibfnamefont {I.}~\bibnamefont {Araya}},
		\bibinfo {author} {\bibfnamefont {M.~G.}\ \bibnamefont {Clerc}}, \bibinfo
		{author} {\bibfnamefont {C.}~\bibnamefont {Falc{\'o}n}}, \bibinfo {author}
		{\bibfnamefont {P.}~\bibnamefont {Gandhi}}, \ and\ \bibinfo {author}
		{\bibfnamefont {E.}~\bibnamefont {Knobloch}},\ }\href@noop {} {\bibfield
		{journal} {\bibinfo  {journal} {Physical Review Fluids}\ }\textbf {\bibinfo
			{volume} {2}},\ \bibinfo {pages} {064401} (\bibinfo {year}
		{2017})}\BibitemShut {NoStop}%
	\bibitem [{\citenamefont {Kramer}\ and\ \citenamefont
		{Kree}(2002)}]{kramer2002pattern}%
	\BibitemOpen
	\bibfield  {author} {\bibinfo {author} {\bibfnamefont {S.~C.}\ \bibnamefont
			{Kramer}}\ and\ \bibinfo {author} {\bibfnamefont {R.}~\bibnamefont {Kree}},\
	}\href@noop {} {\bibfield  {journal} {\bibinfo  {journal} {Physical Review
				E}\ }\textbf {\bibinfo {volume} {65}},\ \bibinfo {pages} {051920} (\bibinfo
		{year} {2002})}\BibitemShut {NoStop}%
	\bibitem [{\citenamefont {Peter}\ and\ \citenamefont
		{Zimmermann}(2006)}]{peter2006traveling}%
	\BibitemOpen
	\bibfield  {author} {\bibinfo {author} {\bibfnamefont {R.}~\bibnamefont
			{Peter}}\ and\ \bibinfo {author} {\bibfnamefont {W.}~\bibnamefont
			{Zimmermann}},\ }\href@noop {} {\bibfield  {journal} {\bibinfo  {journal}
			{Physical Review E}\ }\textbf {\bibinfo {volume} {74}},\ \bibinfo {pages}
		{016206} (\bibinfo {year} {2006})}\BibitemShut {NoStop}%
	\bibitem [{\citenamefont {Coullet}\ and\ \citenamefont
		{Fauve}(1985)}]{coullet1985propagative}%
	\BibitemOpen
	\bibfield  {author} {\bibinfo {author} {\bibfnamefont {P.}~\bibnamefont
			{Coullet}}\ and\ \bibinfo {author} {\bibfnamefont {S.}~\bibnamefont
			{Fauve}},\ }\href@noop {} {\bibfield  {journal} {\bibinfo  {journal}
			{Physical review letters}\ }\textbf {\bibinfo {volume} {55}},\ \bibinfo
		{pages} {2857} (\bibinfo {year} {1985})}\BibitemShut {NoStop}%
	\bibitem [{\citenamefont {Riecke}(1996)}]{riecke1996solitary}%
	\BibitemOpen
	\bibfield  {author} {\bibinfo {author} {\bibfnamefont {H.}~\bibnamefont
			{Riecke}},\ }\href@noop {} {\bibfield  {journal} {\bibinfo  {journal}
			{Physica D: Nonlinear Phenomena}\ }\textbf {\bibinfo {volume} {92}},\
		\bibinfo {pages} {69} (\bibinfo {year} {1996})}\BibitemShut {NoStop}%
	\bibitem [{\citenamefont {Ipsen}\ and\ \citenamefont
		{S{\o}rensen}(2000)}]{ipsen2000finite}%
	\BibitemOpen
	\bibfield  {author} {\bibinfo {author} {\bibfnamefont {M.}~\bibnamefont
			{Ipsen}}\ and\ \bibinfo {author} {\bibfnamefont {P.~G.}\ \bibnamefont
			{S{\o}rensen}},\ }\href@noop {} {\bibfield  {journal} {\bibinfo  {journal}
			{Physical Review Letters}\ }\textbf {\bibinfo {volume} {84}},\ \bibinfo
		{pages} {2389} (\bibinfo {year} {2000})}\BibitemShut {NoStop}%
	\bibitem [{\citenamefont {Winterbottom}\ \emph {et~al.}(2005)\citenamefont
		{Winterbottom}, \citenamefont {Matthews},\ and\ \citenamefont
		{Cox}}]{winterbottom2005oscillatory}%
	\BibitemOpen
	\bibfield  {author} {\bibinfo {author} {\bibfnamefont {D.}~\bibnamefont
			{Winterbottom}}, \bibinfo {author} {\bibfnamefont {P.}~\bibnamefont
			{Matthews}}, \ and\ \bibinfo {author} {\bibfnamefont {S.~M.}\ \bibnamefont
			{Cox}},\ }\href@noop {} {\bibfield  {journal} {\bibinfo  {journal}
			{Nonlinearity}\ }\textbf {\bibinfo {volume} {18}},\ \bibinfo {pages} {1031}
		(\bibinfo {year} {2005})}\BibitemShut {NoStop}%
	\bibitem [{\citenamefont {Hek}\ and\ \citenamefont
		{Valkhoff}(2007)}]{hek2007pulses}%
	\BibitemOpen
	\bibfield  {author} {\bibinfo {author} {\bibfnamefont {G.}~\bibnamefont
			{Hek}}\ and\ \bibinfo {author} {\bibfnamefont {N.}~\bibnamefont {Valkhoff}},\
	}\href@noop {} {\bibfield  {journal} {\bibinfo  {journal} {Physica D:
				Nonlinear Phenomena}\ }\textbf {\bibinfo {volume} {232}},\ \bibinfo {pages}
		{62} (\bibinfo {year} {2007})}\BibitemShut {NoStop}%
	\bibitem [{\citenamefont {Nepomnyashchy}\ and\ \citenamefont
		{Shklyaev}(2016)}]{nepomnyashchy2016longwave}%
	\BibitemOpen
	\bibfield  {author} {\bibinfo {author} {\bibfnamefont {A.}~\bibnamefont
			{Nepomnyashchy}}\ and\ \bibinfo {author} {\bibfnamefont {S.}~\bibnamefont
			{Shklyaev}},\ }\href@noop {} {\bibfield  {journal} {\bibinfo  {journal}
			{Journal of Physics A: Mathematical and Theoretical}\ }\textbf {\bibinfo
			{volume} {49}},\ \bibinfo {pages} {053001} (\bibinfo {year}
		{2016})}\BibitemShut {NoStop}%
	\bibitem [{\citenamefont {Koch}\ and\ \citenamefont
		{Meinhardt}(1994)}]{koch1994biological}%
	\BibitemOpen
	\bibfield  {author} {\bibinfo {author} {\bibfnamefont {A.}~\bibnamefont
			{Koch}}\ and\ \bibinfo {author} {\bibfnamefont {H.}~\bibnamefont
			{Meinhardt}},\ }\href@noop {} {\bibfield  {journal} {\bibinfo  {journal}
			{Reviews of Modern Physics}\ }\textbf {\bibinfo {volume} {66}},\ \bibinfo
		{pages} {1481} (\bibinfo {year} {1994})}\BibitemShut {NoStop}%
	\bibitem [{\citenamefont {Maini}\ \emph {et~al.}(1997)\citenamefont {Maini},
		\citenamefont {Painter},\ and\ \citenamefont {Chau}}]{maini1997spatial}%
	\BibitemOpen
	\bibfield  {author} {\bibinfo {author} {\bibfnamefont {P.}~\bibnamefont
			{Maini}}, \bibinfo {author} {\bibfnamefont {K.}~\bibnamefont {Painter}}, \
		and\ \bibinfo {author} {\bibfnamefont {H.~P.}\ \bibnamefont {Chau}},\
	}\href@noop {} {\bibfield  {journal} {\bibinfo  {journal} {Journal of the
				Chemical Society, Faraday Transactions}\ }\textbf {\bibinfo {volume} {93}},\
		\bibinfo {pages} {3601} (\bibinfo {year} {1997})}\BibitemShut {NoStop}%
	\bibitem [{\citenamefont {Volpert}\ and\ \citenamefont
		{Petrovskii}(2009)}]{volpert2009reaction}%
	\BibitemOpen
	\bibfield  {author} {\bibinfo {author} {\bibfnamefont {V.}~\bibnamefont
			{Volpert}}\ and\ \bibinfo {author} {\bibfnamefont {S.}~\bibnamefont
			{Petrovskii}},\ }\href@noop {} {\bibfield  {journal} {\bibinfo  {journal}
			{Physics of Life Reviews}\ }\textbf {\bibinfo {volume} {6}},\ \bibinfo
		{pages} {267} (\bibinfo {year} {2009})}\BibitemShut {NoStop}%
	\bibitem [{\citenamefont {Baker}\ \emph {et~al.}(2008)\citenamefont {Baker},
		\citenamefont {Gaffney},\ and\ \citenamefont {Maini}}]{baker2008partial}%
	\BibitemOpen
	\bibfield  {author} {\bibinfo {author} {\bibfnamefont {R.~E.}\ \bibnamefont
			{Baker}}, \bibinfo {author} {\bibfnamefont {E.}~\bibnamefont {Gaffney}}, \
		and\ \bibinfo {author} {\bibfnamefont {P.}~\bibnamefont {Maini}},\
	}\href@noop {} {\bibfield  {journal} {\bibinfo  {journal} {Nonlinearity}\
		}\textbf {\bibinfo {volume} {21}},\ \bibinfo {pages} {R251} (\bibinfo {year}
		{2008})}\BibitemShut {NoStop}%
	\bibitem [{\citenamefont {Otsuji}\ \emph {et~al.}(2007)\citenamefont {Otsuji},
		\citenamefont {Ishihara}, \citenamefont {Co}, \citenamefont {Kaibuchi},
		\citenamefont {Mochizuki},\ and\ \citenamefont {Kuroda}}]{otsuji2007mass}%
	\BibitemOpen
	\bibfield  {author} {\bibinfo {author} {\bibfnamefont {M.}~\bibnamefont
			{Otsuji}}, \bibinfo {author} {\bibfnamefont {S.}~\bibnamefont {Ishihara}},
		\bibinfo {author} {\bibfnamefont {C.}~\bibnamefont {Co}}, \bibinfo {author}
		{\bibfnamefont {K.}~\bibnamefont {Kaibuchi}}, \bibinfo {author}
		{\bibfnamefont {A.}~\bibnamefont {Mochizuki}}, \ and\ \bibinfo {author}
		{\bibfnamefont {S.}~\bibnamefont {Kuroda}},\ }\href@noop {} {\bibfield
		{journal} {\bibinfo  {journal} {PLoS Computational Biology}\ }\textbf
		{\bibinfo {volume} {3}} (\bibinfo {year} {2007})}\BibitemShut {NoStop}%
	\bibitem [{\citenamefont {Yochelis}\ \emph
		{et~al.}(2015{\natexlab{a}})\citenamefont {Yochelis}, \citenamefont
		{Ebrahim}, \citenamefont {Millis}, \citenamefont {Cui}, \citenamefont
		{Kachar}, \citenamefont {Naoz},\ and\ \citenamefont
		{Gov}}]{yochelis2015self}%
	\BibitemOpen
	\bibfield  {author} {\bibinfo {author} {\bibfnamefont {A.}~\bibnamefont
			{Yochelis}}, \bibinfo {author} {\bibfnamefont {S.}~\bibnamefont {Ebrahim}},
		\bibinfo {author} {\bibfnamefont {B.}~\bibnamefont {Millis}}, \bibinfo
		{author} {\bibfnamefont {R.}~\bibnamefont {Cui}}, \bibinfo {author}
		{\bibfnamefont {B.}~\bibnamefont {Kachar}}, \bibinfo {author} {\bibfnamefont
			{M.}~\bibnamefont {Naoz}}, \ and\ \bibinfo {author} {\bibfnamefont {N.~S.}\
			\bibnamefont {Gov}},\ }\href@noop {} {\bibfield  {journal} {\bibinfo
			{journal} {Scientific Reports}\ }\textbf {\bibinfo {volume} {5}},\ \bibinfo
		{pages} {13521} (\bibinfo {year} {2015}{\natexlab{a}})}\BibitemShut {NoStop}%
	\bibitem [{\citenamefont {Ishihara}\ \emph {et~al.}(2007)\citenamefont
		{Ishihara}, \citenamefont {Otsuji},\ and\ \citenamefont
		{Mochizuki}}]{ishihara2007transient}%
	\BibitemOpen
	\bibfield  {author} {\bibinfo {author} {\bibfnamefont {S.}~\bibnamefont
			{Ishihara}}, \bibinfo {author} {\bibfnamefont {M.}~\bibnamefont {Otsuji}}, \
		and\ \bibinfo {author} {\bibfnamefont {A.}~\bibnamefont {Mochizuki}},\
	}\href@noop {} {\bibfield  {journal} {\bibinfo  {journal} {Physical Review
				E}\ }\textbf {\bibinfo {volume} {75}},\ \bibinfo {pages} {015203} (\bibinfo
		{year} {2007})}\BibitemShut {NoStop}%
	\bibitem [{\citenamefont {Morita}\ and\ \citenamefont
		{Ogawa}(2010)}]{morita2010stability}%
	\BibitemOpen
	\bibfield  {author} {\bibinfo {author} {\bibfnamefont {Y.}~\bibnamefont
			{Morita}}\ and\ \bibinfo {author} {\bibfnamefont {T.}~\bibnamefont {Ogawa}},\
	}\href@noop {} {\bibfield  {journal} {\bibinfo  {journal} {Nonlinearity}\
		}\textbf {\bibinfo {volume} {23}},\ \bibinfo {pages} {1387} (\bibinfo {year}
		{2010})}\BibitemShut {NoStop}%
	\bibitem [{\citenamefont {Chern}\ \emph {et~al.}(2018)\citenamefont {Chern},
		\citenamefont {Morita},\ and\ \citenamefont {Shieh}}]{chern2018asymptotic}%
	\BibitemOpen
	\bibfield  {author} {\bibinfo {author} {\bibfnamefont {J.-L.}\ \bibnamefont
			{Chern}}, \bibinfo {author} {\bibfnamefont {Y.}~\bibnamefont {Morita}}, \
		and\ \bibinfo {author} {\bibfnamefont {T.-T.}\ \bibnamefont {Shieh}},\
	}\href@noop {} {\bibfield  {journal} {\bibinfo  {journal} {Journal of
				Differential Equations}\ }\textbf {\bibinfo {volume} {264}},\ \bibinfo
		{pages} {550} (\bibinfo {year} {2018})}\BibitemShut {NoStop}%
	\bibitem [{\citenamefont {Kuwamura}\ \emph {et~al.}(2018)\citenamefont
		{Kuwamura}, \citenamefont {Seirin-Lee},\ and\ \citenamefont
		{Ei}}]{kuwamura2018dynamics}%
	\BibitemOpen
	\bibfield  {author} {\bibinfo {author} {\bibfnamefont {M.}~\bibnamefont
			{Kuwamura}}, \bibinfo {author} {\bibfnamefont {S.}~\bibnamefont
			{Seirin-Lee}}, \ and\ \bibinfo {author} {\bibfnamefont {S.-i.}\ \bibnamefont
			{Ei}},\ }\href@noop {} {\bibfield  {journal} {\bibinfo  {journal} {SIAM
				Journal on Applied Mathematics}\ }\textbf {\bibinfo {volume} {78}},\ \bibinfo
		{pages} {3238} (\bibinfo {year} {2018})}\BibitemShut {NoStop}%
	\bibitem [{\citenamefont {Ei}\ and\ \citenamefont {Tzeng}(2020)}]{ei2020spike}%
	\BibitemOpen
	\bibfield  {author} {\bibinfo {author} {\bibfnamefont {S.-I.}\ \bibnamefont
			{Ei}}\ and\ \bibinfo {author} {\bibfnamefont {S.-Y.}\ \bibnamefont {Tzeng}},\
	}\href@noop {} {\bibfield  {journal} {\bibinfo  {journal} {Discrete \&
				Continuous Dynamical Systems-A}\ ,\ \bibinfo {pages} {0}} (\bibinfo {year}
		{2020})}\BibitemShut {NoStop}%
	\bibitem [{\citenamefont {Sakamoto}(2013)}]{sakamoto2013hopf}%
	\BibitemOpen
	\bibfield  {author} {\bibinfo {author} {\bibfnamefont {T.~O.}\ \bibnamefont
			{Sakamoto}},\ }\href@noop {} {\bibfield  {journal} {\bibinfo  {journal}
			{Nonlinearity}\ }\textbf {\bibinfo {volume} {26}},\ \bibinfo {pages} {2027}
		(\bibinfo {year} {2013})}\BibitemShut {NoStop}%
	\bibitem [{\citenamefont {Yochelis}\ \emph {et~al.}(2016)\citenamefont
		{Yochelis}, \citenamefont {Bar-On},\ and\ \citenamefont
		{Gov}}]{yochelis2016reaction}%
	\BibitemOpen
	\bibfield  {author} {\bibinfo {author} {\bibfnamefont {A.}~\bibnamefont
			{Yochelis}}, \bibinfo {author} {\bibfnamefont {T.}~\bibnamefont {Bar-On}}, \
		and\ \bibinfo {author} {\bibfnamefont {N.~S.}\ \bibnamefont {Gov}},\
	}\href@noop {} {\bibfield  {journal} {\bibinfo  {journal} {Physica D}\
		}\textbf {\bibinfo {volume} {318}},\ \bibinfo {pages} {84} (\bibinfo {year}
		{2016})}\BibitemShut {NoStop}%
	\bibitem [{\citenamefont {Zmurchok}\ \emph {et~al.}(2017)\citenamefont
		{Zmurchok}, \citenamefont {Small}, \citenamefont {Ward},\ and\ \citenamefont
		{Edelstein-Keshet}}]{zmurchok2017application}%
	\BibitemOpen
	\bibfield  {author} {\bibinfo {author} {\bibfnamefont {C.}~\bibnamefont
			{Zmurchok}}, \bibinfo {author} {\bibfnamefont {T.}~\bibnamefont {Small}},
		\bibinfo {author} {\bibfnamefont {M.~J.}\ \bibnamefont {Ward}}, \ and\
		\bibinfo {author} {\bibfnamefont {L.}~\bibnamefont {Edelstein-Keshet}},\
	}\href@noop {} {\bibfield  {journal} {\bibinfo  {journal} {Bulletin of
				mathematical biology}\ }\textbf {\bibinfo {volume} {79}},\ \bibinfo {pages}
		{1923} (\bibinfo {year} {2017})}\BibitemShut {NoStop}%
	\bibitem [{\citenamefont {Goryachev}\ and\ \citenamefont
		{Pokhilko}(2008)}]{goryachev2008dynamics}%
	\BibitemOpen
	\bibfield  {author} {\bibinfo {author} {\bibfnamefont {A.~B.}\ \bibnamefont
			{Goryachev}}\ and\ \bibinfo {author} {\bibfnamefont {A.~V.}\ \bibnamefont
			{Pokhilko}},\ }\href@noop {} {\bibfield  {journal} {\bibinfo  {journal} {FEBS
				letters}\ }\textbf {\bibinfo {volume} {582}},\ \bibinfo {pages} {1437}
		(\bibinfo {year} {2008})}\BibitemShut {NoStop}%
	\bibitem [{\citenamefont {Champneys}\ \emph {et~al.}(2007)\citenamefont
		{Champneys}, \citenamefont {Kirk}, \citenamefont {Knobloch}, \citenamefont
		{Oldeman},\ and\ \citenamefont {Sneyd}}]{champneys2007shil}%
	\BibitemOpen
	\bibfield  {author} {\bibinfo {author} {\bibfnamefont {A.~R.}\ \bibnamefont
			{Champneys}}, \bibinfo {author} {\bibfnamefont {V.}~\bibnamefont {Kirk}},
		\bibinfo {author} {\bibfnamefont {E.}~\bibnamefont {Knobloch}}, \bibinfo
		{author} {\bibfnamefont {B.~E.}\ \bibnamefont {Oldeman}}, \ and\ \bibinfo
		{author} {\bibfnamefont {J.}~\bibnamefont {Sneyd}},\ }\href@noop {}
	{\bibfield  {journal} {\bibinfo  {journal} {SIAM Journal on Applied Dynamical
				Systems}\ }\textbf {\bibinfo {volume} {6}},\ \bibinfo {pages} {663} (\bibinfo
		{year} {2007})}\BibitemShut {NoStop}%
	\bibitem [{\citenamefont {Yochelis}\ \emph {et~al.}(2008)\citenamefont
		{Yochelis}, \citenamefont {Knobloch}, \citenamefont {Xie}, \citenamefont
		{Qu},\ and\ \citenamefont {Garfinkel}}]{yochelis2008generation}%
	\BibitemOpen
	\bibfield  {author} {\bibinfo {author} {\bibfnamefont {A.}~\bibnamefont
			{Yochelis}}, \bibinfo {author} {\bibfnamefont {E.}~\bibnamefont {Knobloch}},
		\bibinfo {author} {\bibfnamefont {Y.}~\bibnamefont {Xie}}, \bibinfo {author}
		{\bibfnamefont {Z.}~\bibnamefont {Qu}}, \ and\ \bibinfo {author}
		{\bibfnamefont {A.}~\bibnamefont {Garfinkel}},\ }\href@noop {} {\bibfield
		{journal} {\bibinfo  {journal} {Europhysics Letters}\ }\textbf {\bibinfo
			{volume} {83}},\ \bibinfo {pages} {64005} (\bibinfo {year}
		{2008})}\BibitemShut {NoStop}%
	\bibitem [{\citenamefont {Doedel}\ \emph {et~al.}(1997)\citenamefont {Doedel},
		\citenamefont {Champneys}, \citenamefont {Fairgrieve}, \citenamefont
		{Kuznetsov}, \citenamefont {Sandstede},\ and\ \citenamefont {Wang}}]{auto}%
	\BibitemOpen
	\bibfield  {author} {\bibinfo {author} {\bibfnamefont {E.}~\bibnamefont
			{Doedel}}, \bibinfo {author} {\bibfnamefont {A.~R.}\ \bibnamefont
			{Champneys}}, \bibinfo {author} {\bibfnamefont {T.~F.}\ \bibnamefont
			{Fairgrieve}}, \bibinfo {author} {\bibfnamefont {Y.~A.}\ \bibnamefont
			{Kuznetsov}}, \bibinfo {author} {\bibfnamefont {B.}~\bibnamefont
			{Sandstede}}, \ and\ \bibinfo {author} {\bibfnamefont {X.}~\bibnamefont
			{Wang}},\ }\href@noop {} {\enquote {\bibinfo {title} {{AUTO}: Continuation
				and bifurcation software for ordinary differential equations (with
				{HomCont})},}\ } (\bibinfo {year} {1997}),\ \bibinfo {note}
	{http://cmvl.cs.concordia.ca/auto/}\BibitemShut {NoStop}%
	\bibitem [{\citenamefont {Sherratt}(2012)}]{sherratt2012numerical}%
	\BibitemOpen
	\bibfield  {author} {\bibinfo {author} {\bibfnamefont {J.~A.}\ \bibnamefont
			{Sherratt}},\ }\href@noop {} {\bibfield  {journal} {\bibinfo  {journal}
			{Applied Mathematics and Computation}\ }\textbf {\bibinfo {volume} {218}},\
		\bibinfo {pages} {4684} (\bibinfo {year} {2012})}\BibitemShut {NoStop}%
	\bibitem [{\citenamefont {Uecker}\ \emph {et~al.}(2014)\citenamefont {Uecker},
		\citenamefont {Wetzel},\ and\ \citenamefont
		{Rademacher}}]{uecker2014pde2path}%
	\BibitemOpen
	\bibfield  {author} {\bibinfo {author} {\bibfnamefont {H.}~\bibnamefont
			{Uecker}}, \bibinfo {author} {\bibfnamefont {D.}~\bibnamefont {Wetzel}}, \
		and\ \bibinfo {author} {\bibfnamefont {J.~D.}\ \bibnamefont {Rademacher}},\
	}\href@noop {} {\bibfield  {journal} {\bibinfo  {journal} {Numerical
				Mathematics: Theory, Methods and Applications}\ }\textbf {\bibinfo {volume}
			{7}},\ \bibinfo {pages} {58} (\bibinfo {year} {2014})},\ \bibinfo {note} {see
		also www.staff.uni-oldenburg.de/hannes.uecker/pde2path/}\BibitemShut
	{NoStop}%
	\bibitem [{\citenamefont {Bindel}\ \emph {et~al.}(2014)\citenamefont {Bindel},
		\citenamefont {Friedman}, \citenamefont {Govaerts}, \citenamefont {Hughes},\
		and\ \citenamefont {Kuznetsov}}]{bindel14}%
	\BibitemOpen
	\bibfield  {author} {\bibinfo {author} {\bibfnamefont {D.}~\bibnamefont
			{Bindel}}, \bibinfo {author} {\bibfnamefont {M.}~\bibnamefont {Friedman}},
		\bibinfo {author} {\bibfnamefont {W.}~\bibnamefont {Govaerts}}, \bibinfo
		{author} {\bibfnamefont {J.}~\bibnamefont {Hughes}}, \ and\ \bibinfo {author}
		{\bibfnamefont {Y.}~\bibnamefont {Kuznetsov}},\ }\href@noop {} {\bibfield
		{journal} {\bibinfo  {journal} {J. Comput. Appl. Math.}\ }\textbf {\bibinfo
			{volume} {261}},\ \bibinfo {pages} {232} (\bibinfo {year}
		{2014})}\BibitemShut {NoStop}%
	\bibitem [{\citenamefont {Schilder}\ and\ \citenamefont {Dankowicz}()}]{coco}%
	\BibitemOpen
	\bibfield  {author} {\bibinfo {author} {\bibfnamefont {F.}~\bibnamefont
			{Schilder}}\ and\ \bibinfo {author} {\bibfnamefont {H.}~\bibnamefont
			{Dankowicz}},\ }\href@noop {} {\enquote {\bibinfo {title} {coco},}\ }\bibinfo
	{note} {Http://sourceforge.net/projects/cocotools/}\BibitemShut {NoStop}%
	\bibitem [{\citenamefont {Yochelis}\ \emph
		{et~al.}(2015{\natexlab{b}})\citenamefont {Yochelis}, \citenamefont
		{Knobloch},\ and\ \citenamefont {K{\"o}pf}}]{yochelis2015origin}%
	\BibitemOpen
	\bibfield  {author} {\bibinfo {author} {\bibfnamefont {A.}~\bibnamefont
			{Yochelis}}, \bibinfo {author} {\bibfnamefont {E.}~\bibnamefont {Knobloch}},
		\ and\ \bibinfo {author} {\bibfnamefont {M.~H.}\ \bibnamefont {K{\"o}pf}},\
	}\href@noop {} {\bibfield  {journal} {\bibinfo  {journal} {Physical Review
				E}\ }\textbf {\bibinfo {volume} {91}},\ \bibinfo {pages} {032924} (\bibinfo
		{year} {2015}{\natexlab{b}})}\BibitemShut {NoStop}%
	\bibitem [{\citenamefont {Yochelis}\ \emph {et~al.}(2020)\citenamefont
		{Yochelis}, \citenamefont {Beta},\ and\ \citenamefont
		{Gov}}]{yochelis2020excitable}%
	\BibitemOpen
	\bibfield  {author} {\bibinfo {author} {\bibfnamefont {A.}~\bibnamefont
			{Yochelis}}, \bibinfo {author} {\bibfnamefont {C.}~\bibnamefont {Beta}}, \
		and\ \bibinfo {author} {\bibfnamefont {N.~S.}\ \bibnamefont {Gov}},\
	}\href@noop {} {\bibfield  {journal} {\bibinfo  {journal} {Physical Review
				E}\ }\textbf {\bibinfo {volume} {101}},\ \bibinfo {pages} {022213} (\bibinfo
		{year} {2020})}\BibitemShut {NoStop}%
	\bibitem [{\citenamefont {Zabusky}\ and\ \citenamefont
		{Kruskal}(1965)}]{zabusky1965interaction}%
	\BibitemOpen
	\bibfield  {author} {\bibinfo {author} {\bibfnamefont {N.~J.}\ \bibnamefont
			{Zabusky}}\ and\ \bibinfo {author} {\bibfnamefont {M.~D.}\ \bibnamefont
			{Kruskal}},\ }\href@noop {} {\bibfield  {journal} {\bibinfo  {journal}
			{Physical Review Letters}\ }\textbf {\bibinfo {volume} {15}},\ \bibinfo
		{pages} {240} (\bibinfo {year} {1965})}\BibitemShut {NoStop}%
	\bibitem [{\citenamefont {Scott}\ \emph {et~al.}(1973)\citenamefont {Scott},
		\citenamefont {Chu},\ and\ \citenamefont {McLaughlin}}]{scott1973soliton}%
	\BibitemOpen
	\bibfield  {author} {\bibinfo {author} {\bibfnamefont {A.~C.}\ \bibnamefont
			{Scott}}, \bibinfo {author} {\bibfnamefont {F.}~\bibnamefont {Chu}}, \ and\
		\bibinfo {author} {\bibfnamefont {D.~W.}\ \bibnamefont {McLaughlin}},\
	}\href@noop {} {\bibfield  {journal} {\bibinfo  {journal} {Proceedings of the
				IEEE}\ }\textbf {\bibinfo {volume} {61}},\ \bibinfo {pages} {1443} (\bibinfo
		{year} {1973})}\BibitemShut {NoStop}%
	\bibitem [{\citenamefont {Scott}(1975)}]{scott1975electrophysics}%
	\BibitemOpen
	\bibfield  {author} {\bibinfo {author} {\bibfnamefont {A.~C.}\ \bibnamefont
			{Scott}},\ }\href@noop {} {\bibfield  {journal} {\bibinfo  {journal} {Rev.
				Mod. Phys.}\ }\textbf {\bibinfo {volume} {47}},\ \bibinfo {pages} {487}
		(\bibinfo {year} {1975})}\BibitemShut {NoStop}%
	\bibitem [{\citenamefont {Knobloch}(2015)}]{knobloch2015spatial}%
	\BibitemOpen
	\bibfield  {author} {\bibinfo {author} {\bibfnamefont {E.}~\bibnamefont
			{Knobloch}},\ }\href@noop {} {\bibfield  {journal} {\bibinfo  {journal}
			{Annu. Rev. Condens. Matter Phys.}\ }\textbf {\bibinfo {volume} {6}},\
		\bibinfo {pages} {325} (\bibinfo {year} {2015})}\BibitemShut {NoStop}%
	\bibitem [{\citenamefont {Ablowitz}\ and\ \citenamefont
		{Baldwin}(2012)}]{ablowitz2012nonlinear}%
	\BibitemOpen
	\bibfield  {author} {\bibinfo {author} {\bibfnamefont {M.~J.}\ \bibnamefont
			{Ablowitz}}\ and\ \bibinfo {author} {\bibfnamefont {D.~E.}\ \bibnamefont
			{Baldwin}},\ }\href@noop {} {\bibfield  {journal} {\bibinfo  {journal}
			{Physical Review E}\ }\textbf {\bibinfo {volume} {86}},\ \bibinfo {pages}
		{036305} (\bibinfo {year} {2012})}\BibitemShut {NoStop}%
	\bibitem [{\citenamefont {Santiago-Rosanne}\ \emph {et~al.}(1997)\citenamefont
		{Santiago-Rosanne}, \citenamefont {Vignes-Adler},\ and\ \citenamefont
		{Velarde}}]{santiago1997dissolution}%
	\BibitemOpen
	\bibfield  {author} {\bibinfo {author} {\bibfnamefont {M.}~\bibnamefont
			{Santiago-Rosanne}}, \bibinfo {author} {\bibfnamefont {M.}~\bibnamefont
			{Vignes-Adler}}, \ and\ \bibinfo {author} {\bibfnamefont {M.~G.}\
			\bibnamefont {Velarde}},\ }\href@noop {} {\bibfield  {journal} {\bibinfo
			{journal} {Journal of Colloid and Interface Science}\ }\textbf {\bibinfo
			{volume} {191}},\ \bibinfo {pages} {65} (\bibinfo {year} {1997})}\BibitemShut
	{NoStop}%
	\bibitem [{\citenamefont {Alonso}\ \emph {et~al.}(2016)\citenamefont {Alonso},
		\citenamefont {B{\"a}r},\ and\ \citenamefont
		{Echebarria}}]{alonso2016nonlinear}%
	\BibitemOpen
	\bibfield  {author} {\bibinfo {author} {\bibfnamefont {S.}~\bibnamefont
			{Alonso}}, \bibinfo {author} {\bibfnamefont {M.}~\bibnamefont {B{\"a}r}}, \
		and\ \bibinfo {author} {\bibfnamefont {B.}~\bibnamefont {Echebarria}},\
	}\href@noop {} {\bibfield  {journal} {\bibinfo  {journal} {Reports on
				Progress in Physics}\ }\textbf {\bibinfo {volume} {79}},\ \bibinfo {pages}
		{096601} (\bibinfo {year} {2016})}\BibitemShut {NoStop}%
	\bibitem [{\citenamefont {Petrov}\ \emph {et~al.}(1994)\citenamefont {Petrov},
		\citenamefont {Scott},\ and\ \citenamefont
		{Showalter}}]{petrov1994excitability}%
	\BibitemOpen
	\bibfield  {author} {\bibinfo {author} {\bibfnamefont {V.}~\bibnamefont
			{Petrov}}, \bibinfo {author} {\bibfnamefont {S.~K.}\ \bibnamefont {Scott}}, \
		and\ \bibinfo {author} {\bibfnamefont {K.}~\bibnamefont {Showalter}},\
	}\href@noop {} {\bibfield  {journal} {\bibinfo  {journal} {Philosophical
				Transactions of the Royal Society of London. Series A: Physical and
				Engineering Sciences}\ }\textbf {\bibinfo {volume} {347}},\ \bibinfo {pages}
		{631} (\bibinfo {year} {1994})}\BibitemShut {NoStop}%
	\bibitem [{\citenamefont {Argentina}\ \emph {et~al.}(1997)\citenamefont
		{Argentina}, \citenamefont {Coullet},\ and\ \citenamefont
		{Mahadevan}}]{argentina_colliding_1997}%
	\BibitemOpen
	\bibfield  {author} {\bibinfo {author} {\bibfnamefont {M.}~\bibnamefont
			{Argentina}}, \bibinfo {author} {\bibfnamefont {P.}~\bibnamefont {Coullet}},
		\ and\ \bibinfo {author} {\bibfnamefont {L.}~\bibnamefont {Mahadevan}},\
	}\href {http://link.aps.org/doi/10.1103/PhysRevLett.79.2803} {\bibfield
		{journal} {\bibinfo  {journal} {Physical Review Letters}\ }\textbf {\bibinfo
			{volume} {79}},\ \bibinfo {pages} {2803} (\bibinfo {year}
		{1997})}\BibitemShut {NoStop}%
	\bibitem [{\citenamefont {Nishiura}\ \emph
		{et~al.}(2003{\natexlab{a}})\citenamefont {Nishiura}, \citenamefont
		{Teramoto},\ and\ \citenamefont {Ueda}}]{nishiura_scattering_2003}%
	\BibitemOpen
	\bibfield  {author} {\bibinfo {author} {\bibfnamefont {Y.}~\bibnamefont
			{Nishiura}}, \bibinfo {author} {\bibfnamefont {T.}~\bibnamefont {Teramoto}},
		\ and\ \bibinfo {author} {\bibfnamefont {K.-I.}\ \bibnamefont {Ueda}},\
	}\href {https://link.aps.org/doi/10.1103/PhysRevE.67.056210} {\bibfield
		{journal} {\bibinfo  {journal} {Physical Review E}\ }\textbf {\bibinfo
			{volume} {67}},\ \bibinfo {pages} {056210} (\bibinfo {year}
		{2003}{\natexlab{a}})}\BibitemShut {NoStop}%
	\bibitem [{\citenamefont {Nishiura}\ \emph
		{et~al.}(2003{\natexlab{b}})\citenamefont {Nishiura}, \citenamefont
		{Teramoto},\ and\ \citenamefont {Ueda}}]{nishiura_dynamic_2003}%
	\BibitemOpen
	\bibfield  {author} {\bibinfo {author} {\bibfnamefont {Y.}~\bibnamefont
			{Nishiura}}, \bibinfo {author} {\bibfnamefont {T.}~\bibnamefont {Teramoto}},
		\ and\ \bibinfo {author} {\bibfnamefont {K.-I.}\ \bibnamefont {Ueda}},\
	}\href {https://aip.scitation.org/doi/abs/10.1063/1.1592131} {\bibfield
		{journal} {\bibinfo  {journal} {Chaos}\ }\textbf {\bibinfo {volume} {13}},\
		\bibinfo {pages} {962} (\bibinfo {year} {2003}{\natexlab{b}})}\BibitemShut
	{NoStop}%
	\bibitem [{\citenamefont {Whitelam}\ \emph {et~al.}(2009)\citenamefont
		{Whitelam}, \citenamefont {Bretschneider},\ and\ \citenamefont
		{Burroughs}}]{whitelam2009transformation}%
	\BibitemOpen
	\bibfield  {author} {\bibinfo {author} {\bibfnamefont {S.}~\bibnamefont
			{Whitelam}}, \bibinfo {author} {\bibfnamefont {T.}~\bibnamefont
			{Bretschneider}}, \ and\ \bibinfo {author} {\bibfnamefont {N.~J.}\
			\bibnamefont {Burroughs}},\ }\href@noop {} {\bibfield  {journal} {\bibinfo
			{journal} {Physical Review Letters}\ }\textbf {\bibinfo {volume} {102}},\
		\bibinfo {pages} {198103} (\bibinfo {year} {2009})}\BibitemShut {NoStop}%
	\bibitem [{\citenamefont {Dreher}\ \emph
		{et~al.}(2014{\natexlab{b}})\citenamefont {Dreher}, \citenamefont {Aranson},\
		and\ \citenamefont {Kruse}}]{dreher2014spiral}%
	\BibitemOpen
	\bibfield  {author} {\bibinfo {author} {\bibfnamefont {A.}~\bibnamefont
			{Dreher}}, \bibinfo {author} {\bibfnamefont {I.~S.}\ \bibnamefont {Aranson}},
		\ and\ \bibinfo {author} {\bibfnamefont {K.}~\bibnamefont {Kruse}},\
	}\href@noop {} {\bibfield  {journal} {\bibinfo  {journal} {New Journal of
				Physics}\ }\textbf {\bibinfo {volume} {16}},\ \bibinfo {pages} {055007}
		(\bibinfo {year} {2014}{\natexlab{b}})}\BibitemShut {NoStop}%
	\bibitem [{\citenamefont {Alonso}\ \emph {et~al.}(2018)\citenamefont {Alonso},
		\citenamefont {Stange},\ and\ \citenamefont {Beta}}]{alonso2018modeling}%
	\BibitemOpen
	\bibfield  {author} {\bibinfo {author} {\bibfnamefont {S.}~\bibnamefont
			{Alonso}}, \bibinfo {author} {\bibfnamefont {M.}~\bibnamefont {Stange}}, \
		and\ \bibinfo {author} {\bibfnamefont {C.}~\bibnamefont {Beta}},\ }\href@noop
	{} {\bibfield  {journal} {\bibinfo  {journal} {PloS One}\ }\textbf {\bibinfo
			{volume} {13}},\ \bibinfo {pages} {e0201977} (\bibinfo {year}
		{2018})}\BibitemShut {NoStop}%
	\bibitem [{\citenamefont {Krischer}\ and\ \citenamefont
		{Mikhailov}(1994)}]{krischer1994bifurcation}%
	\BibitemOpen
	\bibfield  {author} {\bibinfo {author} {\bibfnamefont {K.}~\bibnamefont
			{Krischer}}\ and\ \bibinfo {author} {\bibfnamefont {A.}~\bibnamefont
			{Mikhailov}},\ }\href@noop {} {\bibfield  {journal} {\bibinfo  {journal}
			{Physical Review Letters}\ }\textbf {\bibinfo {volume} {73}},\ \bibinfo
		{pages} {3165} (\bibinfo {year} {1994})}\BibitemShut {NoStop}%
	\bibitem [{\citenamefont {B{\"a}r}\ \emph {et~al.}(1994)\citenamefont
		{B{\"a}r}, \citenamefont {Hildebrand}, \citenamefont {Eiswirth},
		\citenamefont {Falcke}, \citenamefont {Engel},\ and\ \citenamefont
		{Neufeld}}]{bar1994chemical}%
	\BibitemOpen
	\bibfield  {author} {\bibinfo {author} {\bibfnamefont {M.}~\bibnamefont
			{B{\"a}r}}, \bibinfo {author} {\bibfnamefont {M.}~\bibnamefont {Hildebrand}},
		\bibinfo {author} {\bibfnamefont {M.}~\bibnamefont {Eiswirth}}, \bibinfo
		{author} {\bibfnamefont {M.}~\bibnamefont {Falcke}}, \bibinfo {author}
		{\bibfnamefont {H.}~\bibnamefont {Engel}}, \ and\ \bibinfo {author}
		{\bibfnamefont {M.}~\bibnamefont {Neufeld}},\ }\href@noop {} {\bibfield
		{journal} {\bibinfo  {journal} {Chaos}\ }\textbf {\bibinfo {volume} {4}},\
		\bibinfo {pages} {499} (\bibinfo {year} {1994})}\BibitemShut {NoStop}%
	\bibitem [{\citenamefont {Mimura}\ and\ \citenamefont
		{Kawaguchi}(1998)}]{mimura1998collision}%
	\BibitemOpen
	\bibfield  {author} {\bibinfo {author} {\bibfnamefont {M.}~\bibnamefont
			{Mimura}}\ and\ \bibinfo {author} {\bibfnamefont {S.}~\bibnamefont
			{Kawaguchi}},\ }\href@noop {} {\bibfield  {journal} {\bibinfo  {journal}
			{SIAM Journal on Applied Mathematics}\ }\textbf {\bibinfo {volume} {59}},\
		\bibinfo {pages} {920} (\bibinfo {year} {1998})}\BibitemShut {NoStop}%
	\bibitem [{\citenamefont {Coombes}\ and\ \citenamefont
		{Owen}(2007)}]{coombes2007exotic}%
	\BibitemOpen
	\bibfield  {author} {\bibinfo {author} {\bibfnamefont {S.}~\bibnamefont
			{Coombes}}\ and\ \bibinfo {author} {\bibfnamefont {M.}~\bibnamefont {Owen}},\
	}in\ \href@noop {} {\emph {\bibinfo {booktitle} {Fluids and Waves: Recent
				Trends in Applied Analysis: Research Conference, May 11-13, 2006, the
				Universtiy of Memphis, Memphis, TN}}},\ Vol.\ \bibinfo {volume} {440}\
	(\bibinfo {organization} {American Mathematical Soc.},\ \bibinfo {year}
	{2007})\ p.\ \bibinfo {pages} {123}\BibitemShut {NoStop}%
	\bibitem [{\citenamefont {Tsyganov}\ \emph {et~al.}(2003)\citenamefont
		{Tsyganov}, \citenamefont {Brindley}, \citenamefont {Holden},\ and\
		\citenamefont {Biktashev}}]{tsyganov2003quasisoliton}%
	\BibitemOpen
	\bibfield  {author} {\bibinfo {author} {\bibfnamefont {M.}~\bibnamefont
			{Tsyganov}}, \bibinfo {author} {\bibfnamefont {J.}~\bibnamefont {Brindley}},
		\bibinfo {author} {\bibfnamefont {A.}~\bibnamefont {Holden}}, \ and\ \bibinfo
		{author} {\bibfnamefont {V.}~\bibnamefont {Biktashev}},\ }\href@noop {}
	{\bibfield  {journal} {\bibinfo  {journal} {Physical Review Letters}\
		}\textbf {\bibinfo {volume} {91}},\ \bibinfo {pages} {218102} (\bibinfo
		{year} {2003})}\BibitemShut {NoStop}%
	\bibitem [{\citenamefont {B{\"a}r}\ \emph {et~al.}(1992)\citenamefont
		{B{\"a}r}, \citenamefont {Eiswirth}, \citenamefont {Rotermund},\ and\
		\citenamefont {Ertl}}]{bar1992solitary}%
	\BibitemOpen
	\bibfield  {author} {\bibinfo {author} {\bibfnamefont {M.}~\bibnamefont
			{B{\"a}r}}, \bibinfo {author} {\bibfnamefont {M.}~\bibnamefont {Eiswirth}},
		\bibinfo {author} {\bibfnamefont {H.-H.}\ \bibnamefont {Rotermund}}, \ and\
		\bibinfo {author} {\bibfnamefont {G.}~\bibnamefont {Ertl}},\ }\href@noop {}
	{\bibfield  {journal} {\bibinfo  {journal} {Physical Review Letters}\
		}\textbf {\bibinfo {volume} {69}},\ \bibinfo {pages} {945} (\bibinfo {year}
		{1992})}\BibitemShut {NoStop}%
	\bibitem [{\citenamefont {Nishiura}\ \emph {et~al.}(2007)\citenamefont
		{Nishiura}, \citenamefont {Teramoto}, \citenamefont {Yuan},\ and\
		\citenamefont {Ueda}}]{nishiura2007dynamics}%
	\BibitemOpen
	\bibfield  {author} {\bibinfo {author} {\bibfnamefont {Y.}~\bibnamefont
			{Nishiura}}, \bibinfo {author} {\bibfnamefont {T.}~\bibnamefont {Teramoto}},
		\bibinfo {author} {\bibfnamefont {X.}~\bibnamefont {Yuan}}, \ and\ \bibinfo
		{author} {\bibfnamefont {K.-I.}\ \bibnamefont {Ueda}},\ }\href@noop {}
	{\bibfield  {journal} {\bibinfo  {journal} {Chaos}\ }\textbf {\bibinfo
			{volume} {17}},\ \bibinfo {pages} {037104} (\bibinfo {year}
		{2007})}\BibitemShut {NoStop}%
	\bibitem [{\citenamefont {Yuan}\ \emph {et~al.}(2007)\citenamefont {Yuan},
		\citenamefont {Teramoto},\ and\ \citenamefont
		{Nishiura}}]{yuan2007heterogeneity}%
	\BibitemOpen
	\bibfield  {author} {\bibinfo {author} {\bibfnamefont {X.}~\bibnamefont
			{Yuan}}, \bibinfo {author} {\bibfnamefont {T.}~\bibnamefont {Teramoto}}, \
		and\ \bibinfo {author} {\bibfnamefont {Y.}~\bibnamefont {Nishiura}},\
	}\href@noop {} {\bibfield  {journal} {\bibinfo  {journal} {Physical Review
				E}\ }\textbf {\bibinfo {volume} {75}},\ \bibinfo {pages} {036220} (\bibinfo
		{year} {2007})}\BibitemShut {NoStop}%
	\bibitem [{\citenamefont {Argentina}\ \emph {et~al.}(2000)\citenamefont
		{Argentina}, \citenamefont {Coullet},\ and\ \citenamefont
		{Krinsky}}]{argentina2000head}%
	\BibitemOpen
	\bibfield  {author} {\bibinfo {author} {\bibfnamefont {M.}~\bibnamefont
			{Argentina}}, \bibinfo {author} {\bibfnamefont {P.}~\bibnamefont {Coullet}},
		\ and\ \bibinfo {author} {\bibfnamefont {V.}~\bibnamefont {Krinsky}},\
	}\href@noop {} {\bibfield  {journal} {\bibinfo  {journal} {Journal of
				Theoretical Biology}\ }\textbf {\bibinfo {volume} {205}},\ \bibinfo {pages}
		{47} (\bibinfo {year} {2000})}\BibitemShut {NoStop}%
	\bibitem [{\citenamefont {Bordyugov}\ and\ \citenamefont
		{Engel}(2008)}]{bordyugov2008anomalous}%
	\BibitemOpen
	\bibfield  {author} {\bibinfo {author} {\bibfnamefont {G.}~\bibnamefont
			{Bordyugov}}\ and\ \bibinfo {author} {\bibfnamefont {H.}~\bibnamefont
			{Engel}},\ }\href@noop {} {\bibfield  {journal} {\bibinfo  {journal} {Chaos}\
		}\textbf {\bibinfo {volume} {18}},\ \bibinfo {pages} {026104} (\bibinfo
		{year} {2008})}\BibitemShut {NoStop}%
	\bibitem [{\citenamefont {Argentina}\ \emph {et~al.}(2004)\citenamefont
		{Argentina}, \citenamefont {Rudzick},\ and\ \citenamefont
		{Velarde}}]{argentina_back-firing_2004}%
	\BibitemOpen
	\bibfield  {author} {\bibinfo {author} {\bibfnamefont {M.}~\bibnamefont
			{Argentina}}, \bibinfo {author} {\bibfnamefont {O.}~\bibnamefont {Rudzick}},
		\ and\ \bibinfo {author} {\bibfnamefont {M.~G.}\ \bibnamefont {Velarde}},\
	}\href {https://aip.scitation.org/doi/abs/10.1063/1.1784911} {\bibfield
		{journal} {\bibinfo  {journal} {Chaos}\ }\textbf {\bibinfo {volume} {14}},\
		\bibinfo {pages} {777} (\bibinfo {year} {2004})}\BibitemShut {NoStop}%
	\bibitem [{\citenamefont {Zimmermann}\ \emph {et~al.}(1997)\citenamefont
		{Zimmermann}, \citenamefont {Firle}, \citenamefont {Natiello}, \citenamefont
		{Hildebrand}, \citenamefont {Eiswirth}, \citenamefont {Bär}, \citenamefont
		{Bangia},\ and\ \citenamefont {Kevrekidis}}]{zimmermann_pulse_1997}%
	\BibitemOpen
	\bibfield  {author} {\bibinfo {author} {\bibfnamefont {M.~G.}\ \bibnamefont
			{Zimmermann}}, \bibinfo {author} {\bibfnamefont {S.~O.}\ \bibnamefont
			{Firle}}, \bibinfo {author} {\bibfnamefont {M.~A.}\ \bibnamefont {Natiello}},
		\bibinfo {author} {\bibfnamefont {M.}~\bibnamefont {Hildebrand}}, \bibinfo
		{author} {\bibfnamefont {M.}~\bibnamefont {Eiswirth}}, \bibinfo {author}
		{\bibfnamefont {M.}~\bibnamefont {Bär}}, \bibinfo {author} {\bibfnamefont
			{A.~K.}\ \bibnamefont {Bangia}}, \ and\ \bibinfo {author} {\bibfnamefont
			{I.~G.}\ \bibnamefont {Kevrekidis}},\ }\href
	{https://linkinghub.elsevier.com/retrieve/pii/S0167278997001127} {\bibfield
		{journal} {\bibinfo  {journal} {Physica D: Nonlinear Phenomena}\ }\textbf
		{\bibinfo {volume} {110}},\ \bibinfo {pages} {92} (\bibinfo {year}
		{1997})}\BibitemShut {NoStop}%
	\bibitem [{\citenamefont {Nishiura}\ and\ \citenamefont
		{Ueyama}(2001)}]{nishiura_spatio-temporal_2001}%
	\BibitemOpen
	\bibfield  {author} {\bibinfo {author} {\bibfnamefont {Y.}~\bibnamefont
			{Nishiura}}\ and\ \bibinfo {author} {\bibfnamefont {D.}~\bibnamefont
			{Ueyama}},\ }\href
	{http://www.sciencedirect.com/science/article/pii/S0167278900002141}
	{\bibfield  {journal} {\bibinfo  {journal} {Physica D: Nonlinear Phenomena}\
		}\textbf {\bibinfo {volume} {150}},\ \bibinfo {pages} {137} (\bibinfo {year}
		{2001})}\BibitemShut {NoStop}%
	\bibitem [{\citenamefont {Bauer}\ \emph {et~al.}(2015)\citenamefont {Bauer},
		\citenamefont {Bonnefont},\ and\ \citenamefont
		{Krischer}}]{bauer_dissipative_2015}%
	\BibitemOpen
	\bibfield  {author} {\bibinfo {author} {\bibfnamefont {P.~R.}\ \bibnamefont
			{Bauer}}, \bibinfo {author} {\bibfnamefont {A.}~\bibnamefont {Bonnefont}}, \
		and\ \bibinfo {author} {\bibfnamefont {K.}~\bibnamefont {Krischer}},\ }\href
	{https://www.nature.com/articles/srep16312} {\bibfield  {journal} {\bibinfo
			{journal} {Scientific Reports}\ }\textbf {\bibinfo {volume} {5}},\ \bibinfo
		{pages} {1} (\bibinfo {year} {2015})}\BibitemShut {NoStop}%
	\bibitem [{\citenamefont {Ge{\ss}ele}\ \emph {et~al.}(2020)\citenamefont
		{Ge{\ss}ele}, \citenamefont {Halatek}, \citenamefont {W{\"u}rthner},\ and\
		\citenamefont {Frey}}]{gessele2020geometric}%
	\BibitemOpen
	\bibfield  {author} {\bibinfo {author} {\bibfnamefont {R.}~\bibnamefont
			{Ge{\ss}ele}}, \bibinfo {author} {\bibfnamefont {J.}~\bibnamefont {Halatek}},
		\bibinfo {author} {\bibfnamefont {L.}~\bibnamefont {W{\"u}rthner}}, \ and\
		\bibinfo {author} {\bibfnamefont {E.}~\bibnamefont {Frey}},\ }\href@noop {}
	{\bibfield  {journal} {\bibinfo  {journal} {Nature communications}\ }\textbf
		{\bibinfo {volume} {11}},\ \bibinfo {pages} {1} (\bibinfo {year}
		{2020})}\BibitemShut {NoStop}%
	\bibitem [{\citenamefont {Siton-Mendelson}\ and\ \citenamefont
		{Bernheim-Groswasser}(2016)}]{siton2016toward}%
	\BibitemOpen
	\bibfield  {author} {\bibinfo {author} {\bibfnamefont {O.}~\bibnamefont
			{Siton-Mendelson}}\ and\ \bibinfo {author} {\bibfnamefont {A.}~\bibnamefont
			{Bernheim-Groswasser}},\ }\href@noop {} {\bibfield  {journal} {\bibinfo
			{journal} {Cell adhesion \& migration}\ }\textbf {\bibinfo {volume} {10}},\
		\bibinfo {pages} {461} (\bibinfo {year} {2016})}\BibitemShut {NoStop}%
	\bibitem [{\citenamefont {Schwille}\ \emph {et~al.}(2018)\citenamefont
		{Schwille}, \citenamefont {Spatz}, \citenamefont {Landfester}, \citenamefont
		{Bodenschatz}, \citenamefont {Herminghaus}, \citenamefont {Sourjik},
		\citenamefont {Erb}, \citenamefont {Bastiaens}, \citenamefont {Lipowsky},
		\citenamefont {Hyman}, \citenamefont {Dabrock}, \citenamefont {Baret},
		\citenamefont {Vidakovic-Koch}, \citenamefont {Bieling}, \citenamefont
		{Dimova}, \citenamefont {Mutschler}, \citenamefont {Robinson}, \citenamefont
		{Tang}, \citenamefont {Wegner},\ and\ \citenamefont
		{Sundmacher}}]{schwille_maxsynbio:_2018}%
	\BibitemOpen
	\bibfield  {author} {\bibinfo {author} {\bibfnamefont {P.}~\bibnamefont
			{Schwille}}, \bibinfo {author} {\bibfnamefont {J.}~\bibnamefont {Spatz}},
		\bibinfo {author} {\bibfnamefont {K.}~\bibnamefont {Landfester}}, \bibinfo
		{author} {\bibfnamefont {E.}~\bibnamefont {Bodenschatz}}, \bibinfo {author}
		{\bibfnamefont {S.}~\bibnamefont {Herminghaus}}, \bibinfo {author}
		{\bibfnamefont {V.}~\bibnamefont {Sourjik}}, \bibinfo {author} {\bibfnamefont
			{T.~J.}\ \bibnamefont {Erb}}, \bibinfo {author} {\bibfnamefont
			{P.}~\bibnamefont {Bastiaens}}, \bibinfo {author} {\bibfnamefont
			{R.}~\bibnamefont {Lipowsky}}, \bibinfo {author} {\bibfnamefont
			{A.}~\bibnamefont {Hyman}}, \bibinfo {author} {\bibfnamefont
			{P.}~\bibnamefont {Dabrock}}, \bibinfo {author} {\bibfnamefont {J.-C.}\
			\bibnamefont {Baret}}, \bibinfo {author} {\bibfnamefont {T.}~\bibnamefont
			{Vidakovic-Koch}}, \bibinfo {author} {\bibfnamefont {P.}~\bibnamefont
			{Bieling}}, \bibinfo {author} {\bibfnamefont {R.}~\bibnamefont {Dimova}},
		\bibinfo {author} {\bibfnamefont {H.}~\bibnamefont {Mutschler}}, \bibinfo
		{author} {\bibfnamefont {T.}~\bibnamefont {Robinson}}, \bibinfo {author}
		{\bibfnamefont {T.-Y.~D.}\ \bibnamefont {Tang}}, \bibinfo {author}
		{\bibfnamefont {S.}~\bibnamefont {Wegner}}, \ and\ \bibinfo {author}
		{\bibfnamefont {K.}~\bibnamefont {Sundmacher}},\ }\href
	{http://doi.wiley.com/10.1002/anie.201802288} {\bibfield  {journal} {\bibinfo
			{journal} {Angewandte Chemie International Edition}\ }\textbf {\bibinfo
			{volume} {57}},\ \bibinfo {pages} {13382} (\bibinfo {year}
		{2018})}\BibitemShut {NoStop}%
\end{thebibliography}
\end{document}